\documentclass{pasa}%
\usepackage{fourier}
\usepackage{xspace}

\usepackage{tablefootnote}
\usepackage{graphicx, color}
\usepackage[space]{grffile}
\usepackage{latexsym}
\usepackage{amsfonts,amsmath,amssymb}
\usepackage{hyperref}
\usepackage{fancyref}
\usepackage{subfig}
\usepackage{balance}

\usepackage{comment}
\usepackage{breakurl}
\usepackage[font=small,labelfont=bf,tableposition=top]{caption}

\newcommand{\lcdm}{{\ensuremath{\Lambda\mathrm{CDM}}}\xspace}

\newcommand{\ee}{\end{equation}}
\newcommand{\ba}{\begin{eqnarray}}

\usepackage{multirow}

\newcommand{\PhilTrans}{{\it Philosophical Transactions of the Royal Society,
London}}

\newcommand{\mnras}{Monthly Notices of the Royal Astronomical Society}
\newcommand{\apj}{Astrophysical Journal}

\newcommand{\apjl}{Astrophysical Journal Letters}
\newcommand{\apjs}{Astrophysical Journal Supplement}
\newcommand{\aap}{Astronomy \& Astrophysics}

\newcommand{\nat}{Nature}

\newcommand{\prd}{Physical Review D}
\newcommand{\aaps}{Astronomy \& Astrophysics, Supplement Series}

\newcommand{\jcap}{J. Cosmology Astropart. Phys.}

\newcommand{\E}{\times10}
\newcommand{\mr}{\mathrm}
\newcommand{\om}{\Omega_\mr m}

\newcommand{\w}{w_0}
\newcommand{\wa}{w_a}

\newcommand{\des}{DES\xspace}

\newcommand{\planck}{\textit{Planck}}

\renewcommand{\comment}[1]{\textcolor{red}{#1}}

\newcommand{\meddeeps}{{\it Medium-Deep Band 2 Survey}}
\newcommand{\wideims}{{\it Wide Band 1 Survey}}
\newcommand{\deeplowims}{{\it Deep SKA1-LOW Survey}}

\title{Cosmology with Phase 1 of the Square Kilometre Array \\
	  {\large {\it Red Book 2018: Technical specifications and performance forecasts}} }

\author{{Square Kilometre Array Cosmology Science Working Group:} David J.\ Bacon$^1$, Richard A.\ Battye$^{2,*}$, Philip Bull$^{3}$, Stefano Camera$^{4,5,6,2}$, Pedro G.\ Ferreira$^7$, Ian Harrison$^{2,7}$, David Parkinson$^{8}$, Alkistis Pourtsidou$^{3}$, M\'ario G.\ Santos$^{9,10,11}$, Laura Wolz$^{12,*}$, 
Filipe Abdalla$^{13,14}$, Yashar Akrami$^{15, 16}$, David Alonso$^7$, Sambatra Andrianomena$^{9,10,17}$, Mario Ballardini$^{9, 18}$, Jos\'e Luis Bernal$^{19, 20}$, Daniele Bertacca$^{21,36}$, Carlos A.P.\ Bengaly$^{9}$, Anna Bonaldi$^{22}$, Camille Bonvin$^{23}$, Michael L.\ Brown$^2$, Emma Chapman$^{24}$, Song Chen$^{9}$, Xuelei Chen$^{25}$, Steven Cunnington$^{1}$, Tamara M.\ Davis$^{27}$, Clive Dickinson$^2$, Jos\'e Fonseca$^{9,36}$, Keith Grainge$^2$, Stuart Harper$^2$, Matt J.\ Jarvis$^{7,9}$, Roy Maartens$^{1,9}$, Natasha Maddox$^{28}$, Hamsa Padmanabhan$^{29}$, Jonathan R.\ Pritchard$^{24}$, Alvise Raccanelli$^{19}$, Marzia Rivi$^{13,18}$, Sambit Roychowdhury$^2$, Martin Sahl\'en$^{30}$, Dominik J.\ Schwarz$^{31}$, Thilo M.\ Siewert$^{31}$, Matteo Viel$^{32}$, Francisco Villaescusa-Navarro$^{33}$, Yidong Xu$^{25}$, Daisuke Yamauchi$^{34}$, Joe Zuntz$^{35}$ \\
\vspace{0.5em}
\textit{Affiliations listed after references}\\
\affil{$^*$ Corresponding Authors: richard.battye@manchester.ac.uk and laura.wolz@unimelb.edu.au}

}

\jid{PASA}
\doi{10.1017/pas.\the\year.xxx}
\jyear{\the\year}

\usepackage{aas_macros}
\usepackage{hyperref} 
\hypersetup{colorlinks,citecolor=blue,linkcolor=blue,urlcolor=blue}

\hypersetup{draft}

\hypersetup{final}
\begin{document}

\begin{frontmatter}
\maketitle
\vfill\eject

\begin{abstract} 
We present a detailed overview of the cosmological surveys that will be carried out with Phase 1 of the Square Kilometre Array (SKA1), and the science that they will enable. We highlight three main surveys: a medium-deep continuum weak lensing and low-redshift spectroscopic HI galaxy survey over 5,000 deg$^2$; a wide and deep continuum galaxy and HI intensity mapping survey over 20,000 deg$^2$ from $z = 0.35 - 3$; and a deep, high-redshift HI intensity mapping survey over 100 deg$^2$ from $z = 3 - 6$. Taken together, these surveys will achieve an array of important scientific goals: measuring the equation of state of dark energy out to $z \sim 3$ with percent-level precision measurements of the cosmic expansion rate; constraining possible deviations from General Relativity on cosmological scales by measuring the growth rate of structure through multiple independent methods; mapping the structure of the Universe on the largest accessible scales, thus constraining fundamental properties such as isotropy, homogeneity, and non-Gaussianity; and measuring the HI density and bias out to $z = 6$. These surveys will also provide highly complementary clustering and weak lensing measurements that have independent systematic uncertainties to those of optical surveys like LSST and Euclid, leading to a multitude of synergies that can improve constraints significantly beyond what optical or radio surveys can achieve on their own. 
This document, the {\it 2018 Red Book}, provides reference technical specifications, cosmological parameter forecasts, and an overview of relevant systematic effects for the three key surveys, and will be regularly updated by the Cosmology Science Working Group in the run up to start of operations and the Key Science Programme of SKA1.

\end{abstract}


\end{frontmatter}

\tableofcontents

\section{Introduction and rationale}
Recent progress in defining the standard cosmological model - known as $\Lambda$CDM - has been dominated by observations of the Cosmic Microwave Background (CMB, \citealt{2013ApJS..208...19H,2016A&A...594A..13P,2018arXiv180706209P}). Maps of the microwave sky made by the {\it Planck} satellite between 30 and 857~GHz, have allowed almost cosmic variance limited measurements of the temperature anisotropy spectrum out to multipoles in excess of $\ell=1000$ as well as high fidelity measurements of the polarization of the CMB. These measurements have constrained the five of the standard six parameters $\Lambda$CDM to 1\% precision and the final one (the optical depth to reionization) to 10\%. The parameter constraints from CMB observations are broadly compatible with other cosmological indicators such as measurements of the cosmic distance scale using standard candles (Cepheids and Supernovae, \citealt{2006A&A...447...31A}) and number counts of clusters of galaxies \citep{2016A&A...594A..24P}.

A wide range of physical phenomena can be probed beyond the $\Lambda$CDM model. These include the dark sector which is responsible for cosmic acceleration, massive neutrinos and primordial non-Gaussianity. Although these phenomena can be constrained with further observations of the CMB, probes of large scale structure, mapping the Universe at relatively lower redshifts, are essential to break some of the degeneracies inherent in CMB observations.

Measurements of the matter power spectrum through galaxy redshift surveys have been around for some time \citep{2005MNRAS.362..505C}, indeed before the detection of the CMB anisotropies, and have played a significant role in defining $\Lambda$CDM \citep{1990Natur.348..705E}. The next two decades will see rapid progress in the field of Large Scale Structure (LSS) surveys with the advent of the {\it Euclid} Satellite \citep{2011arXiv1110.3193L}, the Large Synoptic Survey Telescope (LSST, \citealt{2009arXiv0912.0201L}) and the Dark Energy Spectroscopic Instrument (DESI, \citealt{2016arXiv161100036D}) which will create large scale maps of the Universe. In particular they will use measurements of the angular positions and redshifts of galaxies to infer the matter power spectrum, facilitating measurements of Baryonic Acoustic Oscillations (BAOs) and Redshift Space Distortions (RSDs), and measurements of cosmic shear power spectrum by estimation of galaxy shapes. 
There are many challenges in achieving the fantastic levels of statistical precision which will be possible with these instruments, notably reducing the levels of observational systematic errors.

The Square Kilometre Array\footnote{https://www.skatelescope.org} (SKA) is an international project to build a next generation radio observatory which will ultimately have a collecting area of $10^6\,{\rm m}^2$, i.e. the collecting area necessary to detect the neutral hydrogen (HI) emission at 21cm from an  $L_*$ galaxy at $z\sim 1$ in a few hours \citep{1991ASPC...19..428W}. The SKA will comprise of two telescopes: a dish array (SKA-MID) based in the Northern Cape province of South Africa, and an array of dipole antennas (SKA-LOW) based near Geraldton in Western Australia, with the international headquarters on the Jodrell Bank Observatory Site in the United Kingdom. There will be two phases to the project dubbed SKA1 and SKA2 with a cost cap of $\sim$675~MEuros being set for the SKA1. Only when SKA2 is built will the SKA live up to its name.

The science case for the SKA has been presented in some detail in two volumes produced in 2015 \citep{2015aska.confE.174B}, with 18 separate chapters presenting the cosmology science case for the SKA (see \citealt{2015aska.confE..16M} for the overview chapter). The aim of this {\it Red Book} is to present the status of this science case, with updated forecasts based on the now agreed instrumental design of SKA1, to the cosmology community and beyond. We will not attempt to make detailed forecasts for SKA2 since its precise configuration is yet to be decided; suffice to say that it will have a significant impact on cosmology when it comes online. Furthermore, this is not intended to be a complete review of the subject area, rather it is a summary of the main science goals. We refer the reader to the individual papers for many of the details of the individual science cases.

The observations we will focus on here are:
\begin{itemize}
\item Continuum emission largely due to synchrotron emission from electrons moving in the magnetic field of galaxies.  Selecting galaxies in this way will allow the measurements of the positions and shapes of galaxies.
\item Line emission due to the spin-flip transition between the hyperfine states of neutral hydrogen (HI) at 21cm. Using the redshifted HI line, it is possible to perform spectroscopic galaxy redshift surveys and also to use a new technique called Intensity Mapping (IM) whereby one  measures the large-scale correlations in the HI brightness temperature without detecting individual galaxies.
\end{itemize}
Note that it should be possible to perform continuum and line surveys at the same time and that it may be possible to use the line emission of the galaxies to deduce redshifts, at least statistically, for the continuum galaxy samples\footnote{Furthermore, the same surveys are compatible with the aims of many of the other science goals of the SKA related to extragalactic astronomy including understanding star formation and galaxy evolution, cosmic magnetism and neutral hydrogen in galaxies.}.

We have already pointed out that the next generation of LSS surveys such as those made by {\it Euclid}, LSST and DESI could suffer from observational systematics. The addition of radio observations by the SKA could be crucial to achieving their ultimate goals, as cross-correlating the distribution and shapes of galaxies in two different wavebands will heavily suppress systematic effects. This is because one only expects weak correlations between the contaminants in the different wavebands. Furthermore, additional wavebands can lead to a host of other synergies, a topic we will return to in the discussion section.

\section{Cosmological surveys with SKA1}

In this section we will present the specifications of SKA1 telescopes required for forecasting cosmological parameters, adopting the SKA1 Design Baseline in accordance with SKA-TEL-SKO-0000818\footnote{To be found under  https://astronomers.skatelescope.org/documents/} (\textit{Anticipated SKA1 Science Performance}).
In addition we will define the fiducial cosmological model. 

\begin{table}
\caption{Summary of the array properties of SKA1-MID which will comprise purpose-built SKA dishes and those from the South African precursor instrument, MeerKAT.}
\begin{center}
\begin{tabular}{|l|c|}
\hline
SKA dishes & 133\\
SKA dish diameter & 15\,{\rm m}\\
MeerKAT dishes &  64 \\
MeerKAT dish diameter &13.5\,{\rm m}\\
Maximum Baseline & 150\,{\rm km}\\
Resolution at 1.4\,{\rm GHz}& 0.3\,{\rm arcsec}\\
\hline
\end{tabular}
\end{center}
\label{tab:ska1-mid-tel}
\end{table}

\subsection{SKA1-MID}
SKA1-MID will be a dish array consisting of a set of sub-arrays. The first is the South African SKA precursor MeerKAT which has 64 13.5~m diameter dishes which will be supplemented by 133 SKA1 dishes with 15~m diameter. These will be configured with a compact core and three logarithmically spaced spiral arms with a maximum baseline of $150\,{\rm km}$ which corresponds to an angular resolution $\sim 0.3\,{\rm arcsecs}$ at frequency of $1.4\,{\rm GHz}$. The details of the telescope configuration are presented in Table~\ref{tab:ska1-mid-tel}. It is planned that ultimately these dishes will be equipped with receivers sensitive to 5 different frequency ranges or {\it bands}. The frequency ranges and, where appropriate, the redshift range for HI line observations are tabulated in Table~\ref{tab:ska1-mid-bands}\footnote{The situation is somewhat complicated by the fact that the relevant MeerKAT bands do not have the same boundaries as the SKA Bands 1 and 2. They are: the UHF band $580-1015\,{\rm MHz}$ ($0.4<z<1.45$) and L-band $900-1670\,{\rm MHz}$ ($0<z<0.58$). The table only refers to the SKA dishes.}. In the present SKA baseline configuration there are only sufficient funds to deploy Bands 1 and 2, which are most relevant to cosmology, and Band 5. 

The overall system temperature for the SKA1-MID array
can be calculated using
\begin{equation}
T_{\rm sys}=T_{\rm rx}+T_{\rm spl}+T_{\rm CMB}+T_{\rm gal}\,,
\end{equation}
where we have ignored contributions from the atmosphere. $T_{\rm spl}\approx 3\,{\rm K}$ is the contribution from spill-over, $T_{\rm CMB}\approx 2.73\,{\rm K}$ is the temperature of the CMB, $T_{\rm gal}\approx 25\,{\rm K}(408\,{\rm MHz}/f)^{2.75}$ is the contribution of our own galaxy at frequency $f$ and $T_{\rm rx}$ is the receiver noise temperature. In Band 1 we will assume 
\begin{equation}
T_{\rm rx}=15\,{\rm K}+30\,{\rm K}\left(\frac{f}{\rm GHz}-0.75\right)^2\,,
\end{equation}
and in Band 2 $T_{\rm rx}=7.5\,{\rm K}$.

\begin{table}
\caption{Receiver bands on SKA1-MID. Included also is the range of redshift these receiver bands will probe using the 21~cm spectral line.}
\begin{center}
\begin{tabular}{|c|c|c|}
\hline
Band & $\nu/{\rm GHz}$ & z range \\
\hline
1 & 0.35-1.05 & 0.35-3\\
2 & 0.95-1.75 & 0-0.5\\
3 & 1.65-3 & N/A\\
4 & 3-5.2 & N/A\\
5 & 4.6-15.8 & N/A\\
\hline
\end{tabular}
\end{center}
\label{tab:ska1-mid-bands}
\end{table}

\subsection{SKA1-LOW}
The SKA1-LOW interferometer array will consist of 512 stations, each containing 256 dipole antennas observing in one band at $0.05{\rm GHz}<\nu<0.35 {\rm GHz}$. Most of the large-scale sensitivity comes from the tightly packed ``core" configuration of the array with $N_{\rm d} = 224$ stations, however the long baselines will be crucial for calibration and foreground removal. We assume that the core stations are uniformly distributed out to a $500$ m radius, giving a maximum baseline $D_{\rm max} = 1\, {\rm km}$. The station size is $D =40\, {\rm m}$, the area per antenna is $3.2 \, {\rm m}^2$ at $110$ MHz, and the instantaneous field of view is $(1.2\lambda/D)^2$ sr, with $\lambda = 21(1+z) \, {\rm cm}$. Although multi-beaming should be possible, we consider the conservative case of one beam only. The system temperature is given by $T_{\rm sys} = T_{\rm rx}+T_{\rm gal}$, with the receiver temperature $T_{\rm rx} = 0.1T_{\rm gal}+40 \, {\rm K}$, and $T_{\rm gal}$ defined as for SKA-MID.

\subsection{Proposed cosmology surveys}
\label{sec:surveys}
In this document we will refer to the following surveys targeting cosmology with the SKA: 
\begin{itemize}
\item \meddeeps~: SKA1-MID in Band 2 covering $5,000\,{\rm deg}^2$ and an integration time of approximately $t_{\rm tot}= 10,000$ hrs on sky.  Main goals: a continuum weak lensing survey and an HI galaxy redshift survey out to $z\sim 0.4$ (see sections \ref{weak_lensing} and \ref{HI_gal}).
\item \wideims~: SKA1-MID in Band 1 covering $20,000\,{\rm deg}^2$ and an integration time of approximately $t_{\rm tot}= 10,000$ hrs on sky. Main goals:  a wide continuum galaxy survey and HI intensity mapping in the redshift range $z=0.35-3$ (see sections \ref{sect:angcorrisw}, \ref{sect:dipole} and \ref{HI_IM}).
\item \deeplowims~: This survey will naturally follow the Epoch of Reionisation (EoR) survey strategy. Currently, a three-tier survey consisting of a wide-shallow, a medium-deep, and a deep survey is planned. For our forecasts in this paper we have assumed a deep-like survey with $100 {\rm \, deg}^2$ sky coverage and an integration time of approximately $t_{\rm tot}= 5,000$ hrs on sky using data from sub-bands at frequencies $200-350\rm MHz$, equivalent to $3<z<6$ (see section~\ref{HI_IM}). 
\end{itemize}

\subsection{Survey Processing Requirements}

The production of SKA data products will be performed by the Science Data Processor (SDP) element through High Performance Computer facilities at Perth and Cape Town for SKA1-LOW and SKA1-MID respectively. The SKA1 Design Baseline for the telescope will deliver a compute power of 260 PFLOPs to deliver the science data products that will be transported to Regional Data Centres for further analysis. However, in order to meet the overall telescope cost cap a Deployment Baseline has been defined which will deliver only 50 PFLOPs of compute power when telescope operations start, with a plan to increase to the full capability then being delivered over a 5-year period. Although it is already planned that scientific programmes will be scheduled to spread the computational load across a period defined by the SDP ingest buffer, here we assess the computational load that will result from the surveys defined in section~\ref{sec:surveys}. This assessment is based upon document SKA-TEL-SKO-0000941\footnote{To be found under  https://astronomers.skatelescope.org/documents/} (\textit{Anticipated SKA1 HPC Requirements}).

\meddeeps: This survey will require approximately 2 hours of observing time on each individual field. Since the survey is assumed to be commensal with the project to create and an all sky rotation measure map to probe the galactic magnetic field, data products for all 4 polarisations will be required. The weak lensing experiment (section~\ref{weak_lensing}) requires use of the longest baselines (150km). The HI galaxy redshift survey requires that spectral line data products are generated in addition to the continuum ones needed for other purposes. Although combining these various requirements would seem to imply a maximally difficult data processing task, one of the key findings of SKA-TEL-SKO-0000941 is that the dominant computational cost is driven by the calibration step and that after this has been achieved the delivery of multiple different science products to address their differing requirements at minimal incremental cost. Assuming that observations are only required in sub-band Mid sb4 (as defined in SKA-TEL-SKO-0000941) we therefore estimate that the computational cost of this experiment is approximately 75 PFLOPs (assuming 10\% efficiency). While sb4 observations are sufficient for most continuum science goals, note that this would only cover $z>0.2$ for HI galaxy surveys, and additional sb5 observations doubling the computational cost might be necessary.

\wideims: The primary data products required for the HI IM experiment (section~\ref{HI_IM}) are the antenna auto-correlations, potentially complemented with additional calibration derived from the shortest interferometer baselines. The compute power needed for processing autocorrelation data is negligible compared with that for visibility data. This survey will also be used to generate the Band 1 continuum source sample discussed in section~\ref{continuum}. The total observing time on each individual field is around 1 hour, so the analysis in SKA-TEL-SKO-0000941 suggests that the computational cost of this survey is approximately 50 PFLOPs (assuming 10\% efficiency) for each of the three sub-bands in Band 1 that are desired. However, as discussed in section~\ref{HI_IM}, in order to beat down systematic errors on the autocorrelation measurements, a fast scanning strategy may be adopted for this survey. Commensality with the continuum survey will then require an on-the-fly observing mode for the interferometer\footnote{Such observing mode is currently available with the VLA}. Although it seems technically feasible to implement such mode with SKA1-MID up to scanning speeds of 1 deg/s, further assessments are still needed on the calibration requirements for the continuum survey and on the extra computational costs.

\deeplowims: This survey consists of more than 1000 hour integrations on a small number of individual fields with observations being commensal with the EoR Key Science Project (KSP). The computational load of calibrating such deep observations is severe, but is also a strong function of frequency across the SKA1-LOW band, with 200-350 MHz being substantially easier than 50-200 MHz. Although the signal of interest resides on the shortest baselines, it is likely that high angular resolution image data products will be required in order to remove the effects of contamination of discrete radio sources in the field, so we assume that baselines out to 65km will need to be processed. We therefore estimate that the computational load for the 200-350 MHz survey is approximately 130 and 70 PFLOPs (assuming 10\% efficiency) for sub-bands LOW sb5 and sb6. It should be noted that if these observations are performed commensally with the EoR, the requirement for the Low sb 1,2,3,4 data are approximately 200, 300, 200 and 200 PFLOPs (assuming 10\% efficiency) respectively.

In conclusion, if balanced against other projects with low computational demands such as the pulsar search and timing, then both the \meddeeps~ should be feasible to conduct even with the reduced capability offered by the Deployment Baseline. 
The Wide intensity mapping survey by itself will not be constrained by computational demands, but
commensality with the Wide continuum source survey requires further assessments depending on the scanning strategy. Observing a single sub-band of the \wideims~ should be feasible with the initial HPC capability, but processing all three sub-bands simultaneously will be challenging until the HPC capability increases.
The \deeplowims~ will be more problematic and may need to wait until the HPC capability increases. A caveat to this is that the EoR observing is planned to be conducted in only the best ionospheric conditions, or approximately 15\% of the total available time, so potentially this work can start before the full Design Baseline capability is realised.

\subsection{Synergies with other surveys}
\label{sec:synergysurveys}

SKA cosmology will greatly benefit from synergies with optical surveys. Throughout this paper we refer to the classification of surveys in the report of the Dark Energy Task Force (DETF, \citealt{Albrecht:2009ct}), which describes dark energy research developing in stages. Stage III comprises current and near-term projects, which improve the dark energy figure of merit by at least a factor of 3 over previous measurements; representatives of cosmic shear and galaxy clustering Stage III DETF experiments are, respectively, the Dark Energy Survey (DES) and SDSS Baryon Oscillation Spectroscopic Survey (BOSS). It is also customary to categorize Phase 1 of the SKA as Stage III. Stage IV experiments increase the dark energy figure of merit by at least a factor of 10 over previous measurements; Euclid, LSST and the full SKA stand as Stage IV observational campaigns. In the following, we outline various optical experiments suggested for synergies with the SKA1 throughout this document.

The Stage III Dark Energy Survey (DES) explores the cosmic acceleration via four distinct cosmological probes: type Ia supernovae, galaxy clusters, Baryon Acoustic Oscillations, and weak gravitational lensing. Over a 5 year programme it is covering $5,000 {\rm deg}^2$ in the Southern hemisphere, with a median redshift $z\approx$0.7 \citep{2016MNRAS.460.1270D}.

DESI (Dark Energy Spectroscopic Instrument) is a Stage IV ground-based spectroscopic survey with 14,000 {\rm deg}$^2$ sky coverage \citep{Aghamousa:2016zmz}. It will use a number of tracers of the underlying dark matter field: luminous red galaxies (LRGs) up to $z=1$; emission line galaxies (ELGs) up to $z=1.7$; and  quasars and Ly-$\alpha$ features up to $z=3.5$. It plans to measure around $30$ million galaxy and quasar redshifts and obtain extremely precise measurements of the Baryon Acoustic Oscillation features and matter power spectrum in order to constrain dark energy and gravity, as well as inflation and massive neutrinos.

The {\it Euclid} satellite is a European Space Agency's medium class astronomy and astrophysics space mission. It comprises of two different instruments: a high quality panoramic visible imager (VIS); and a near infrared 3-filter (Y, J and H) photometer (NISP-P) together with a slitless spectrograph (NISP-S) (see \citet{Markovic:2016qzf} for details on the survey strategy). With these instruments, {\it Euclid} will probe the expansion history of the Universe and the evolution of cosmic structures, by measuring the modification of shapes of galaxies induced by gravitational lensing, and the three-dimensional distribution of structures from spectroscopic redshifts of galaxies and clusters of galaxies \citep{Laureijs:2011gra,Amendola:2012ys,Amendola:2016saw} 

The Large Synoptic Survey Telescope (LSST) is a forthcoming ground based, wide field survey telescope. It will examine several probes of dark energy, including weak lensing tomography and baryon acoustic oscillations. The LSST survey will cover $18,000\,{\rm deg}^2$, with a number density of galaxies $40\,{\rm arcmin}^{-2}$, redshift range $0 < z < 2$ with median redshift $z \approx 1$ \citep{2012arXiv1211.0310L}.\footnote{Note that these numbers, used also in forecasts in the present work, have recently been updated to more conservative values, such as $14,300\,{\rm deg}^2$ for the area, and smaller number densities and median redshift \citep{Mandelbaum:2018ouv}.}

\subsection{Fiducial cosmological model and extensions}
\label{sec:cosmoparams}

The standard cosmological model that we have used is a $\Lambda$CDM model based on the the parameters preferred by the 2015 {\it Planck} analysis (TTTEEE+lowP) . In particular the physical baryon and cold dark matter (CDM) densities are $\Omega_{\rm b}h^2=0.02225$ and $\Omega_{\rm c}h^2=0.1198$, the value of the Hubble constant is $H_0=100h\,{\rm km}\,{\rm sec}^{-1}\,{\rm Mpc}^{-1}=67.27\,{\rm km}\,{\rm sec}^{-1}\,{\rm Mpc}^{-1}$, the amplitude and spectral index of density fluctuations are given by $\log(A_{\rm S})=3.094$ and $n_{\rm S}=0.9645$, and the optical depth to reionisation is $\tau=0.079$. We note that these parameter constraints were derived under the assumption that the sum of the neutrino masses is fixed to $\sum m_\nu=0.06\,{\rm eV}$ and therefore we use this in the definition of our fiducial model.

We also consider extensions to the standard model, focusing on those where addition of information from SKA1 can have an impact. Specifically we will consider the following possibilities.

\begin{itemize}
\item Curvature: parameterized by $\Omega_{\rm k}$.

\item Massive neutrinos: parameterized by the sum of the masses $M_{\nu}=\sum m_{\nu}$.

\item Modifications to the dark sector equation of state: using the CPL parameterization~\citep{CPL_w0wa}, $P/\rho=w(a)=w_0+(1-a)w_a$.

\item Modified gravity: deviations from General Relativity (GR) can be encoded by an effective description of the relation between the metric potentials of the form
\begin{equation} \label{eq:mgmu}
-2k^2\Psi=8\pi G_Na^2 \mu(a,k) \rho \Delta,
\end{equation}
\begin{equation} \label{eq:mggamma}
\frac{\Phi}{\Psi} = \gamma(a,k),
\end{equation}
where the GR limit is $\mu=\gamma=1$ and $\Delta$ is the comoving density perturbation. We consider scale independent deviations from GR which emerge at late times (we neglect the effect at $z>5$), hence we assume they are proportional to the dark energy density parameter:
\begin{equation}
\mu(a,k) = 1+\mu_0\frac{\Omega_\Lambda(a)}{\Omega_{\Lambda,0}}, \qquad \gamma(a,k) = 1+\gamma_0\frac{\Omega_\Lambda(a)}{\Omega_{\Lambda,0}}.
\end{equation}
$\mu_0$ and $\gamma_0$  are the free parameters in our analysis.

\item Non-Gaussianity: this is parameterised using the local $f_{\rm NL}$ defined in terms of the amplitude of the quadratic contributions to the metric potential $\Phi$ as a local function of a single Gaussian field $\phi$,
\begin{equation}
\Phi(x) = \phi(x) + f_{\rm NL}\left(\phi^2(x)-\langle\phi^2\rangle\right) + \ldots\,.
\end{equation}

\end{itemize}


At various stages during the analysis we have imposed a {\it Planck} prior on our forecast cosmological parameter constraints. Unless stated otherwise, this is based on the {\it Planck} 2015 CMB + BAO + lensing results presented in \citet{2016A&A...594A..13P}. This was implemented by taking published MCMC chains\footnote{\texttt{base\_w\_wa\_plikHM\_TT\_lowTEB\_BAO\_post\_lensing}} and calculating the covariance matrix for the following extended set of cosmological parameters: $n_s$, $\sigma_8$, $\Omega_b h^2$, $\Omega_m h^2$, $h$, $w_0$, and $w_a$. The covariance matrix was then inverted to obtain an effective Fisher matrix for the prior, which is marginalised over all other parameters (including nuisance parameters) that were included in the {\it Planck} analysis. Applying the prior is then simply a matter of adding it to the forecast Fisher matrix for the survey of interest. While this method is approximate (e.g. it discards non-Gaussian information from the {\it Planck} posterior), it is sufficiently accurate for forecasting.

\section{Continuum galaxy surveys}
\label{continuum}

\subsection{Modeling the continuum sky}
\label{continuumsky}

In this section, we outline how to model the continuum sky and the science cases for the \wideims~ and \meddeeps. The continuum flux density limit of the \meddeeps~ is estimated to be 8.2$\mu$Jy assuming a 10$\sigma$ r.m.s. detection threshold  whereas the \wideims~ will cover four times the area, to approximately slightly less than half the depth and the flux density limit is predicted to be more than double the \meddeeps, at $22.8\,\mu$Jy assuming a 10$\sigma$ r.m.s. detection threshold. Note that this is not exactly a factor of two different to that for the \meddeeps since the overall sensitivity of the array varies with frequency.

\begin{figure}
\centering
\includegraphics[width=0.45\textwidth]{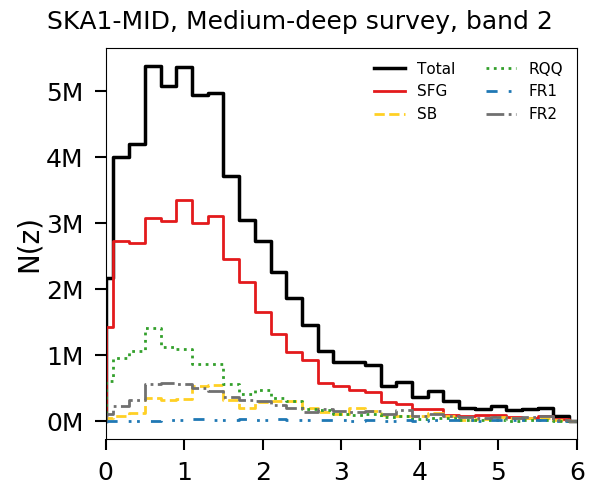}\\
\includegraphics[width=0.45\textwidth]{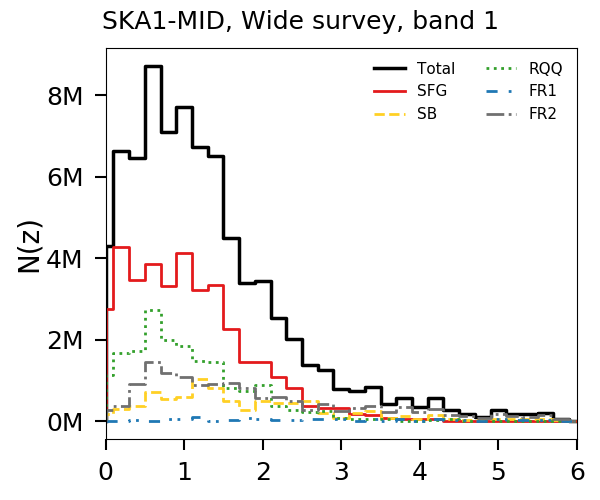}
\caption{\label{fig:surveynz} The total and number of each galaxy species as function of redshift $N(z)$ for a $5,000\,{\rm deg}^2$ survey (\textit{above}) and a $20,000\,{\rm deg}^2$ survey (\textit{below}) on SKA1-MID, assuming a flux limit of 8.2$\mu$Jy (for the \meddeeps) and 22.8$\mu$Jy (for the \wideims), both assuming 10$\sigma$ detection.  The galaxy types are star forming galaxy (SFG), starburst (SB), Fanaroff-Riley type-I and type-II radio galaxies (FR1 \& FR2), and radio-quiet quasars (RQQ).}
\end{figure}

In Fig.~\ref{fig:surveynz} we plot the expected number distribution as a function of redshift of all radio galaxies as well as split by galaxy type, for the two different surveys in the top and bottom panel respectively. These distributions are generated using the SKA Simulated Skies ($S^3$) simulations\footnote{http://s-cubed.physics.ox.ac.uk/}, based on \cite{Wilman2008}. 

\begin{figure}
\begin{center}
\includegraphics[width=0.45\textwidth]{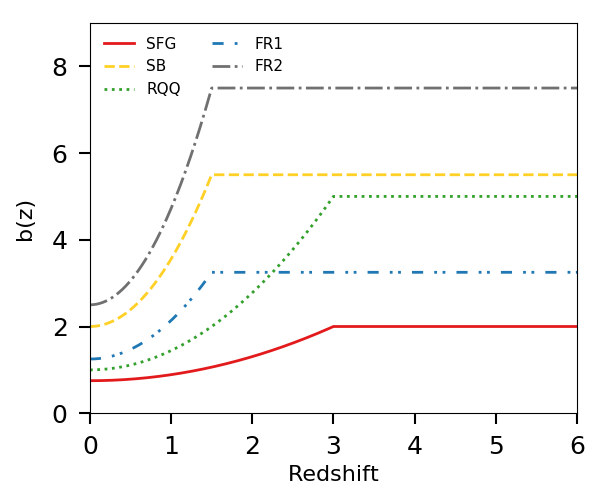}
\caption{\label{fig:bias_model}Bias as a function of redshift for the different  source types, as
following the simulated $S^3$ catalogues of \cite{Wilman2008} including the cut-off above some redshift as described in the text.}
\end{center}
\end{figure}

We also need to choose a model for the galaxy bias. Each of the species of source (i.e.\ starburst, star-forming galaxy, FRI-type radio galaxy, etc.) from the $S^3$ simulation has a  different bias model, as described in \cite{Wilman2008}. The bias in these models increases continuously with redshift, which is unphysical at high redshift; to avoid this we follow the approach of \cite{Raccanelli2011} holding the bias constant above a cut-off redshift (see Fig.~\ref{fig:bias_model}). Having a handle on the redshift evolution of bias and structure will represent a strong improvement for radio continuum galaxy surveys, thanks to the high-redshift tail of continuum sources and will translate into tighter constraints on dark energy parameters compared to the unbinned case, as shown in \citet{Camera:2012ez}. The true nature of the bias for high-redshift, low-luminosity radio galaxies, remains currently unknown; the choice of a bias model therefore remains a source of uncertainty, but one that the SKA will be able to resolve. 

As well as predicting the number and bias of the galaxies for the two strategies, we also use the fluxes from the $S^3$ simulation to predict values for the slope of the source-flux to number density power law, which couples the observed number density to the magnification (magnification bias), given by
\begin{equation}
\alpha_{\rm mag}(S) = -\frac{d(\log n)}{d(\log S)}\,,
\end{equation} 
where $S$ is the flux density and $n$ is the unmagnified number density \citep{Bartelmann:2001}. Magnification bias arises because faint objects are more likely to be seen if they are magnified by gravitational lenses due to overdensities along the line of sight. This changes the clustering properties of the sample, and thus contains cosmological information.  

Finally, we will be able to divide our sample into redshift bins, based on photometric or statistical information \citep{Kovetz2017,2017arXiv170408278H}. While these bins will not be as accurate as spectroscopic redshifts, they will still allow us to recover some of the 3D information from the distribution of galaxies. The \meddeeps~ will have  cross-identifications from other wave-bands (optical from the Dark Energy Survey, for example) over its smaller area, allowing for  accurate photometric redshift bins, whereas the \wideims~ will have limited all sky optical/IR information. We assume nine photo-$z$ bins for \meddeeps, and five for \wideims. The assumed redshift bin distribution, as well as the number of galaxies, bias and slope of the source count power-law, is given in Table~\ref{tab:contredshiftbins}. 

\begin{table}
\centering
\begin{tabular}{|l|c|c|c|c|c|} \hline
Bin & $z_{\rm min}$ & $z_{\rm max}$  & $N/10^6$ &  bias & $\alpha_{\rm mag}$  \\ \hline
 \multicolumn{6}{|c|}{\wideims} \\ \hline
1 & 0.0 & 0.5  & 17.53 	 & 0.94 & 0.95  \\
2 & 0.5 & 1.0  & 23.98 & 1.26 & 1.31 \\
3 & 1.0 & 1.5  & 22.80 & 1.85 & 1.48 \\
4 & 1.5 & 2.0  & 13.20 & 2.26 & 1.34 \\
5 & 2.0 & 6.0  & 20.30 & 3.72 & 1.26  \\ \hline
\multicolumn{3}{|l|}{Total} & 97.81 & & \\ \hline
 \multicolumn{6}{|c|}{\meddeeps} \\ \hline
1 & 0.0  & 0.3 & 4.14 & 0.86 & 0.76  \\
2 & 0.3 & 0.6  & 6.25 & 0.86 & 1.04 \\
3 & 0.6 & 0.9  & 8.06 & 0.90 & 1.05 \\
4 & 0.9 & 1.2  & 7.78 & 1.21 & 1.19 \\
5 & 1.2 & 1.5  & 7.85 & 1.52 & 1.30  \\
6 & 1.5 & 1.8  & 5.77 & 1.58 & 1.22 \\
7 & 1.8 & 2.1  & 4.54 & 2.09 & 1.46 \\
8 & 2.1 & 3.0  & 7.90 & 2.39 & 1.25 \\
9 & 3.0 & 6.0  & 6.12 & 2.85 & 1.25  \\ \hline
\multicolumn{3}{|l|}{Total} & 58.41 &  &  \\ \hline
\end{tabular}
\caption{\label{tab:contredshiftbins} For each redshift bins used in our analysis we present the redshift range, expected number of galaxies, galaxy bias, and magnification bias ($\alpha_{\rm mag}$), for the two continuum surveys. The bias refers to the number-weighted average of the bias of all galaxies in the bin. These surveys are expected to have a total angular number density $n\approx  1.4 {\rm arcmin}^{-2}$ for the \wideims~ and $\approx 3.2 {\rm arcmin}^{-2}$ for the \meddeeps. }
\end{table}

\subsection{Weak lensing}
\label{weak_lensing}

A statistical measurement of the shapes of millions of galaxies as a function of sky position and redshift enables us to measure the gravitational lensing effect of all matter - dark and baryonic - along the line of sight between us and those galaxies. Weak lensing shear measurements are insensitive to factors such as galaxy bias. A number of studies have made marginal detections of the radio weak lensing signal \citep{2004ApJ...617..794C} and radio-optical cross correlation signals \citep{2016MNRAS.456.3100D, 2018MNRAS.473..937D}, but convincing detections have not yet been possible due to a lack of high number densities of resolved, high redshift sources \citep[see][]{2010MNRAS.401.2572P, 2016MNRAS.463.3339T, 2018arXiv181001220H}.

Here, we demonstrate the capabilities of SKA1 as a weak lensing experiment, both alone and in cross-correlation with optical lensing experiments. We consider only a total intensity continuum lensing survey, but note that useful information could also be gained on the important intrinsic alignment astrophysical systematic by using polarisation \citep{Brown:2010rr,Brown:2011db,Thomas:2016xhb} and resolved rotational velocity \citep[e.g.][]{Morales:2006fq} measurements.

\subsubsection{Cosmic shear simulations for SKA}

We create forecasts for the SKA1 \meddeeps. This survey is very similar to the optimal observing configuration found from catalogue-level simulations in \cite{2016MNRAS.463.3686B}. We assume the survey will use the lower $1/3$ of Band 2 and the weak lensing data will be weighted to give an image plane point spread function (PSF) width of $0.55\,\mathrm{arcsec}$, with the source population cut to include all sources which have flux $>10\sigma$ and a size $>1.5\times$ the PSF size. These source populations are also rescaled, as in \cite{2016MNRAS.463.3686B}, to more closely match more recent data and the T-RECS simulation \citep{2018arXiv180505222B}. For comparison to a similar Stage III optical weak lensing experiment, and for use in shear cross-correlations, we take the DES with expectations for the full 5-year survey. The assumed parameters of the two surveys are fully specified in Table~\ref{tab:experiments}. For the \meddeeps~ we assume a sensitivity corresponding to baseline weighting resulting in an image plane PSF with a best-fitting Gaussian FWHM of $0.55\,\mathrm{arcsec}$.
\begin{table*}
\centering
\begin{tabular}{|l|c|c|c|c|c|c|c|c|c|}
\hline 
Weak lensing& $A_{\rm sky} $ & $n \, $ & $z_{m}$ & $\gamma$ & $f_{\textrm{spec-}z}$ & $z_{\textrm{spec-max}}$ & $\sigma_{\textrm{photo-}z}$ & $z_{\textrm{photo-max}}$ & $\sigma_{\textrm{no-}z}$ \\
experiment & $[\mathrm{deg}^2]$ & $[\mathrm{arcmin}^{-2}]$ & & & & & & & \\
\hline
SKA1 Medium-Deep & 5,000 & 2.7 & 1.1  & 1.25 & 0.15 & 0.6 & 0.05 & 2.0 & 0.3 \\
\des & 5,000 & 12 & 0.6  & 1.5 & 0.0 & N/A & 0.05 & 2.0 & 0.3 \\
\hline
\end{tabular}
\caption{Parameters used in the creation of simulated weak lensing data sets for SKA1 \meddeeps~ and DES 5-year survey considered in this section.}
\label{tab:experiments}
\end{table*}

We assume redshift distributions for weak lensing galaxies follow a distribution for the number density of the form
\begin{equation}
\label{eqn:nofz}
\frac{\mathrm{d} n}{\mathrm{d} z} \propto z^{2} \exp\left( -(z/z_0)^{\gamma} \right),
\end{equation}
where $z_0 = z_{m} / \sqrt{2}$ and $z_{m}$ is the median redshift of sources using best fitting parameters for the SKA1-MID \meddeeps~ population and DES survey given in Table~\ref{tab:experiments}. Sources are split into ten tomographic redshift bins, with equal numbers of sources in each bin and each source is attributed an error as follows. A fraction $f_{\textrm{spec-}z}$ out to a redshift of $z_{\textrm{spec-max}}$ are assumed to have spectroscopic errors, in line with the predictions of \cite{2015MNRAS.450.2251Y,2017arXiv170408278H}. The remainder are given photometric redshift errors with a Gaussian distribution (constrained with the physical prior $z>0$) of width $(1+z)\sigma_{\textrm{photo-}z}$ out to a redshift of $z_{\textrm{photo-max}}$. Beyond $z_{\textrm{photo-max}}$ we assume very poor redshift information, with $(1+z)\sigma_{\textrm{no-}z}$.

Of crucial importance to weak lensing cosmology is precise, accurate measurement of source shapes in order to infer the shear transformation resulting from gravitational lensing. For our forecasts, we assume systematic errors due to shear measurement will be sub-dominant to statistical ones. For the \meddeeps, the formulae of \cite{2008MNRAS.391..228A} allow us to calculate requirements on the multiplicative shear bias of $\sigma_{m} < 6.4\E^{-3}$ and additive shear bias of $\sigma_{c} < 8.0\E^{-4}$. These requirements are of the same order of magnitude as those achieved in current optical weak lensing surveys such as DES and the Kilo-Degree Survey\footnote{http://kids.strw.leidenuniv.nl}, but tighter (by an order of magnitude in the case of multiplicative bias) than current methods for radio interferometer to date \citep{2018MNRAS.tmp..362R, 2018arXiv180506799R}. We assume that in the period to 2028, when observations are currently expected to begin, sufficient progress will be made in radio shear measurement methods such that biases are comparable to those achievable in optical surveys today. Previous work has shown this is highly unlikely to be possible with images created with the \textsc{CLEAN} algorithm \citep{1974A&AS...15..417H} meaning access to lower level data products such as gridded visibilities (or equivalently dirty images) will be essential \citep[see also][]{2015aska.confE..30P, 2015arXiv150706639H}. 
For the intrinsic ellipticity distribution of galaxies we use a shape dispersion of $\sigma_{g_i} = 0.3$.

There are significant advantages to forming cosmic shear power spectra by cross-correlating shear maps made using two different experiments. In such power spectra, wavelength-dependent additive and multiplicative systematics can be removed \citep{Camera:2016owj} and almost all of the statistical constraining power on cosmological parameters is retained \citep{2016MNRAS.463.3674H}. Care must be taken in identifying the noise power spectra in the case of cross-power spectra; it will be affected by the overlap in shape information between cross-experiment bins.
We note that constraints are relatively insensitive to the number of galaxies which are present in both bins, being degraded by only $4\%$ when the fraction of overlap is varied between zero and one \citep[see][Fig.~1]{2016MNRAS.463.3674H}.

\subsubsection{Results from autocorrelation}
\label{sec:shearresults}
\begin{figure}
 \centering
 \includegraphics[width=0.45\textwidth]{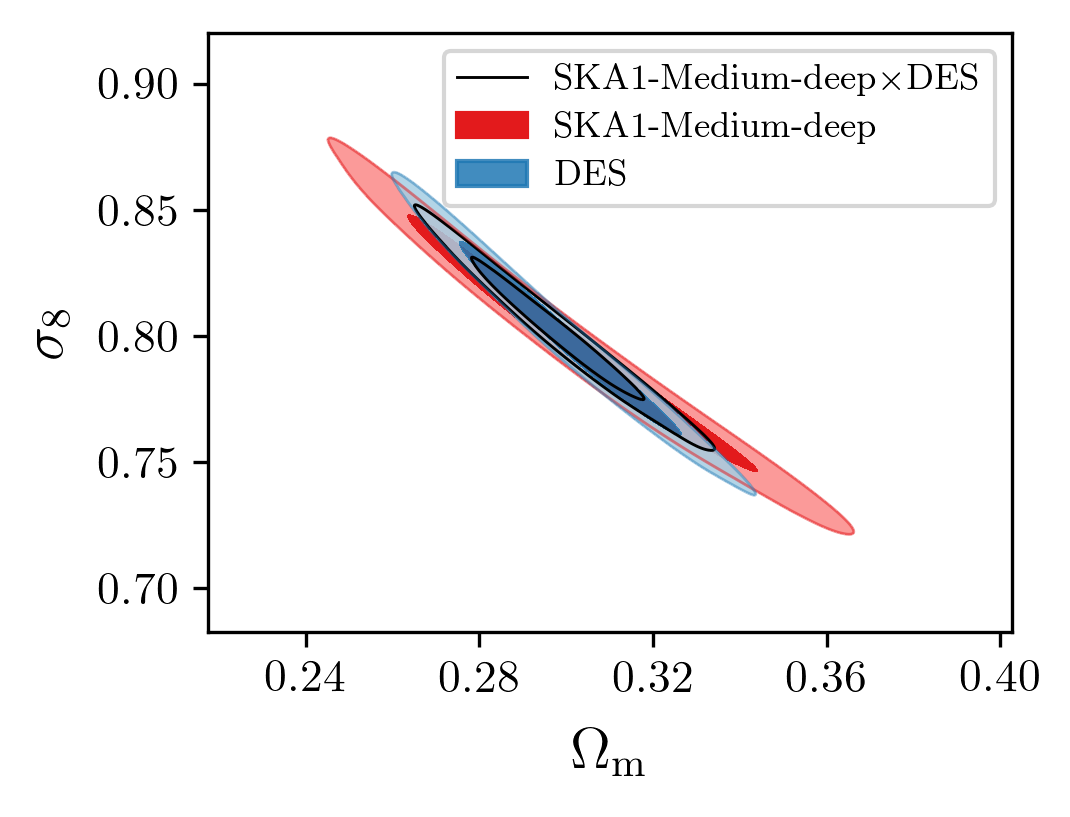}\\
 \includegraphics[width=0.45\textwidth]{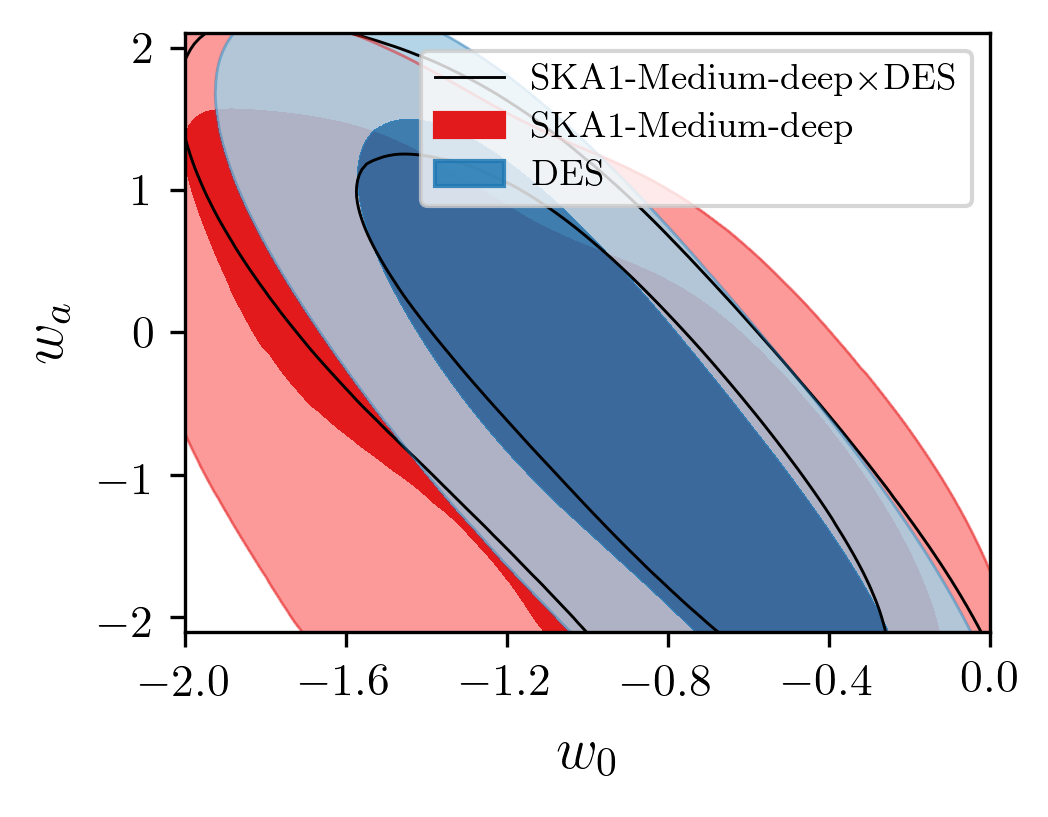}\\
  \includegraphics[width=0.45\textwidth]{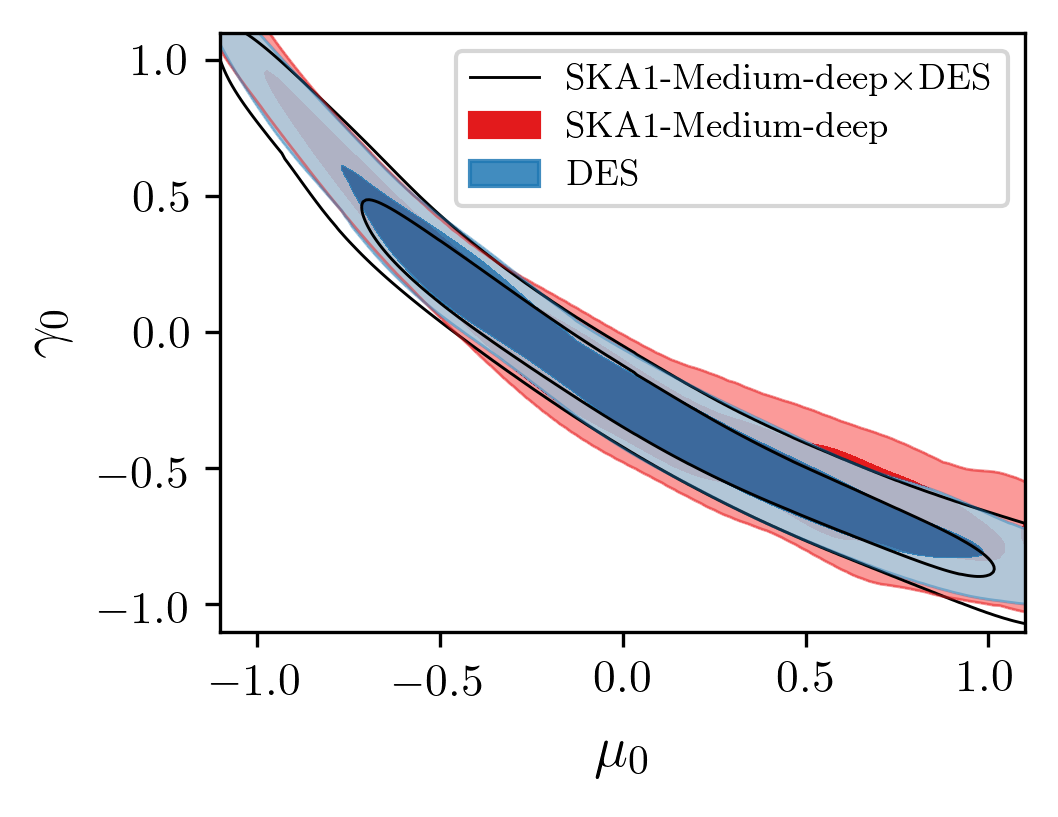}
 \caption{Forecast constraints for weak lensing with the SKA1 \meddeeps~ as specified in the text, compared to the Stage III optical weak lensing DES and including cross-correlation constraints.}
 \label{fig:wl-shear-forecast} 
\end{figure}

\begin{figure}
 \centering
 \includegraphics[width=0.95\columnwidth]{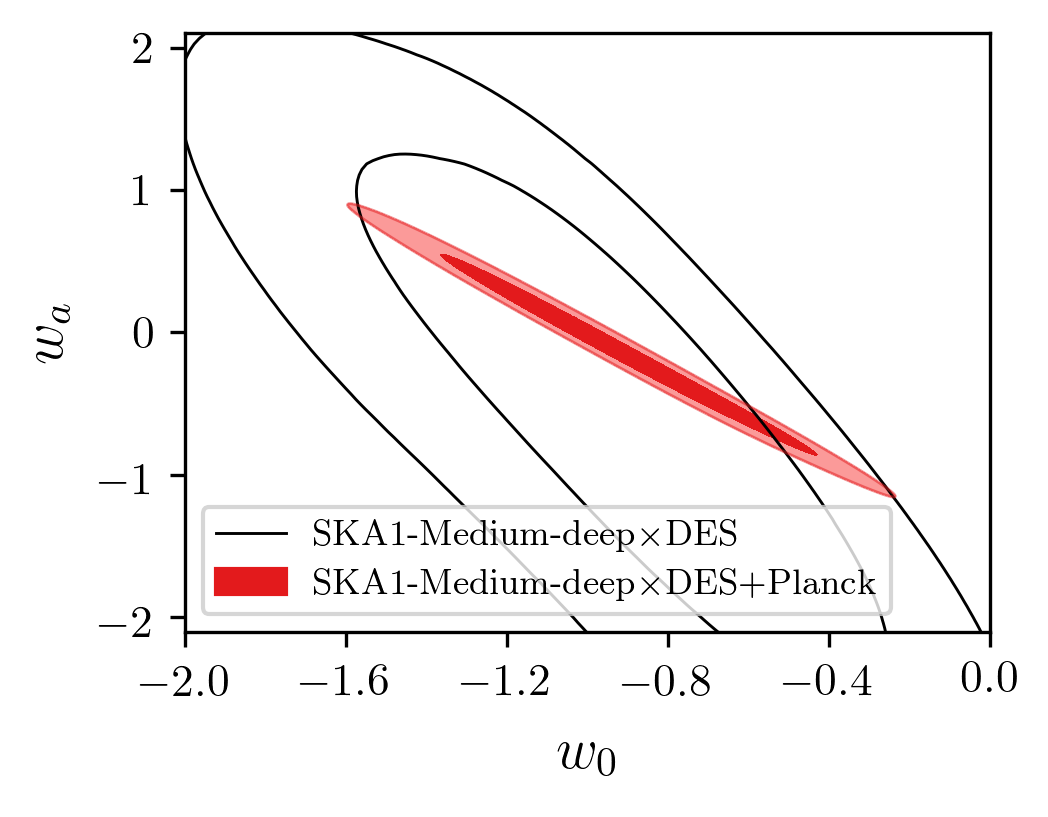}
 \caption{The effect of including a prior from the \planck\ satellite ({\it Planck} 2015 CMB + BAO + lensing as described in section~\ref{sec:cosmoparams}) on the forecast Dark Energy constraints for the specified cross-correlation weak lensing experiment (note that constraints in the other two parameter spaces on not significantly affected).}
 \label{fig:wl-shear-forecast-planck}
\end{figure}

We show forecast constraints in three cosmological parameter spaces in Fig.~\ref{fig:wl-shear-forecast}: matter ($\om$-$\sigma_8$), Dark Energy equation of state in the CPL parameterisation ($\w$-$\wa$) and modified gravity modifications to the Poisson equation and Gravitational slip ($\mu_0$-$\gamma_0$). Our results show that the SKA1 \meddeeps~ will be capable of comparable constraints to other DETF Stage III surveys such as DES and also, powerfully, that cross-correlation constraints (which are free of wavelength-dependent systematics) retain almost all of the statistical power of the individual experiments.
In Fig.~\ref{fig:wl-shear-forecast-planck} we also present forecast constraints in the Dark Energy parameter space including priors from the \planck\ CMB experiment, specifically a Gaussian approximation to the \planck\ 2015 CMB + BAO + lensing likelihood as described in section~\ref{sec:cosmoparams} with constraints on the other parameters considered not significantly affected by application of the \planck\ prior.
We also display tabulated summaries of the one dimensional marginalised uncertainties on these parameters in Table~\ref{tab:marginals}.

\begin{figure}
\centering
\includegraphics[width=0.39\textwidth]{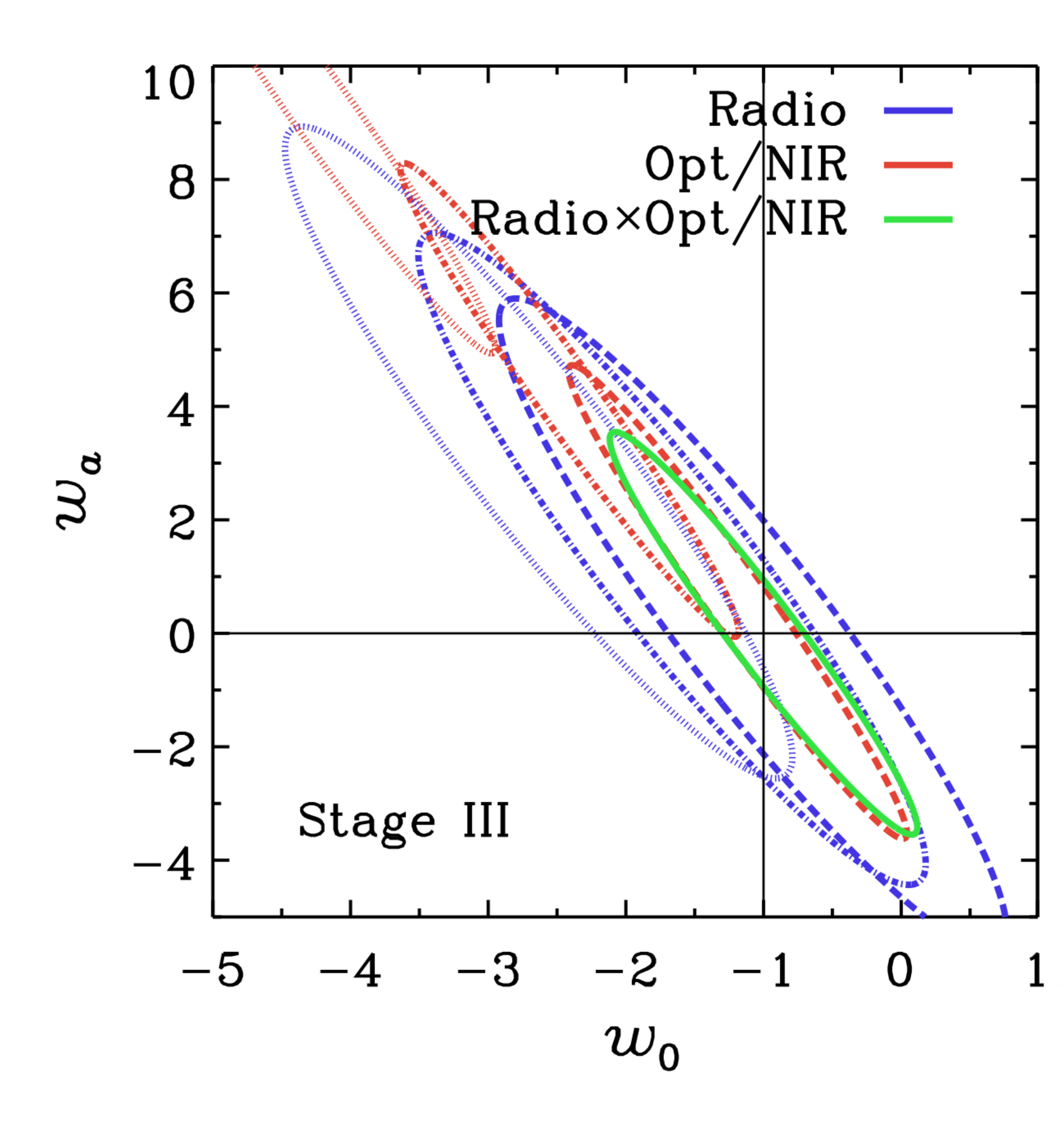}\\
\includegraphics[width=0.40\textwidth]{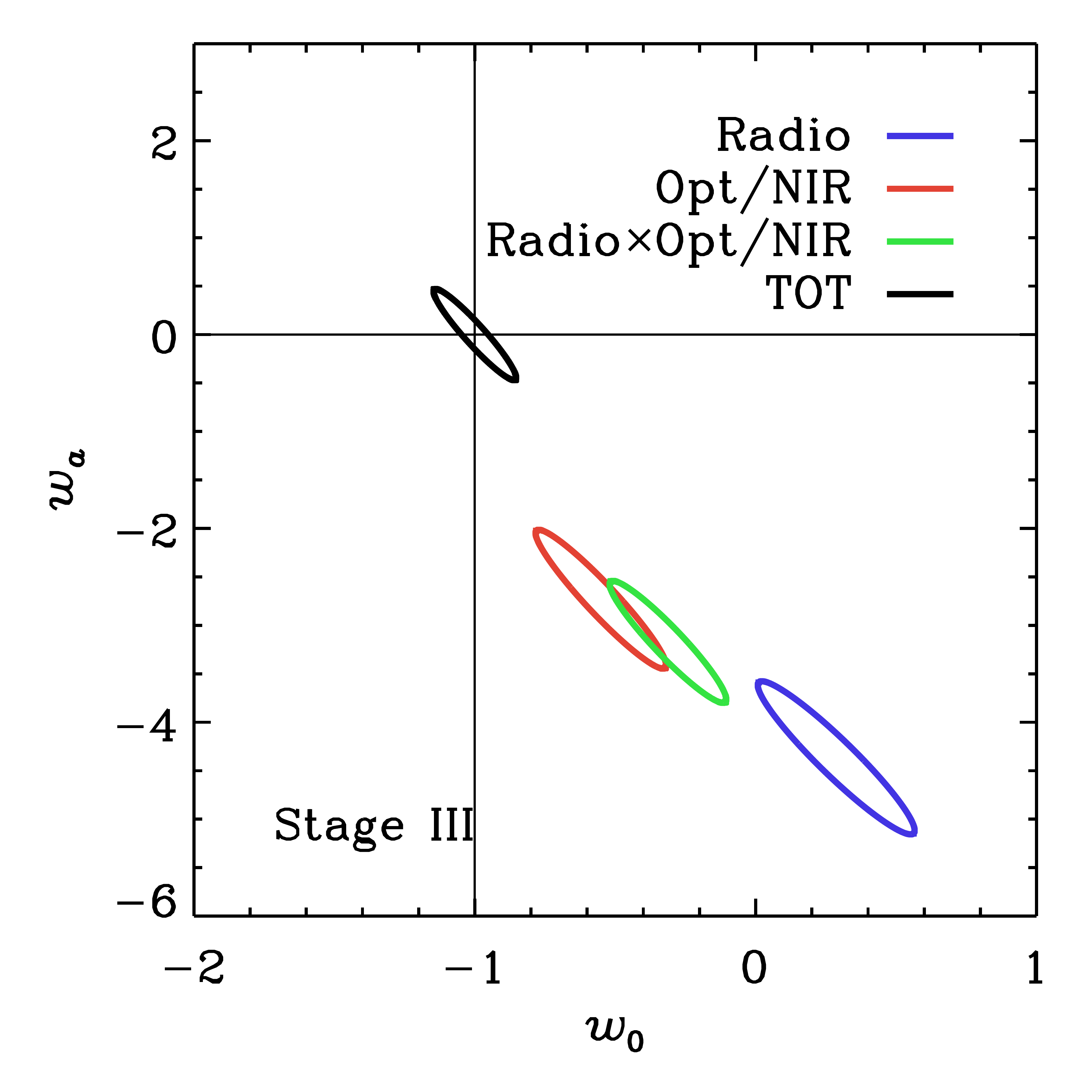}
    \caption{Weak lensing marginal joint 1$\sigma$ error contours in the dark energy equation-of-state parameter plane with additive (\emph{left}) and multiplicative (\emph{right}) systematics on the shear power spectrum measurement. The black cross indicates the \lcdm\ fiducial values for dark energy parameters. Blue, red and green ellipses are for radio and optical/near-IR surveys and their cross-correlation, respectively. (Details in the text.)}\label{fig:sys_add}
\end{figure}
\begin{table*}
\begin{center}
\begin{tabular}{|l|rl|rl|rl|c|}
\hline 
Experiment & $\sigma(\om)/\om$, & $\sigma(\sigma_{8})/\sigma_{8}$ & $ \sigma(\w)$, & $\sigma(\wa)$ & $\sigma(\mu_{0})$, & $\sigma(\gamma_{0})$ & DETF FoM\\
\hline
SKA1-Medium-deep & 0.083 & 0.040 & 0.52 & 1.6 & 0.77 & 0.63 & 1.6\\
SKA1-Medium-deep + \planck\ & 0.084 & 0.040 & 0.28 & 0.43 & - & - & 77\\
\des & 0.056 & 0.032 & 0.43 & 1.4 & 0.64 & 0.52 & 3.5\\
\des + \planck\ & 0.058 & 0.033 & 0.22 & 0.33 & - & - & 89\\
SKA1-Medium-deep$\times$\des & 0.046 & 0.024 & 0.45 & 1.3 & 0.59 & 0.48 & 3.3\\
SKA1-Medium-deep$\times$\des + \planck\ & 0.046 & 0.024 & 0.23 & 0.36 & - & - & 106\\
\hline
\end{tabular}
\end{center}
\caption{One dimensional marginalised constraints, from weak lensing alone and in combination with \emph{Planck} CMB ({\it Planck}CMB2015 + BAO + lensing as described in section~\ref{sec:cosmoparams}), on the parameters considered, where all pairs (indicated by brackets) are also marginalised over the base \lcdm parameter set.}
\label{tab:marginals}
\end{table*}

\subsubsection{Results from radio-optical cosmic shear cross-correlations}
A key consideration in weak lensing surveys are the systematics induced by the instrument on galaxy shape measurements, which must be controlled to high levels in order to ensure unbiased constraints on cosmological parameters. In contrast with the optical weak lensing surveys conducted to date, radio weak lensing surveys will measure galaxy shapes from $uv$-data, allowing for direct Fourier plane measurement, as well as measurement in images reconstructed by deconvolving the interferometer PSF. The systematics from these shape measurements will be very different, and uncorrelated with, those from measuring shapes from CCD images. In \cite{2018MNRAS.tmp..362R}, the authors adapted the optical method \emph{lensfit} to shape measurement on Fourier-domain interferometer data which is capable of satisfying the requirements for the SKA1 \meddeeps~ on sources with $\mathrm{SNR}>18$. Residual systematics are typically modelled as linear in the shear and shear power spectrum, with an additive and multiplicative component. In Fig.~\ref{fig:wl-shear-forecast} \citep[and][]{2016MNRAS.463.3674H} the unfilled black contours show the constraints from cross-correlating radio and optical weak lensing experiments, demonstrating that nearly all of the statistical constraining power remains. In Fig.~\ref{fig:sys_add} \citep[and][]{Camera:2016owj} we show explicitly how multi-wavelength cosmic shear analyses measuring the dark energy equation of state parameters $\lbrace w_0, w_{\rm a}\rbrace$ can be made free of both additive and multiplicative systematics. The left panel shows how cross-correlation of radio and optical (green) experiments directly removes additive systematics in both radio-radio (blue) and optical-optical (red) experiments. Solid, dashed, dash-dotted and dotted ellipses represent increasing values for the residual systematics power spectrum. Also in the right panel of Fig.~\ref{fig:sys_add} we show constraints for systematics which are multiplicative on shear power spectrum measurements. Here the radio-radio, optical-optical and radio-optical measurements are all biased away from the input cosmology individually, but may be used in self-calibration to recover it correctly. Mitigation of such multiplicative systematics is expected to be extremely important even at the level of Stage III surveys and represents a powerful argument for performing weak lensing in the radio band.

\subsection{Angular Correlation Function and Integrated Sachs Wolfe Effect}
\label{sect:angcorrisw}

The angular distribution of galaxies and the cross-correlation of the galaxy positions with other tracers can yield important cosmological tests. The two-point distribution of radio galaxy positions in angle space can be represented by the  angular correlation power spectrum $C_{\ell}^{i,j}$, where $\ell$ is the multipole number and  $i$, $j$ label redshift bins with the galaxies distributed across these bins defined by window functions, $W_i(z)$. This statistic encodes the density distribution projected on to the sphere of the sky, and so smooths over structure along the line of sight. This can dampen the effect of Redshift Space Distortions (RSDs) on the angular power spectrum for broad redshift distributions, but these can become important as the distributions narrow \citep{Padmanabhan2007}. 

When two non-overlapping redshift bins are considered the cross-correlation of density perturbations between these two bins measured through $C_{\ell}^{i,j}$ will be negligible in the absence of lensing. However, the observed galaxy distribution is also affected by gravitational lensing through magnification, which can induce a correlation between the two bins, creating an observed correlation between the positions of some high redshift galaxies and the distribution of matter at low redshift.

The distribution of matter in the Universe can also be measured by the effect on the CMB temperature anisotropies, through the Integrated Sachs-Wolfe effect (ISW), where the redshifting and blueshifting of CMB photons by the intervening gravitational potentials generates an apparent change in temperature \citep{ISW_paper}. Since the distribution of matter (which generates the gravitational potentials) can be mapped through the distribution of tracer particles, such as galaxies,  the effect is detected by cross-correlating the positions of galaxies and temperature anisotropies on the sky. For a more detailed description of the use of the ISW with SKA continuum surveys, see \citet{Raccanelli2014}.

Here, we demonstrate the capabilities of SKA for using the angular correlation function and relevant cross-correlations as a cosmological probe.

\subsubsection{Forecasting}

In order to estimate the effectiveness of the surveys and make predictions for the constraints on the cosmological parameters, we simulate the auto- and cross-correlation galaxy clustering angular power spectra, including the effects of cosmic magnification and the ISW. As only the observed galaxy distributions (which are affected by gravitational lensing) can be measured, it is impossible to measure the galaxy angular power spectrum decoupled from magnification. Hence, the galaxy clustering angular power spectrum contains both the density and magnification perturbations. 

We use the simulated source count and galaxy bias model from section~\ref{continuumsky} to simulate the angular correlation and cross-correlation functions $C_\ell$, and the relevant measurement covariance matrices, for the \wideims~ and \meddeeps. In the case of galaxy clustering and ISW, we limit the analysis to the multipoles $\ell_{\rm min}\leq \ell\leq 200$, where $\ell_{\rm min} = \pi/(2f_{\rm sky})$ and $f_{\rm sky}$ is the fraction of sky surveyed.

When making our forecasts, we also compare to and combine with current constraints from {\it Planck} CMB 2015, BAO and RSD observations, as described in \ref{sec:cosmoparams} (with additional relevant information for the extension parameters under consideration). We also assume that the overall bias for a particular redshift bin to be unknown, and so marginalised over. As such there are five (or nine, depending on the number of photometric bins for the given survey) extra parameters being considered in the Fisher matrix, which will degrade the performance of these cosmological probes.

\subsubsection{Results}

\begin{table*}
\caption{\label{tab:cont-big} Predicted constraints from continuum galaxy clustering measurements using the two different survey strategies (\wideims~ and \meddeeps). These are 68\% confidence levels on each of the parameters of the four different cosmological models we tested. The three main columns show results of galaxy clustering (GC) by itself (left), GC combined with Integrated Sachs Wolfe (ISW) constraints (centre), and when {\it Planck} priors from {\it Planck} CMB 2015+BAO are added to GC+ISW (right). Note that these cases assume that the overall bias in each of the photometric redshift bins is unknown, and needs to be marginalised over.}
\centering
\begin{tabular}{|l|l|c|c|c|c|c|c|} \hline
& &\multicolumn{6}{|c|}{Data combination and parameters} \\
Survey& Model & \multicolumn{2}{|c|}{Galaxy Clustering (GC)} & \multicolumn{2}{|c|}{GC+ISW} & \multicolumn{2}{|c|}{GC+ISW+{\it Planck}} \\ \hline
& & \hspace{3mm} $\sigma(w_0)$\hspace{3mm} & \hspace{3mm}$\sigma(w_a)$\hspace{3mm} & \hspace{3mm}$\sigma(w_0)$\hspace{3mm} & \hspace{3mm}$\sigma(w_a)$\hspace{3mm} & \hspace{3mm}$\sigma(w_0)$\hspace{3mm} & \hspace{3mm}$\sigma(w_a)$\hspace{3mm} \\ \hline
SKA1-Wide & $(w_0w_a)$CDM & 1.8 & 6.3 & 1.3 & 3.8 & 0.29 & 0.79\\
SKA1-Medium-Deep & $(w_0w_a)$CDM & 1.6 & 4.4 & 1.5 & 4.1 & 0.28 & 0.77\\ \hline \hline
& & $\sigma(\mu_0)$ & $\sigma(\gamma_0)$ & $\sigma(\mu_0)$ & $\sigma(\gamma_0)$& $\sigma(\mu_0)$ & $\sigma(\gamma_0)$ \\ \hline
{SKA1-Wide} & $\Lambda$CDM+$\mu_0$+$\gamma_0$ & 2.6 & 6.0 & 0.88 & 1.9 & 0.15 & 0.35 \\
{SKA1-Medium-Deep} & $\Lambda$CDM+$\mu_0$+$\gamma_0$ &  3.8 & 8.8 & 1.8 & 4.1 & 0.16 & 0.37 \\
\hline\hline
& & \multicolumn{2}{|c|}{$\sigma(\Omega_k)$} & \multicolumn{2}{|c|}{$\sigma(\Omega_k)$} & \multicolumn{2}{|c|}{$\sigma(\Omega_k)$} \\ \hline
SKA1-Wide & $\Lambda$CDM+$\Omega_k$ & \multicolumn{2}{|c|}{$18\times 10^{-2}$}  & \multicolumn{2}{|c|}{$14\times 10^{-2}$} & \multicolumn{2}{|c|}{$2.0\times 10^{-3}$}\\
SKA1-Medium-Deep & $\Lambda$CDM+$\Omega_k$ & \multicolumn{2}{|c|}{$12\times 10^{-2}$} & \multicolumn{2}{|c|}{$12\times 10^{-2}$} &\multicolumn{2}{|c|}{$2.0\times 10^{-3   }$}\\ \hline\hline
& & \multicolumn{2}{|c|}{$\sigma(f_{\rm NL})$} & \multicolumn{2}{|c|}{$\sigma(f_{\rm NL})$} & \multicolumn{2}{|c|}{$\sigma(f_{\rm NL})$} \\ \hline
SKA1-Wide & $\Lambda$CDM+$f_{\rm NL}$ & \multicolumn{2}{|c|}{5.2}  & \multicolumn{2}{|c|}{5.2} & \multicolumn{2}{|c|}{3.4}\\
SKA1-Medium-Deep & $\Lambda$CDM+$f_{\rm NL}$ & \multicolumn{2}{|c|}{13 } & \multicolumn{2}{|c|}{12 } &\multicolumn{2}{|c|}{5.1}\\ \hline
\end{tabular}
\end{table*}

The 68\% confidence level constraints on  the different parameters described in section~\ref{sec:cosmoparams} for the \wideims~ and the \meddeeps~ are given Table~\ref{tab:cont-big}.

\begin{figure}[htp]
\centering
\includegraphics[width=0.45\textwidth]{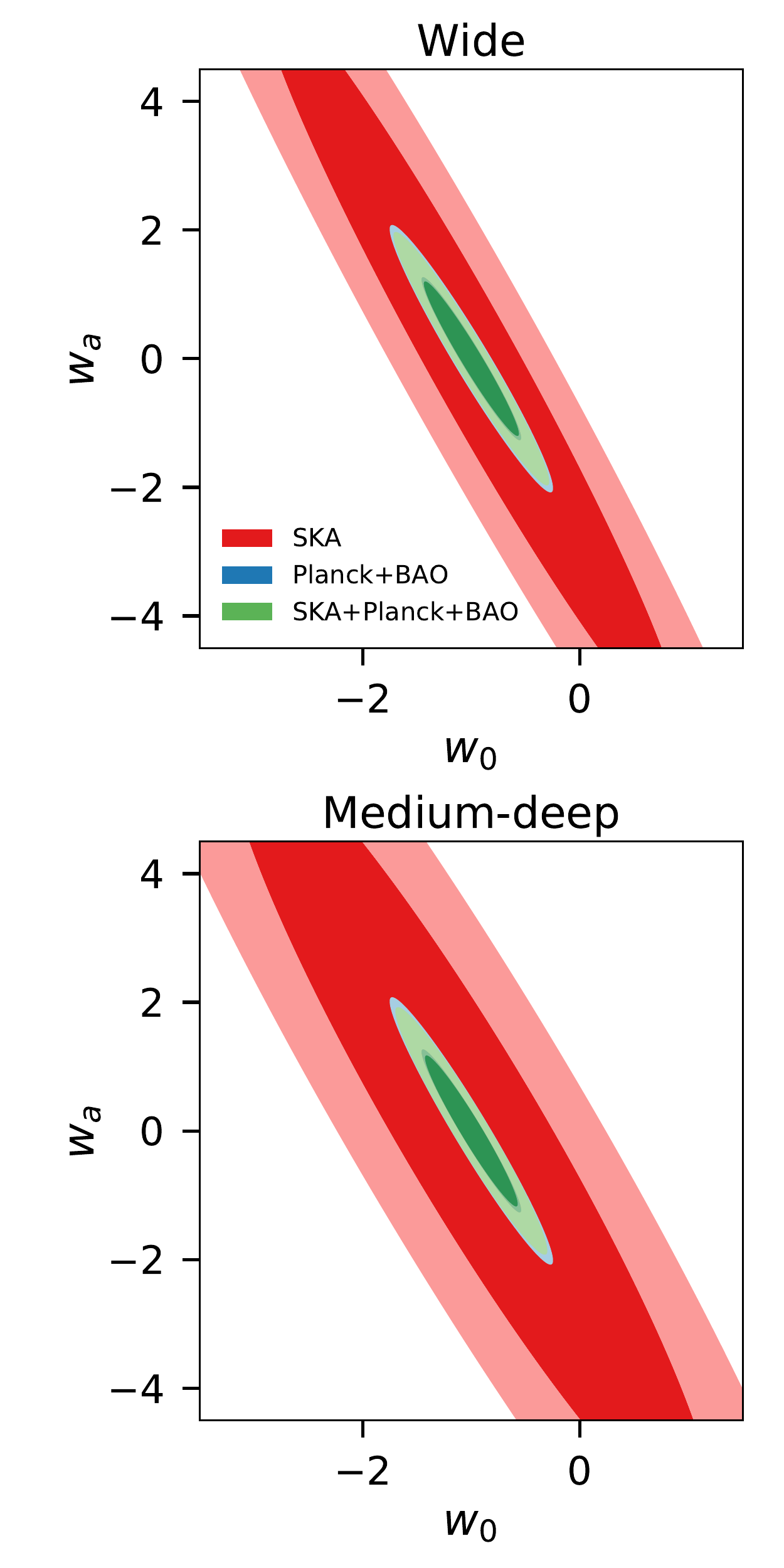}
\caption{68\% and 95\% confidence level forecast constraints on the deviation of the dark energy parameters $w_0,w_a$  from their fiducial values for the \wideims~ (top) and \meddeeps~ (bottom), using galaxy clustering data, including the effects of cosmic magnification. We show constraints from {\it Planck} CMB 2015 and BAO and RSD observations, as described in \ref{sec:cosmoparams} in blue,  SKA1 forecasts in red and the constraints for the combination of both experiments in green. We show here that for the dark energy parameters, the continuum data adds little to the existing constraints, owing to the uncertainty in the bias in each redshift bin. }
\label{fig:Gaussian_bins_w0wa}
\end{figure}

\begin{figure}[htp]
\centering
\includegraphics[width=0.45\textwidth]{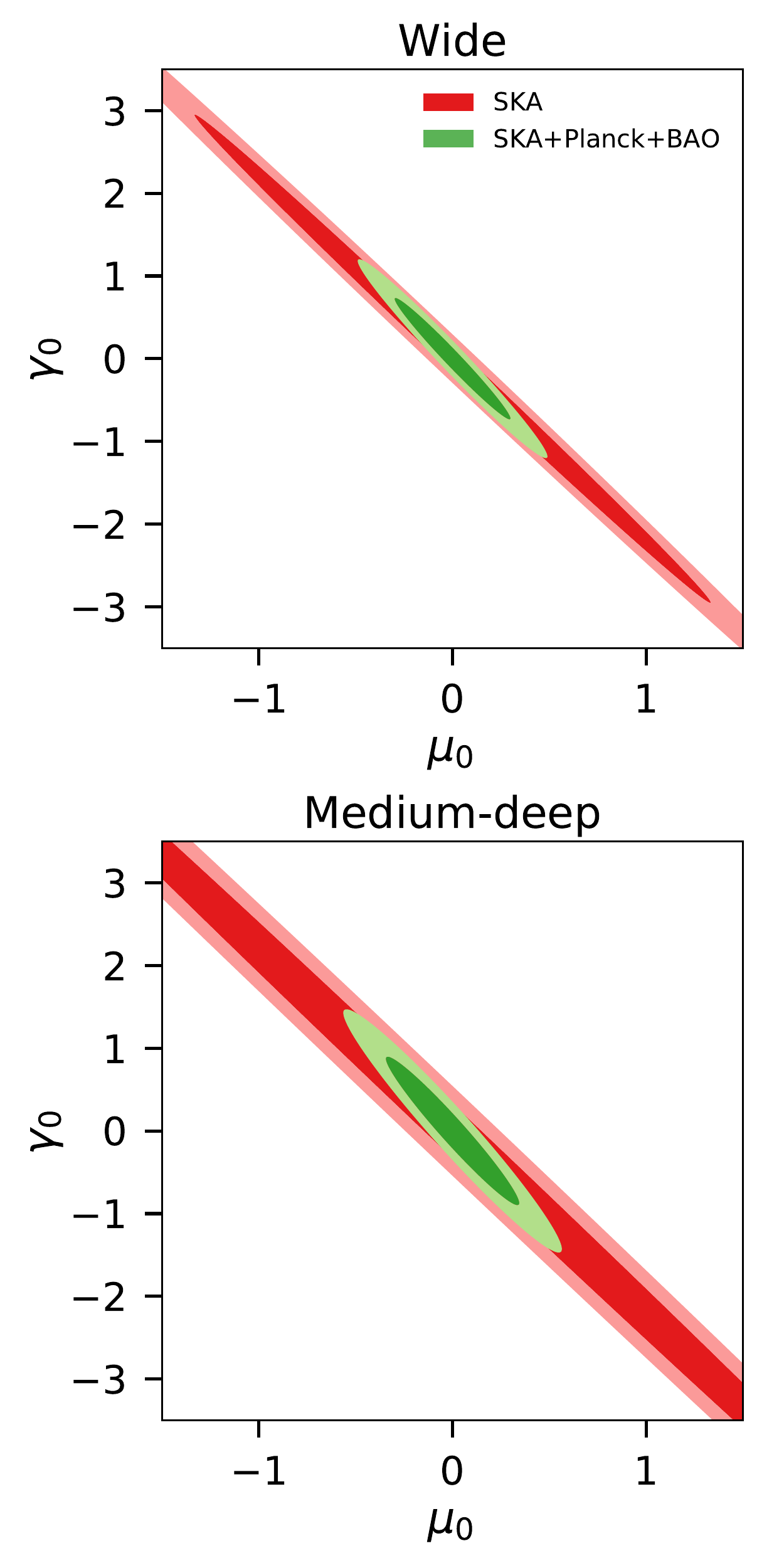}
\caption{68\% and 95\% confidence level forecast constraints on the deviation of the  modified gravity parameters $\mu_0,\gamma_0$ from their fiducial values for the \wideims~ (top) and \meddeeps~ (bottom), using galaxy clustering data, including the effects of cosmic magnification. We show  SKA forecasts constraints in red and the constraints for the combination of SKA1 with {\it Planck} CMB 2016 and BAO in green.  }
\label{fig:Gaussian_bins_mu0gam0}
\end{figure}

We show the predicted 68\% and 95\% confidence level constraints as a 2D contour, for the dark energy parameters $w_0$ and $w_a$ in Fig.~\ref{fig:Gaussian_bins_w0wa}, and the modified gravity parameters $\mu_0$ and $\gamma_0$ in  Fig.~\ref{fig:Gaussian_bins_mu0gam0}.  These constraints are shown for the \wideims~ and the \meddeeps~ in red, combining measurements from all photometric redshift bins, and including constraints from the ISW. In the dark energy case, we also show current constraints from Planck in blue, but for the modified gravity case the Planck MCMC chains for these models are not public.

The predicted constraints on the dark energy parameters do not improve significantly on those presently available. This is also somewhat the case for the modified gravity parameters and the curvature, in the case of the \meddeeps, though the \wideims~ does improve on current knowledge. However, such constraints will improve with a better knowledge of the bias (decreasing the number of extra parameters to be marginalised over) and with a larger number of photometric redshift bins.

Constraints on $f_{\rm NL}$ from the \meddeeps~ will not be significantly better than those currently made by the Planck surveyor, $f_{\rm NL} = 2.5 \pm 5.7$ \citep{PlanckNG2016}. In contrast the \wideims~ is capable of improving the constraint, with further potential gain from an increased number of redshift bins \citep{Raccanelli2017}.
Finally, more competitive constraints on all parameters, but especially for $f_{NL}$, may be achievable through the use of different radio galaxy populations as tracers of different
mass halos, as described in \cite{2014MNRAS.442.2511F}.

\subsection{Cosmic dipole}\label{sect:dipole}

The standard model of cosmology predicts that that the radio sky should be isotropic on large scales. Deviations from isotropy are expected to arise from proper motion of the Solar system with respect to the isotropic CMB (the cosmic dipole), the formation of large scale structures and light propagation effects like gravitational lensing. 

The CMB dipole is normally associated with the proper motion of the Sun with respect to the cosmic heat bath at $T_0 = 2.725$K. 
However, the CMB dipole could also contain other contributions, e.g.\ a primordial temperature dipole or an integrated Sachs-Wolfe (ISW) effect, 
and measurements using only  CMB  data are limited by cosmic variance.

The extragalactic radio sky offers an excellent opportunity to perform an independent test of the origin of the cosmic dipole. It is expected that the radio dipole is dominated by the kinematic dipole,  as radio continuum surveys have median redshifts well above one (unlike  visible or infrared surveys). Current estimates of the radio dipole show good agreement with the CMB dipole direction, but find a dipole amplitude that is a factor of 2 to 5 larger than expected \citep{2002Natur.416..150B, 2011ApJ...742L..23S, 2013A&A...555A.117R, 2017MNRAS.471.1045C, 2017arXiv171008804B}. See also \cite{Bengaly:2018ykb} for a study on dipole measurements with the SKA1 and SKA2. 

\subsubsection{Forecasting}
In this section, we estimate the ability of SKA1 continuum surveys to measure the cosmic radio dipole using realistic mock catalogues, which include the effects of large scale structure and the kinematic dipole. Details of that study will be published elsewhere. Briefly, the mock catalogues assumed an angular power spectrum of the radio galaxies generated by {\sc CAMB sources}~\citep{2011PhRvD..84d3516C}, assuming the Planck best-fit flat $\Lambda$CDM model~\citep{2016A&A...594A..13P}. 
The  redshift distribution $N(z)$ is shown in Fig.~\ref{fig:surveynz}, and the bias $b(z)$ follows~\cite{2015ApJ...814..145A}.
The available sky area is $f_{\rm sky} \approx 0.52$ due to the removal of the galactic plane on low latitudes ($|b| \leq 10^{\circ}$). Using the lognormal code {\sc flask}~\citep{2016MNRAS.459.3693X}, we produced ensembles of $100$ catalogues each, where the radio source positions follow the expected clustering distribution. 
 
The effect of the kinetic dipole is implemented by boosting the maps of galaxy number densities according to the theoretical expectation \citep{1984MNRAS.206..377E}, 
\begin{equation}
\left[\frac{\mathrm{d} N}{\mathrm{d}\Omega} (S, \mathbf{n})\right]_{\mathrm{obs}} = \left[\frac{\mathrm{d} N}{\mathrm{d}\Omega} (S, \mathbf{n})\right]_{\mathrm{rest}} \left( 1 + [2 + \beta (1+\alpha)] \frac{\mathbf{n}\cdot\mathbf{v}} {c}\right),
\end{equation}
where $S$ denotes the flux density threshold of the survey, $\mathbf{n}$ the direction on the sky and $\mathbf{v}$ the Sun's proper motion. This expression assumes that radio sources follow a power-law spectral energy distribution, $S \propto \nu^{-\beta}$ with $\beta = 0.75$. 
The source counts are assumed to scale with $S$ as $dn/dS\propto S^{-\alpha}$, and we assume $\alpha = 1$ (which is very similar to the values of $\alpha$ for the individual redshift bins from simulations given in Table~\ref{tab:contredshiftbins}). 

Here we show results from estimations of the radio dipole direction and amplitude,
$A = [2+ \beta(1+\alpha)]|\mathbf{v}|/c$, of the generated mock catalogues by means of a quadratic estimator in pixel space on a HEALPix\footnote{HEALPix package \url{http://healpix.sourceforge.net/}} grid with Nside = 64.  Using pixel space has the advantage that incomplete sky coverage does not bias the results.

\subsubsection{Results}

\begin{figure}
\centerline{
\includegraphics[width=0.95\columnwidth]{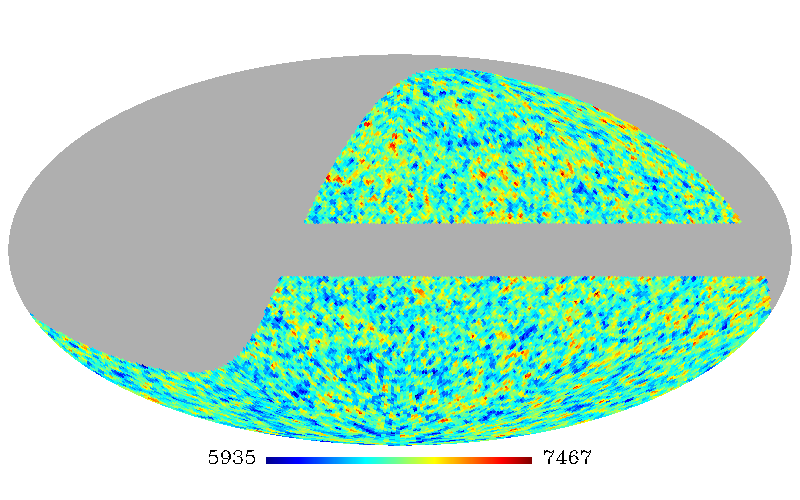}}
\caption{Simulated source count per pixel for the SKA1-MID \wideims~ at a central frequency of 700 MHz and a flux threshold of $22.8 \mu$Jy in galactic coordinates and Mollweide projection at HEALPix resolution  $N_{\rm side}=64$ including the kinematic dipole and cosmic structure up to multipole moment $\ell_{\rm max}=128$. This shows the effect of the dipole on the source counts, as the southern sky appears here slightly bluer than the northern hemisphere.} 
\label{mock_example}
\end{figure}

\begin{figure*}
\centerline{
\includegraphics[width=0.65\linewidth]{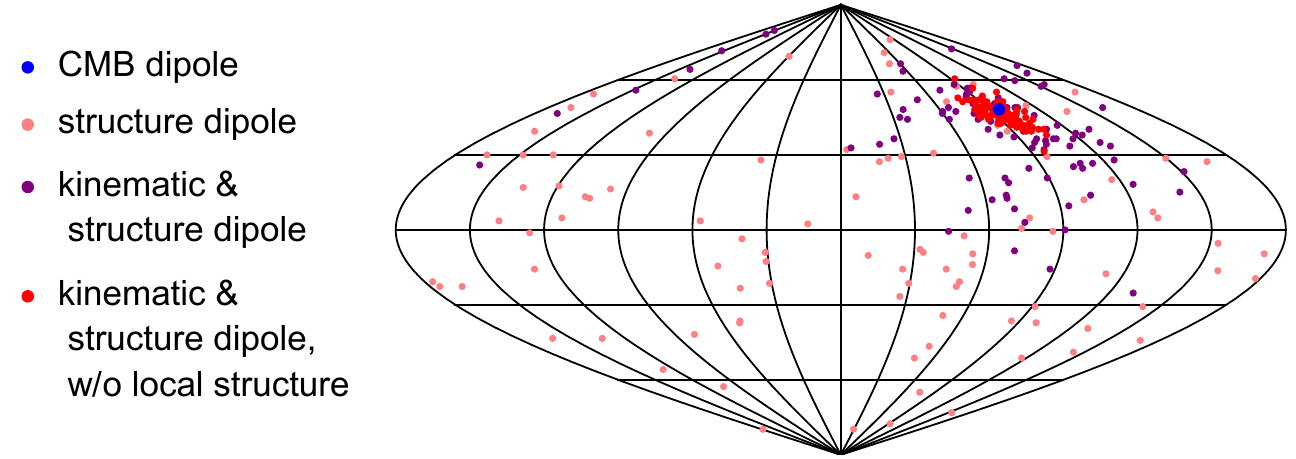}
\includegraphics[width=0.35\linewidth]{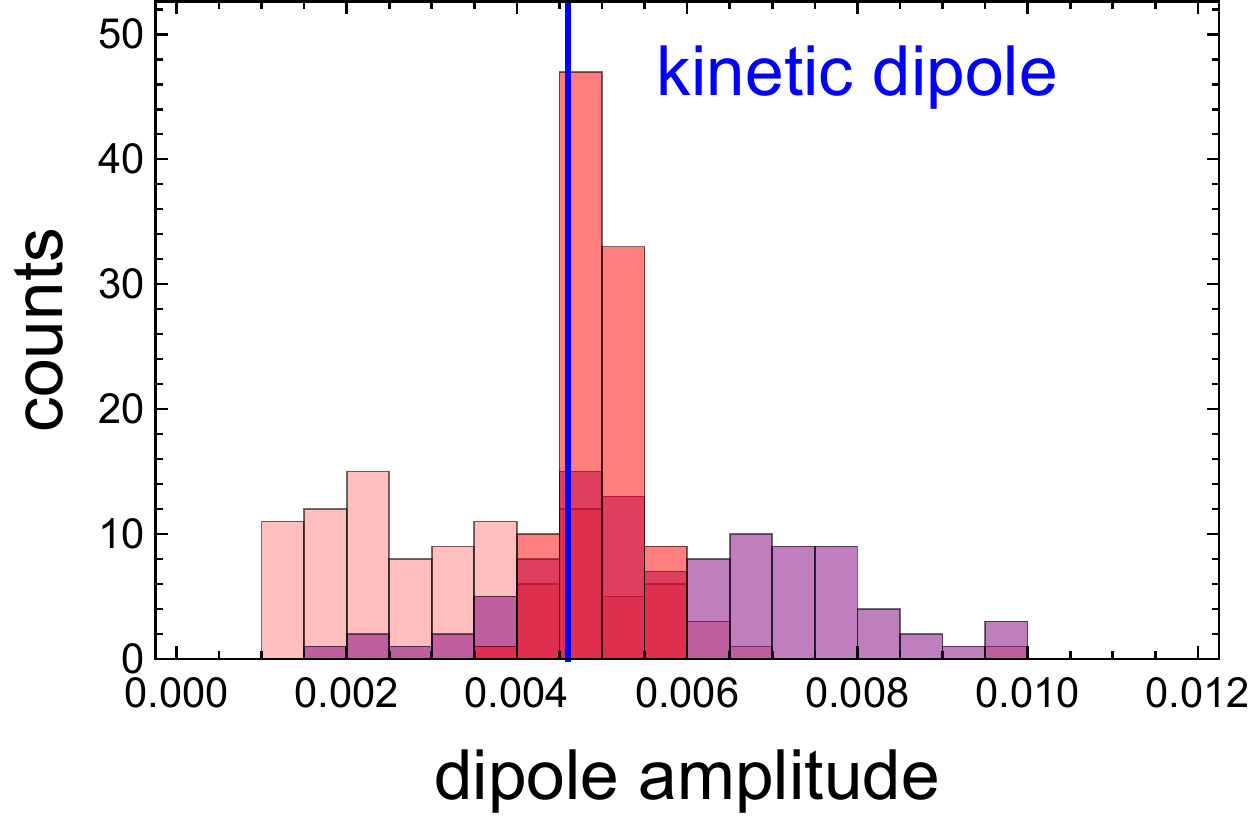}}
\caption{Dipole directions (left) and histogram of dipole amplitudes (right) based on 100 large scale structure simulations each for a flux density threshold of $22.8 \mu$Jy at 700 MHz without kinetic dipole (pink), with kinetic dipole (purple) and with the contribution from the local structure dipole removed (red). The blue dot shows the direction of the CMB dipole. The results are displayed in galactic coordinates and in stereographic projection.}
\label{dipole_direction22}
\end{figure*}

Fig.~\ref{mock_example} shows an example of a simulated sky for a flux density threshold of $22.8 \mu$Jy at a central frequency of $700$ MHz (Band 1), demonstrating  the effect of the dipole on the source counts, as the southern sky appears to be  slightly more dominated by blue than the northern hemisphere. The results from a set of $100$ such simulations is shown in Fig.~\ref{dipole_direction22}. Given the assumptions, we would expect our mocks to produce a kinematic radio dipole amplitude of $A = 0.0046$, pointing to the CMB dipole direction.  The large scale structure contributes a dipole with a mean amplitude of $A = 0.0031 \pm 0.0016$. This prediction depends on the assumed luminosity functions, spectral energy distributions, bias, redshift and luminosity evolution of radio sources, see e.g.\ \cite{2016JCAP...03..062T}. 

Fig.~\ref{dipole_direction22}  shows the expected total radio dipole, which comprises contributions from large scale structure and the proper motion of the solar system. The expected kinematic contribution dominates the structure contribution and the measured amplitude is $A = 0.0056 \pm 0.0017$ in direction $(l,b) = (263.5\pm 28.0, 38.8\pm 19.7)$ deg. The distribution  of dipole directions from the mocks is centered on the CMB dipole direction, but with some scatter due to the large scale structure. 

The structure dipole is in fact dominated by contributions from local structure. Removing the low-redshift structure dipole ($z < 0.5$), which might be possible using  optical or infra-red catalogues, or by means of the HI redshift measured by the SKA, we measure the dipole direction $(l,b) = (265.3 \pm 4.9, 46.4 \pm 4.3)$ deg, in excellent agreement with the simulated dipole direction, with an amplitude of $A = 0.0047 \pm 0.0004$, also agreeing with the input value. The distributions of dipole amplitudes  is shown in the right panel of Fig.~\ref{dipole_direction22}.

We also simulated catalogues with $S = 5, 10$ and $16 \mu$Jy, which show that the structure dipole depends on the flux density threshold, providing an extra handle to separate them from the kinematic dipole. In none of our simulations was shot noise a limitation, in contrast to contemporary  radio continuum surveys \citep{2015aska.confE..32S}. \comment{}

\section{HI galaxy redshift survey}
\label{HI_gal}
The HI galaxy redshift survey mode involves detecting the redshifted 21cm emission from many individual galaxies above the confusion limit, predominantly at low redshift ($z \lesssim 0.4$). At a minimum, the positions and spectroscopic redshifts of the detected galaxies will be available. The 21cm line widths and angular sizes of some subset of the galaxies will also be measured, allowing direct estimates of peculiar velocities to be made via the Tully-Fisher relation and Doppler magnification effects respectively.

The galaxies detected in this survey mode will not necessarily be well-resolved, but resolved detections can be used to study galaxy dynamics. The variation of the HI content of galaxies over cosmic time is also an important observable for studies of galaxy formation and evolution. All galaxies with a detectable 21cm line are expected to have strong continuum detections, and so this survey is expected to be carried out commensally with a continuum galaxy survey. In fact, characterising the continuum emission along the line of sight to HI-emitting galaxies may be a necessary step in detecting the 21cm line.

In this section, we describe the properties of a HI galaxy redshift survey using the SKA1 \meddeeps~, and the main cosmological applications of the resulting dataset.

\subsection{Survey characteristics}

The HI galaxy sample from the SKA1-MID \meddeeps~ will be sample variance-limited out to $z_{\rm max} \sim 0.4$. It will be significantly oversampled (i.e. $n(z) P(k) \gg 1$ where here $n(z)$ is the comoving number density of galaxies in this context) at $z\lesssim 0.2$, which provides an opportunity for multi-tracer studies, in which the uncertainty on certain cosmological quantities is dominated by shot noise rather than sample variance. Similarly, procedures such as void detection will be more robust thanks to the high number density. Note that Band 1 is expected to yield too few galaxies for a cosmological survey, but deep and narrow surveys may be carried out in this band to characterise the evolution of HI galaxies.

\begin{table*}[t]
\caption{Fitting coefficients for $dn/dz$ and $b(z)$ for a HI galaxy sample from the SKA1 \meddeeps, for two detection thresholds. $z_{\rm max}$ is the maximum redshift at which $n(z) P(k_{\rm NL}) > 1$, where $k_{\rm NL}$ is the non-linear scale.}
\centering
\begin{tabular}{|l|r|rrr|rr|r|r|}
\hline
{\bf Survey} & \multicolumn{1}{c|}{Thres.} & \multicolumn{1}{c}{$c_1$} & \multicolumn{1}{c}{$c_2$} & \multicolumn{1}{c|}{$c_3$} & \multicolumn{1}{c}{$c_4$} & \multicolumn{1}{c|}{$c_5$} & \multicolumn{1}{c|}{$z_{\rm max}$} & \multicolumn{1}{c|}{$N_{\rm gal} / 10^6$} \\
\hline
\multirow{2}{*}{SKA1 \meddeeps{}} & $5\sigma$~~ & 5.450 & 1.310 & 14.394 & 0.616 & 1.017 & 0.391 & 3.49~~~ \\
 & $8\sigma$~~ & 4.939 & 1.027 & 14.125 & 0.913 & -0.153 & 0.329 & 2.04~~~ \\
\hline
\end{tabular} 
\label{tab:dndzfits}
\end{table*}

\begin{table}[t]
\caption{Binned number density and bias of HI galaxies, and corresponding flux r.m.s. sensitivity, for the SKA1 \meddeeps. The assumed detection threshold is 5$\sigma$.} 

\begin{center}
\begin{tabular}{|c|c|ccr|}
\hline
 $z_{\rm min}$ & $z_{\rm max}$ & $n(z)$ [Mpc$^{-3}$] & $b(z)$ & $S_{\rm rms}$ $[\mu{\rm Jy}]$ \\
\hline
0.0 & 0.1 & $2.73 \times 10^{-2}$ & 0.657 & 117.9~~~~ \\
0.1 & 0.2 & $4.93 \times 10^{-3}$ & 0.714 & 109.6~~~~ \\
0.2 & 0.3 & $9.49 \times 10^{-4}$ & 0.789 & 102.9~~~~ \\
0.3 & 0.4 & $2.23 \times 10^{-4}$ & 0.876 & 97.5~~~~ \\
0.4 & 0.5 & $6.44 \times 10^{-5}$ & 0.966 & 93.1~~~~ \\
\hline
\end{tabular}
\label{tab:nzska1}
\end{center}
\end{table}

Basic predictions for the number density (and corresponding bias) of galaxies that will be detected by a blind SKA1 HI galaxy survey were made in \cite{2015MNRAS.450.2251Y} for the original SKA1 specifications, and \cite{Camera:2014bwa} provided a companion fitting function for the estimated magnification bias. These calculations were based on the S$^3$-SAX simulations \citep{Obreschkow:2009ha}, and assumed that any galaxy with an integrated line flux above a given (linewidth-dependent) SNR threshold would be detectable. This detection criterion implicitly assumes that a matched filter has been applied to the sources (e.g. so the total detected flux of galaxies is taken into account, even if it is spread across multiple resolution elements). \cite{2015MNRAS.450.2251Y} also includes fitting functions that can be used to rescale the number density and bias for different instrumental specifications.

Updated number density and bias predictions for the current SKA1 specifications were presented in \cite{Bull2016}, and are reproduced in Table~\ref{tab:dndzfits}, using the following fitting functions:
\begin{eqnarray} \label{eq:hi_dndz}
\frac{dn}{dz} &=& 10^{c_1}{\rm deg}^{-2}\, z^{c_2} \exp({-c_3 z})\,, \\
b(z) &=& c_4 \exp({c_5 z})\,.
\end{eqnarray}
Redshift-binned numerical values of the number density and bias are given in Table~\ref{tab:nzska1}. \cite{Bull2016} also included a survey optimisation study to establish the optimal survey area as a function of total survey time, finding that the \wideims~ would optimise the survey volume that is sample variance limited, while the \meddeeps~ would provide a reasonable trade-off between total volume and maximum redshift.

\begin{figure*}[t]
 \centering
 \includegraphics[width=0.49\textwidth]{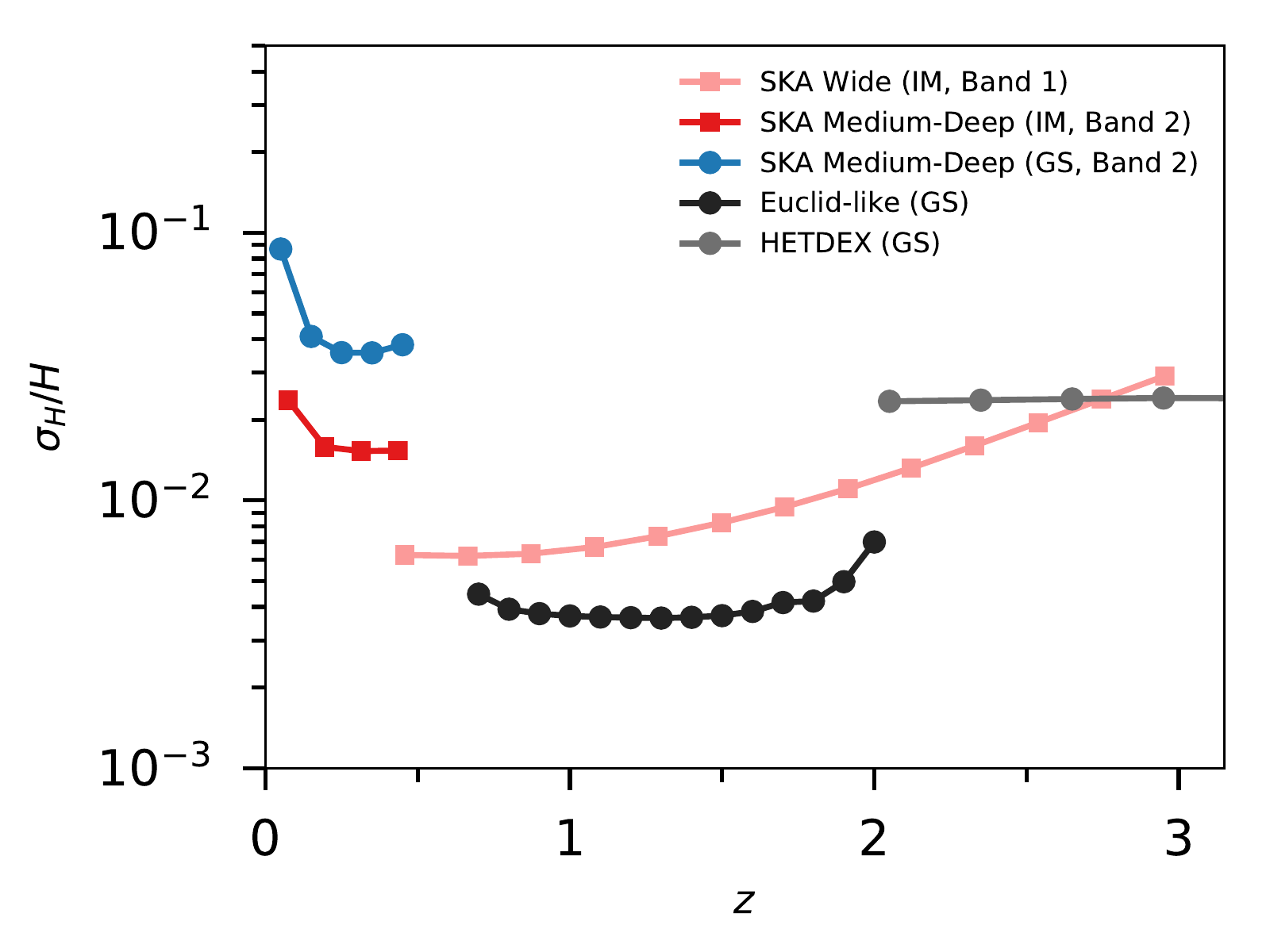}
 \includegraphics[width=0.49\textwidth]{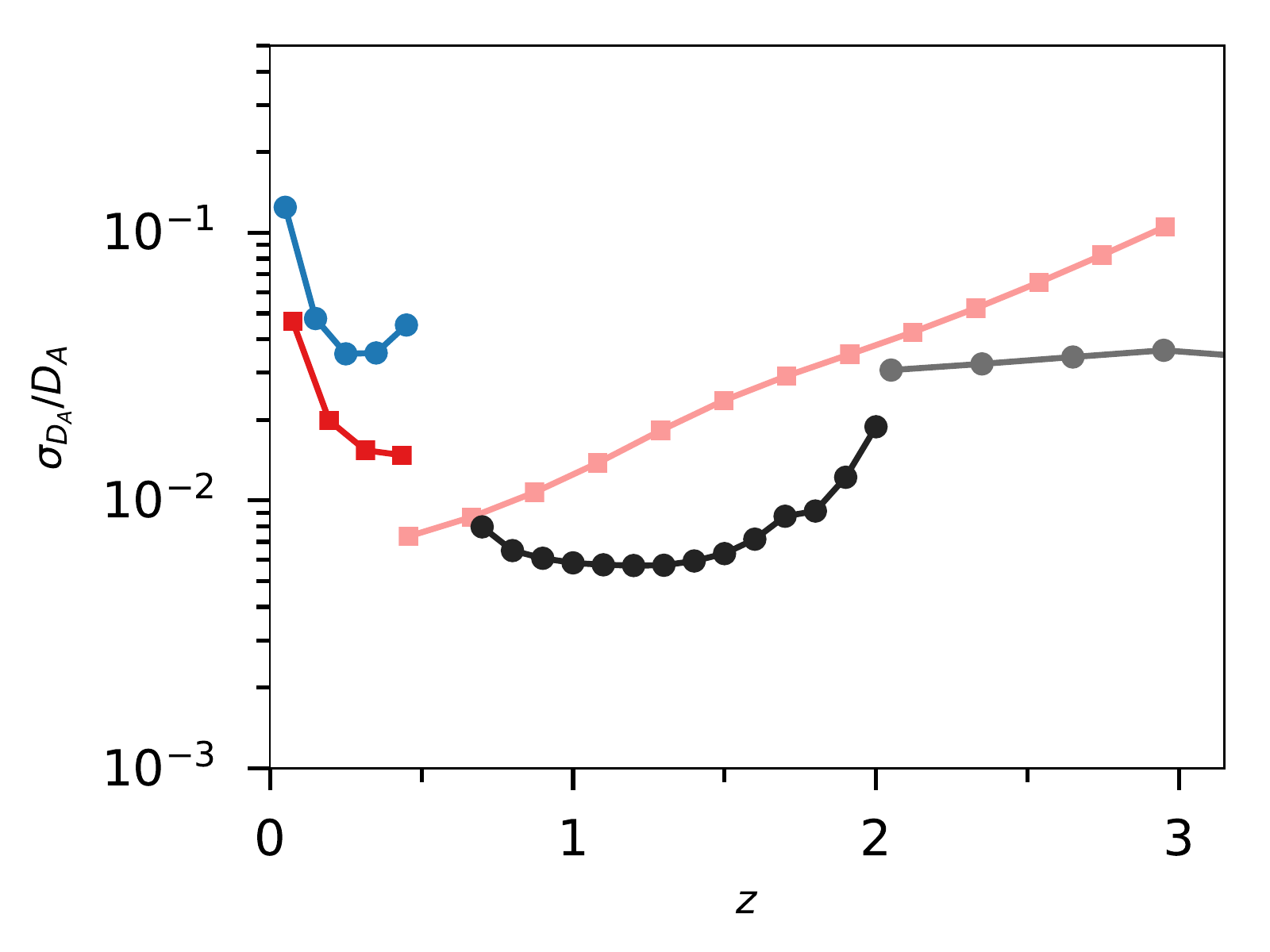}
 \caption{Forecast constraints on the cosmic expansion rate, $H$, (left panel) and angular diameter distance, $D_A(z)$, (right panel) for several different experiments, following the forecasting methodology described in \cite{Bull2016}. The SKA1 \meddeeps~ for HI galaxy redshifts is shown in light blue, HI intensity mapping are shown in red/pink (see Sec.~\ref{HI_IM} for details), and optical/NIR spectroscopic galaxy surveys are shown in black/grey.}
 \label{fig:hi-forecast} 
\end{figure*}

\begin{figure}[t]
 \centering
 \includegraphics[width=0.49\textwidth]{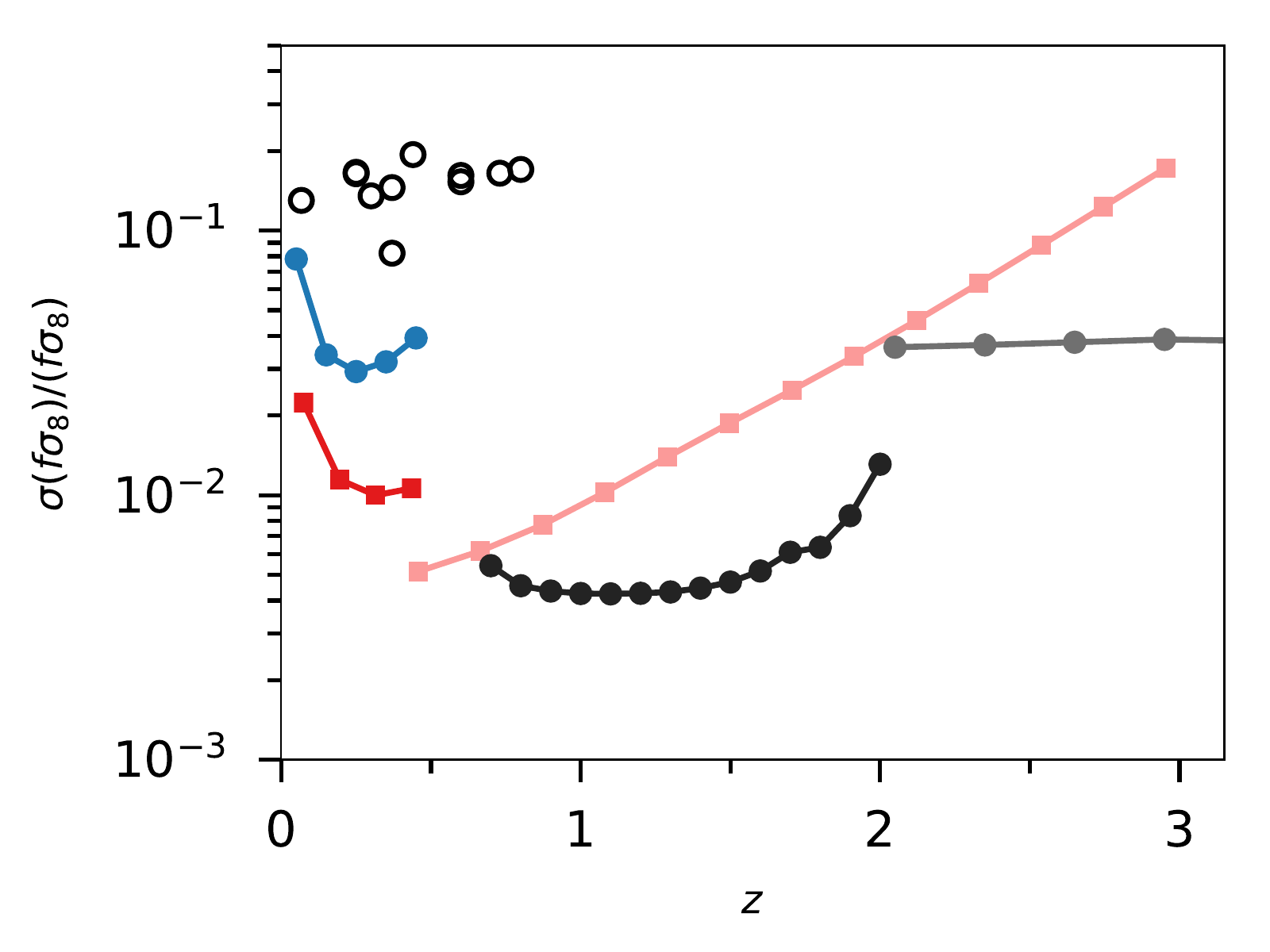}
 \caption{Forecast constraints on the linear growth rate of large-scale structure, $f\sigma_8$, for the same surveys as in Fig.~\ref{fig:hi-forecast}. Open circles show a compilation of current constraints on $f\sigma_8$ from {\cite{Macaulay:2013swa}}.}
 \label{fig:hi-forecast-fs8} 
\end{figure}

Alternative number density predictions were made in \cite{2017arXiv170408278H}, using a Bayesian line-fitting method on simulated spectra for continuum-selected galaxies (i.e. a non-blind survey). The population of galaxies that is selected by this method is quite different to those selected using the SNR threshold of \cite{2015MNRAS.450.2251Y} but, coincidentally, the predicted number density curves are very similar. Typically $\sim 10\%$ of continuum galaxies (for the \meddeeps) will have significant detections of the 21cm line using this method.

We note that bright RFI from navigation satellites is expected to impact our ability to detect HI galaxies in the redshift range from approximately $0.09 \lesssim z \lesssim 0.23$, corresponding to $1164-1300$ MHz. Terrestrial RFI is also expected to be present elsewhere in the band, but at a much lower level thanks to the excellent radio-quietness of the SKA1-MID site. Source detection algorithms can also incorporate features to reject RFI.

\subsection{Cosmological probes}

The primary purpose of spectroscopic galaxy redshift surveys is generally to measure the 3D clustering of galaxies, particularly the Baryon Acoustic Oscillation scale and Redshift Space Distortion features in the galaxy 2-point function, which we discuss below. Several other probes will be supported by the HI galaxy survey, however, providing additional information about galaxy velocities, weak lensing convergence, and the distribution of cosmic voids. Each of these will require alternative analysis pipelines to be developed, with the ability to measure marked correlation functions, galaxy sizes, and 21cm line widths, in addition to the usual 3D position information. While these probes will not drive the survey optimization, they provide new information that will enable a number of novel cosmological analyses, and hence it is important to make sure that they are accommodated in the survey specifications. It is also important to ensure appropriate sky overlap with other surveys that provide complementary information, such as optical images (for lensing studies) and $\gamma$-ray maps (for detecting dark matter annihilation in cross-correlation).

\subsubsection{Baryon Acoustic Oscillations and Redshift Space Distortions}
\label{HIGal_BAO}
The Baryon Acoustic Oscillation (BAO) feature is a preferred scale in the clustering of galaxies, set by sound waves emitted in the early Universe when photons and baryons were coupled. Since the true physical scale of the BAO is known from CMB observations, we can use the feature as a `standard ruler' to measure the cosmological expansion rate and distance-redshift relation. This is achieved by separately measuring the apparent size of the BAO feature in the transverse and radial directions on the sky, and comparing with its known physical size (set by the size of the comoving sound horizon during the baryon drag epoch, $r_s(z_d)$). The radial BAO scale is sensitive to the expansion rate, $H(z)$, while the transverse BAO scale is sensitive to the angular diameter distance, $D_A(z)$.

The HI galaxy \meddeeps~ will be able to detect and measure the BAO feature at low redshift \citep{2015MNRAS.450.2251Y, 2015aska.confE..17A, Bull2016}. This measurement has already been performed by optical spectroscopic experiments, such as BOSS and WiggleZ \citep{2017MNRAS.470.2617A, 2014MNRAS.441.3524K}, but over different redshift ranges and patches of the sky. An SKA1 HI galaxy redshift survey will add independent data points at low redshift, $z \lesssim 0.3$, which will help to better constrain the time evolution of the energy density of the various components of the Universe -- particularly dark energy. The expected constraints on $H(z)$ and $D_A(z)$ are shown in Fig.~\ref{fig:hi-forecast}, and are typically a few percent for the HI galaxy survey. While this is not competitive with the precision of forthcoming optical/near-IR spectroscopic surveys such as DESI and  {\it Euclid}, it will be at lower redshift than these experiments can access, and so is  complementary to them.

Another feature that is present in the clustering pattern of galaxies are Redshift Space Distortions (RSDs), a characteristic squashing of the 2D correlation function caused by the peculiar motions of galaxies \citep{1987MNRAS.227....1K, 2004PhRvD..70h3007S, 2011RSPTA.369.5058P}. Galaxies with a component of motion in the radial direction have their spectral line emission Doppler shifted, making them appear closer or further away than they actually are according to their observed redshifts. This results in an anisotropic clustering pattern as seen in redshift space. The degree of anisotropy is controlled by several factors, including the linear growth rate of structure, $f(z)$, and the clustering bias of the galaxies with respect to the underlying cold dark matter distribution, $b(z)$. The growth rate in particular is valuable for testing alternative theories of gravity, which tend to enhance or suppress galaxy peculiar velocities with respect to the GR prediction \citep{2008PhRvD..78f3503J, Baker:2013hia}. RSDs occur on smaller scales than the BAO feature, but can also be detected by an HI galaxy redshift survey as long as the shot noise level is sufficiently low. The SKA1 HI galaxy survey will be able to measure the normalised linear growth rate, $f\sigma_8$, to $\sim 3\%$ at $z \approx 0.3$ (see Fig.~\ref{fig:hi-forecast-fs8}). This is roughly in line with what existing optical experiments can achieve at similar redshifts (see \cite{Macaulay:2013swa} for a summary).

\begin{figure}[t]
\centering
\includegraphics[width=1.0\columnwidth]{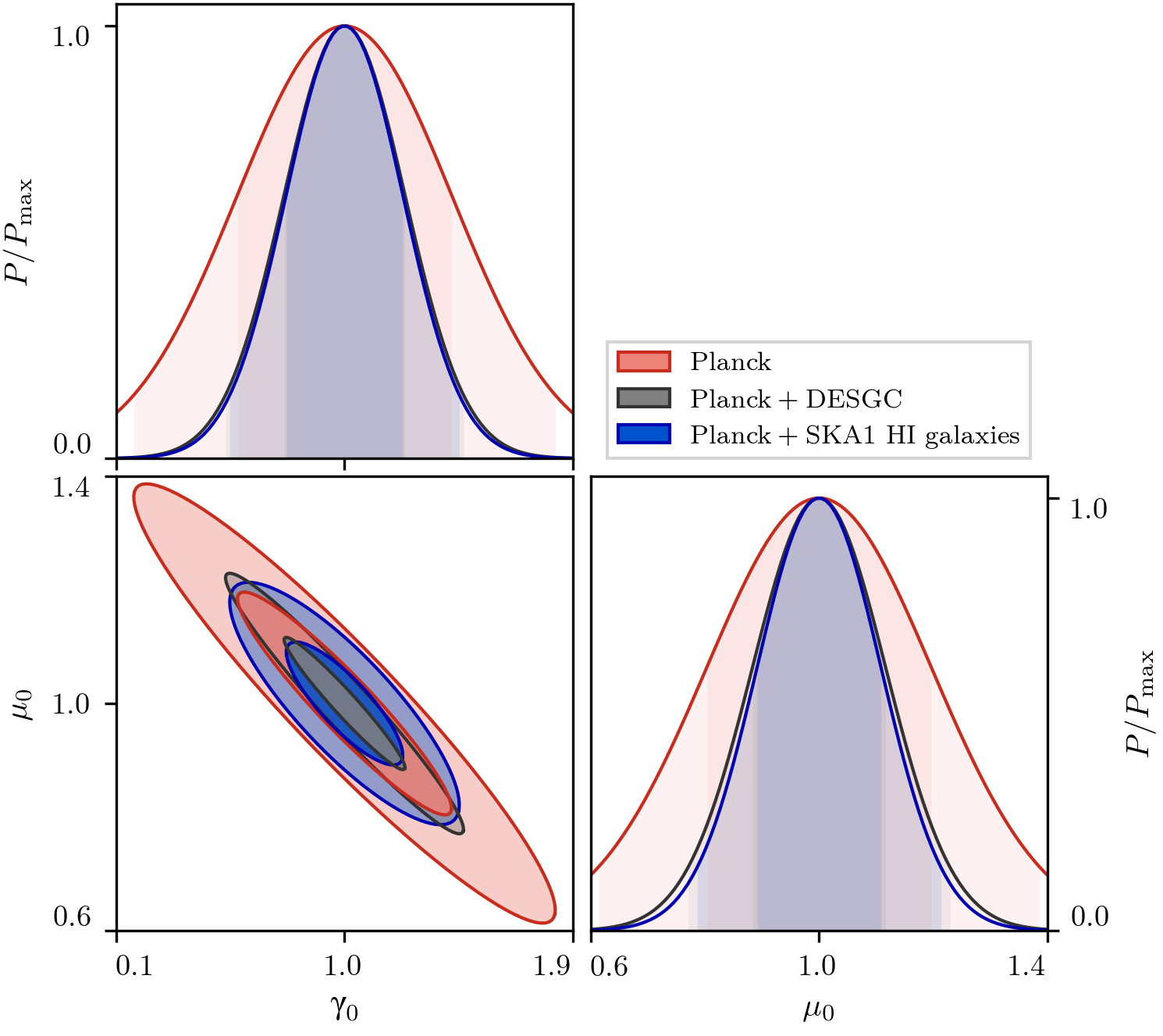}
\caption{\label{fig:mgparams} Forecast constraints on phenomenological modified gravity parameters using the broadband shape of the power spectrum, detected using the HI galaxy sample of the \meddeeps. {\it Planck} and DES (galaxy clustering only) constraints are included for comparison. The improvement from adding SKA1 is comparable to DES. Specifications for DES were taken from \cite{Lahav:2009zr}.}
\end{figure}

Fig.~\ref{fig:mgparams} shows results for when the growth rate constraints are mapped onto the phenomenological modified gravity parametrisation defined in Eqs.~(\ref{eq:mgmu}) and (\ref{eq:mggamma}).\footnote{The results in Fig.~\ref{fig:mgparams} used the forecasting code and Planck prior described in \cite{2016arXiv160606268R, 2016arXiv160606273R}.} The constraints on both $\mu_0$ and $\gamma_0$ are improved by roughly a factor of two over {\it Planck} - comparable to what can be achieved with DES (galaxy clustering only). This is not competitive with bigger spectroscopic galaxy surveys like {\it Euclid} or DESI, but does provide an independent datapoint at low redshift.

\subsubsection{Doppler magnification}
\begin{figure}[t]
\centering
\includegraphics[width=\columnwidth]{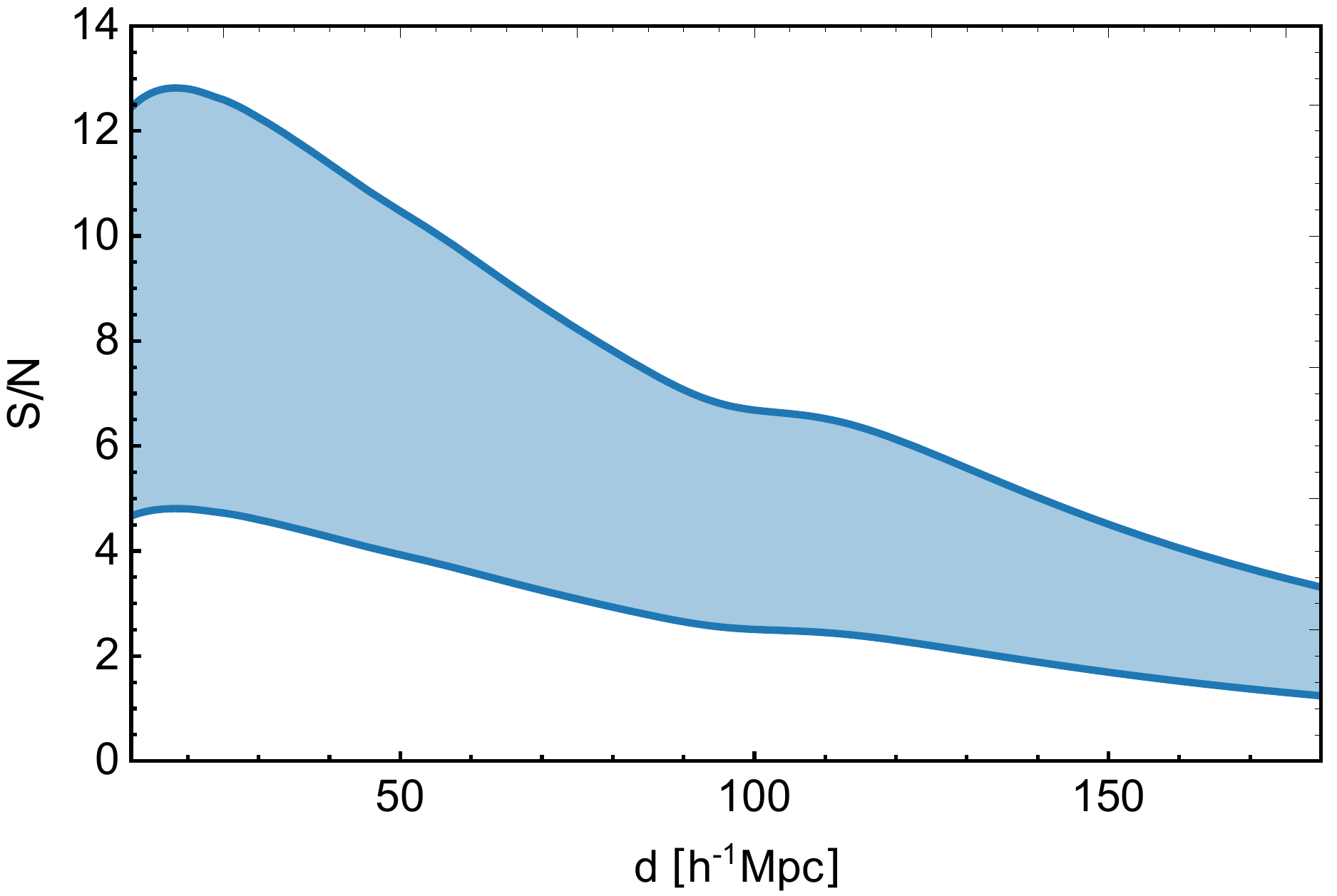}
\caption{\label{fig:doppmag} The signal-to-noise ratio of the Doppler magnification dipole for SKA1 as a function of separation $d$ at $z=0.15$ (the redshift bin in which the SNR is largest). A pixel size of $4h^{-1}\,{\rm Mpc}$ has been assumed. The upper bound and lower bounds are for convergence errors (size noise) of $\sigma_\kappa=0.3$ and $\sigma_\kappa=0.8$ respectively.}
\end{figure}

There is a contribution to the apparent magnification of galaxies due to their peculiar motion, as well as weak gravitational lensing \citep{2008PhRvD..78l3530B}. The motion of the galaxies causes a shift in their apparent radial position (as seen in redshift space), while their angular size depends only on the actual (real space) angular diameter distance. As such, a galaxy that is moving away from us will maintain fixed angular size while appearing to be further away than it really is (and thus `bigger' than it should be for a galaxy at that apparent distance). This effect has been called Doppler magnification, and dominates the weak lensing convergence at low redshift \citep{2014MNRAS.443.1900B, 2017MNRAS.471.3899B, 2017MNRAS.472.3936B, 2018arXiv181012793A}. It can be detected statistically through the dipolar pattern it introduces in the density-convergence cross-correlation, $\langle \kappa \delta_g \rangle$. The galaxy density, $\delta_g$, can be measured from the 3D galaxy positions, while the convergence, $\kappa$, can be estimated from the angular sizes of the galaxies. 

\begin{figure}[t]
\hspace{-1.5em}\includegraphics[width=1.08\columnwidth]{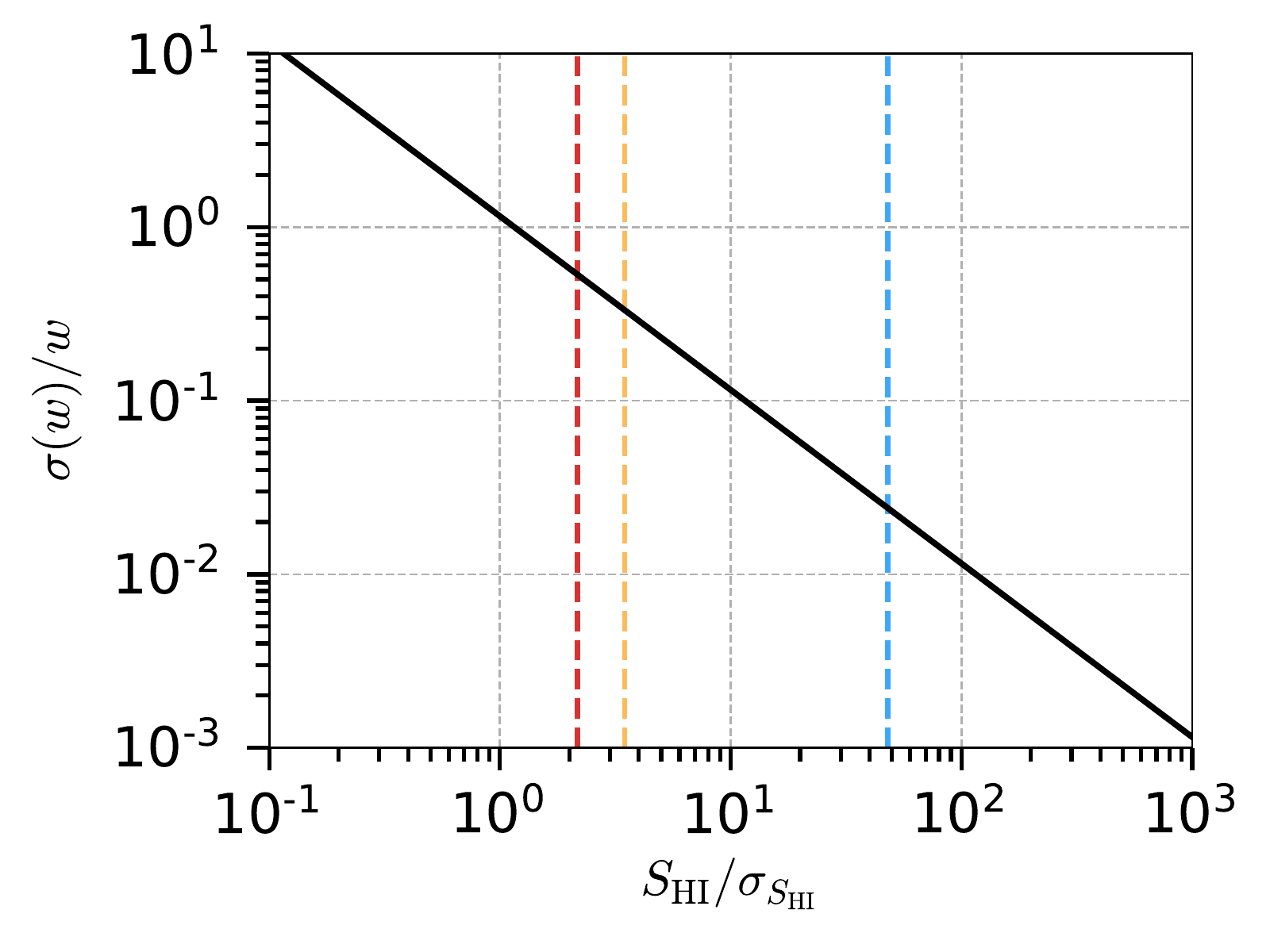}
\caption{\label{fig:linewidth} Expected fractional error on the width of the 21cm line, as a function of the signal-to-noise ratio on the integrated line flux. The vertical dashes lines show three different detection thresholds: (red) $5\sigma$ threshold on the peak per channel SNR; (yellow) $8\sigma$ threshold on the peak per-channel SNR; and (blue) threshold corresponding to $\sigma(v_{\rm pec}) < 0.2 c$.}
\end{figure}

As discussed above, an SKA1 HI galaxy redshift survey will yield high number densities of galaxies with spectroscopic redshifts at $z \lesssim 0.4$, approximately covering the redshift range where Doppler magnification dominates the weak lensing convergence. If the HI-emitting galaxies can be resolved, their sizes can also be measured (e.g. from their surface brightness profile in continuum emission), making it possible to measure the Doppler magnification signal using a single survey. Galaxy size estimators often suffer from large scatter, and it remains an open question as to how well SKA1 will be able to measure sizes. This scatter has a significant effect on the expected SNR of the Doppler magnification signal. There is a known relation between the size of an HI disk and the HI mass \citep{2016MNRAS.460.2143W} that shows very little scatter over several orders of magnitude, however. For objects that are spatially resolved in HI, their expected sizes can be computed from their HI masses, and compared with their apparent sizes.

Following the forecasting methodology of \cite{2017MNRAS.472.3936B}, we expect SKA1 to achieve a signal-to-noise ratio of $\approx  8$ on the Doppler magnification dipole for galaxies separated by $\sim 100$ $h^{-1}$Mpc (Fig.~\ref{fig:doppmag}), assuming a size scatter of $\sigma(\kappa) = 0.3$ (comparable to what optical surveys can achieve). The cumulative SNR over $0.1 \le z \le 0.5$, for the full range of separations, is $\approx 40$.

\subsubsection{Direct peculiar velocity measurements}

The Tully-Fisher relation \citep{1977A&A....54..661T} can be used to infer the intrinsic luminosity of a galaxy from its 21cm line width, which is a proxy for rotational velocity. Combined with the redshift of the line and a measurement of the galaxy inclination, this makes it possible to measure the galaxy's peculiar velocity in the line-of-sight direction. The statistics of the peculiar velocity field, sampled by a large set of galaxies, can then be used to measure various combinations of cosmological quantities. Peculiar velocity statistics are particularly sensitive to the growth rate of structure, and so can be used as powerful probes of modified gravitational physics \citep[e.g.][]{Hellwing:2014nma, 2014MNRAS.445.4267K, Ivarsen:2016xre}.

Measuring the width of the 21cm line requires line detections with significantly better signal-to-noise than would be needed to measure redshift alone. Fig.~\ref{fig:linewidth} shows the expected fractional error on the 21cm linewidth of a galaxy as a function of the signal-to-noise ratio on the integrated flux of the line, assuming a simplified Gaussian line profile model. The $5\sigma$ and $8\sigma$ thresholds (on the peak per-channel SNR, not the integrated flux) from Table~\ref{tab:dndzfits} are shown as red and yellow dashed lines respectively. These are the thresholds we assumed for 21cm line detection for a redshift-only survey. To measure the peculiar velocity to better than $20\%$ of the speed of light \citep[as required by the analysis in][]{2014MNRAS.445.4267K}, a fractional measurement precision of $\sim 2.4\%$ is required on the linewidth, which translates to a peak per-channel SNR of $\sim 110\sigma$ according to Fig.~\ref{fig:linewidth}. As such, the number density of galaxies for which peculiar velocity measurements are available will be significantly lower than for the redshift-only sample. Some way of measuring the inclination (e.g. from continuum or optical/NIR images) for all galaxies in the sample is also required. Nevertheless, direct measurements of the peculiar velocity field are sensitive cosmological probes, so the constraining power of even relatively small peculiar velocity samples can be substantial. Forecasts for SKA precursor experiments were presented in \cite{2014MNRAS.445.4267K}, and showed that a $\sim 3\%$ measurement of $f\sigma_8$ should be achievable at $z \simeq 0.025$ with a combined redshift + velocity survey, for example.

\subsubsection{Void statistics}

Future large-area galaxy surveys will offer an unprecedented spectroscopic view of both large and small scales in the cosmic web of structure. Thanks to its high galaxy density and low bias, the SKA1 HI galaxy survey will allow unusually small voids and comoving scales to be probed compared to other spectroscopic surveys. 

The number counts \citep{Pisani:2015jha, Sahlen:2015wpc}, shapes \citep{Massara:2015msa}, RSDs \citep{Sutter:2014oca}, and lensing properties \citep{Spolyar:2013maa} of voids are examples of sensitive void-based probes of cosmology. Voids are particularly sensitive to the normalization and shape of the matter power spectrum, its growth rate, and the effects of screened theories of gravity which exhibit modifications to General Relativity in low-density environments \citep{Voivodic:2016kog}. This is because void distributions contain objects ranging from the linear to the non-linear regime, across scale, density and redshift \citep{Sahlen:2016kzx}. 

\begin{figure}[t]
\centering
\includegraphics[trim={13.2cm 0.3cm 0cm 0.3cm},clip,width=0.5\textwidth]{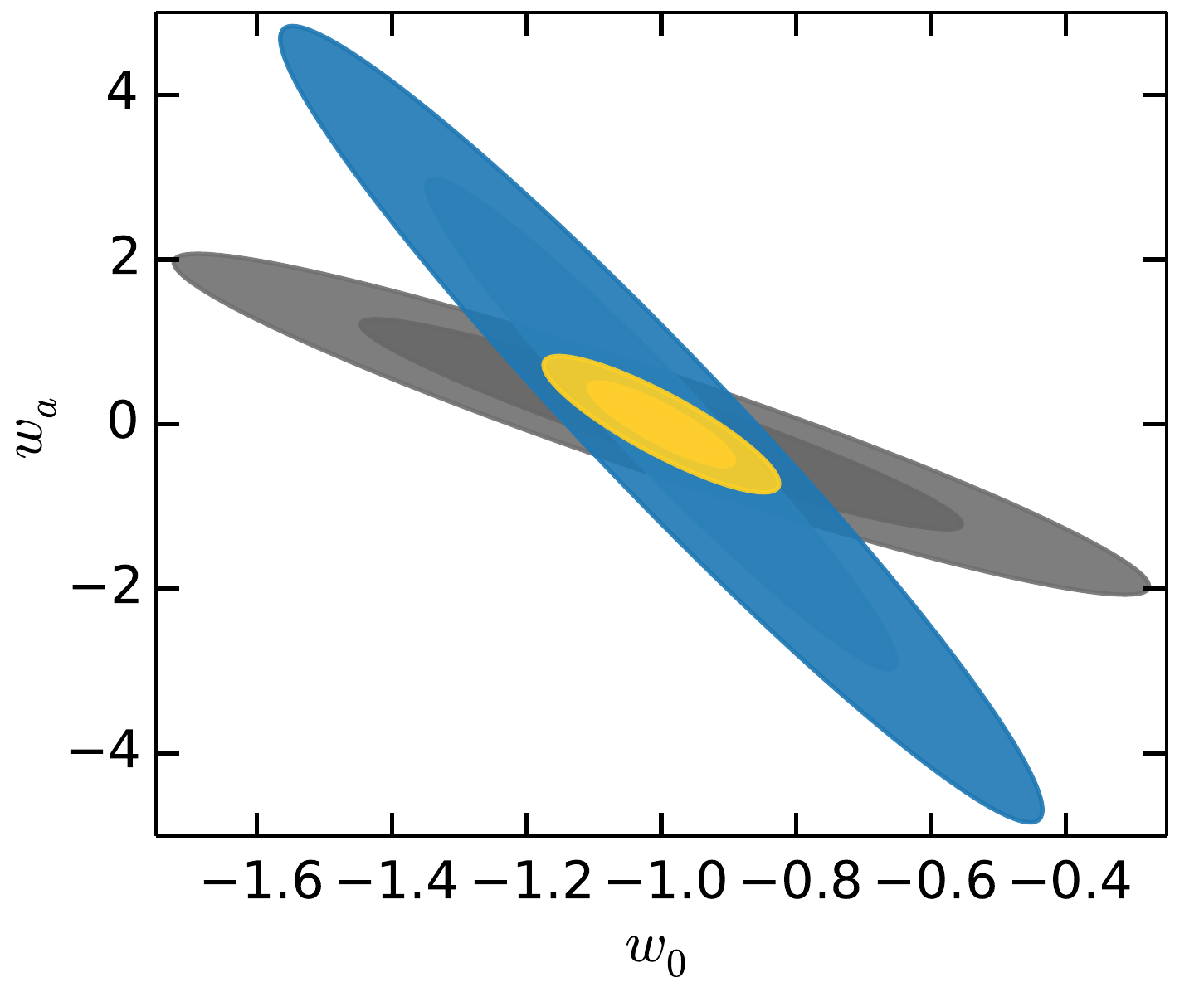}
\caption{\label{fig:voidw0wa} Forecast marginalized parameter constraints for $w_0$ and $w_a$ from the void counts  of the HI galaxy \meddeeps~(grey), {\it Planck} (blue), and both combined (yellow). Apart from the cosmological parameters, we have also marginalized over uncertainty in void radius \citep{Sahlen:2016kzx}, and in the theoretical void distribution function \citep{Pisani:2015jha}.}
\end{figure}

\begin{table}[t]
\vspace{-5pt}
\caption{Forecast dark energy constraints for void counts.}
\label{tab:voidw0wa}
\begin{center}
\begin{tabular}{|c|c|c|c|c|}
\hline
Survey & $\sigma(w_0)$ & $\sigma(w_a)$ & FoM \\
\hline
SKA1 HI galaxy void counts & $0.22$ & $1.84$ & 9 \\
{\it Planck}+lensing+BAO & $0.30$ & $0.85$ & 13 \\
Joint SKA1 + Planck & $0.07$ & $0.34$  & 84 \\ 
\hline
\end{tabular}
\end{center}
\vspace{-12pt}
\end{table}%

We forecast cosmological parameter constraints from the HI galaxy \meddeeps~ in our fiducial cosmology, using a Fisher matrix method. The void distribution is modeled following \citep{Sahlen:2015wpc, Sahlen:2016kzx} using an approximate modeling scheme to incorporate the effects of massive neutrinos on the void distribution \citep{2018arXiv180702470S}. We also take into account the galaxy density and bias for the survey. Below $z \approx 0.18$, the survey is limited by the void-in-cloud limit. Voids smaller than this limit tend to disappear due to collapse of the overdensity cloud within which they are situated.

We expect to find around $4 \times 10^4$ voids larger than $10\,h^{-1}$Mpc. The marginalised constraints on $w_0$ and $w_a$ inferred from void abundances are shown in Fig.~\ref{fig:voidw0wa} and Table~\ref{tab:voidw0wa}. The SKA1 void counts and {\it Planck}+lensing+BAO parameter constraints offer similar but complementary constraining power, and their combination can strengthen the Figure of Merit by a factor of $\sim 6-10$. Also including the sum of neutrino masses as a free parameter only marginally weakens the void constraints \citep{2018arXiv180702470S}. Recalling that additional cosmological information is also available in e.g. shapes/profiles, voids are therefore a promising application of an SKA1 HI galaxy survey. 

\subsubsection{Particle dark matter searches in cross-correlation with \texorpdfstring{$\gamma$}-ray maps}

\citet{Camera:2012cj} and subsequent studies \citep{Fornengo:2013rga, Camera:2014rja} proposed a new technique for indirect particle dark matter detection, based on the cross-correlation of direct gravitational probes of dark matter, such as weak gravitational lensing or the clustering of galaxies. A cross correlation between the unresolved $\gamma$-ray background seen by the \textit{Fermi} Large Area Telescope \citep[LAT;][]{2009ApJ...697.1071A} and various cosmological observables has already been detected \citep{Fornengo:2014cya, Xia:2015wka, Cuoco:2015rfa, Branchini:2016glc}. Currently, the vast majority of the $\gamma$-ray sky is unresolved and only a few thousand $\gamma$-ray sources are known. On large scales, non-thermal emission mechanisms are expected to greatly exceed any other process in the low-frequency radio band and the $\gamma$-ray range. Thus, radio data is expected to correlate with the $\gamma$-ray sky and can be exploited to filter out the information concerning the composition of the $\gamma$-ray background contained in maps of the unresolved $\gamma$-ray emission.

\begin{figure}[t]
\centering
\includegraphics[width=0.95\columnwidth]{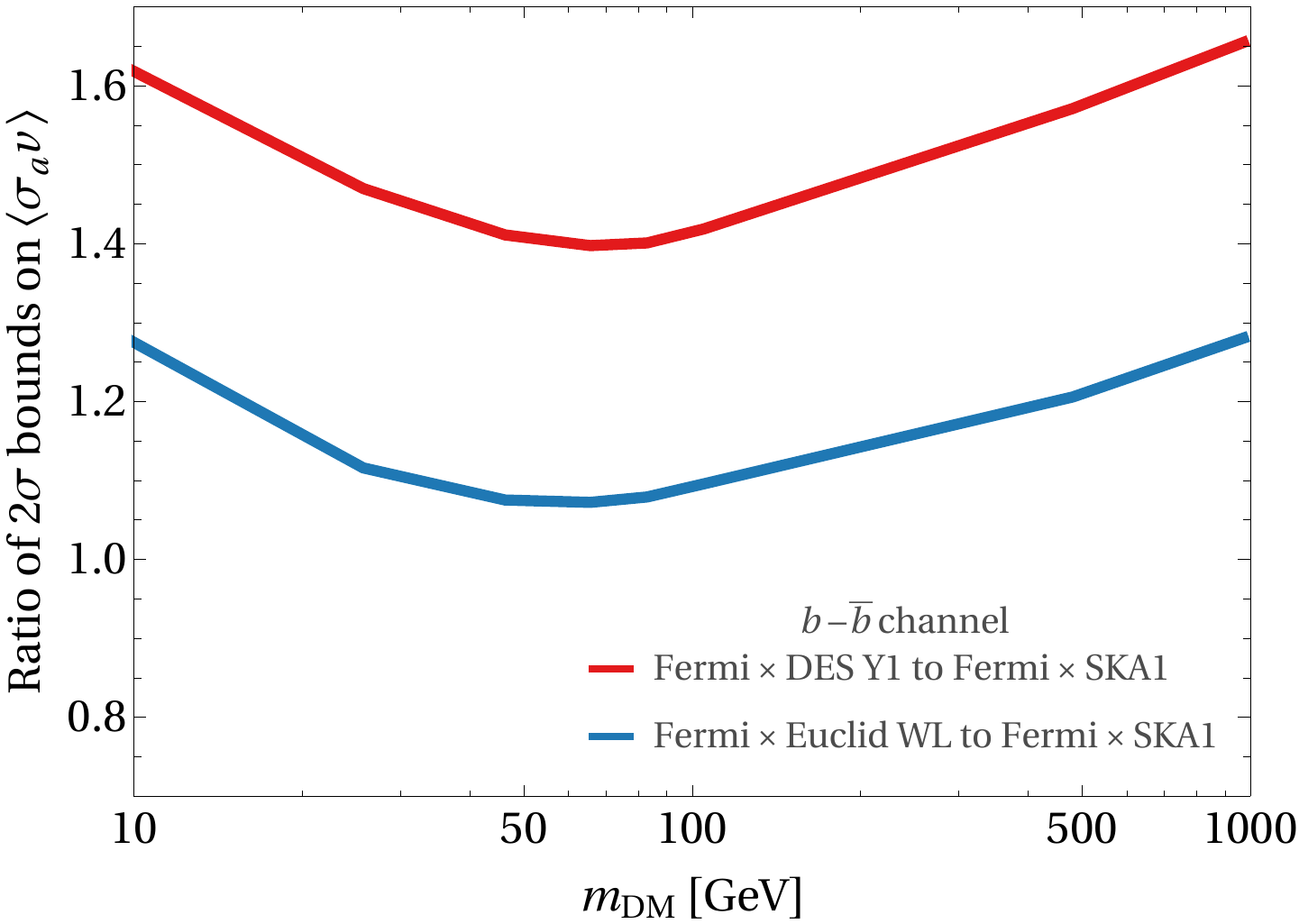}
\caption{Improvement factor in constraints on the dark matter annihilation cross-section as a function of particle dark matter mass, when an SKA1 HI galaxy survey is used for the cross-correlation with \textit{Fermi}-LAT data, instead of DES year 1 (blue) or \textit{Euclid} (orange).}\label{fig:ska1-fermi}
\vspace{-1em}
\end{figure}

Here, we present forecasts for the cross-correlation of SKA1 HI galaxies and the $\gamma$-ray sky from \textit{Fermi}. A major added value of SKA1 HI galaxies is that is their redshift distribution peaks at low redshift and has an extremely low shot noise \citep[see][Fig.~4]{2015MNRAS.450.2251Y}. This is the very regime where the non-gravitational dark matter signal is strongest. Specifically, we adopt an SKA1 HI galaxy survey with specifics given in \citet{2015MNRAS.450.2251Y} for the baseline configuration. We consider only galaxies in the redshift range $0<z\le0.5$, which we further subdivide into 10 narrow spectroscopic redshift bins. For the $\gamma$-ray angular power spectrum, we employ the fitting formul\ae\ found by \citet{Troster:2016sgf} for Pass-8 \textit{Fermi}-LAT events gathered until September 2016 (i.e. over eight years of data taking). This is a conservative choice, as by the time the SKA1 HI galaxy catalogue will be available, a much larger amount of \textit{Fermi}-LAT data will be available. Fig.~\ref{fig:ska1-fermi} shows the improvement on bounds on particle dark matter cross-section as a function of dark matter mass when SKA1 HI galaxies are used, compared with the two main probes studied in \citet{Camera:2014rja}, i.e.\ cosmic shear from DES (Year 1 data only) and \textit{Euclid}. The high density of spectroscopically detected HI galaxies from the HI galaxy survey provides constraints on particle dark matter properties that are $10-60\%$ tighter than with state-of-the-art and even future experiments.

\subsubsection{Cross-correlation with gravitational wave sources}

Gravitational wave (GW) experiments are expected to directly detect tens to thousands of binary black holes (BBHs) and neutron star (NS) coalescence events per year over the coming decade \citep[e.g.][]{2018PhRvD..97b3012N}, depending on the natural rate of mergers and how detector sensitivity improves with time. As the number of known events increases, and the accuracy of source localisation improves, large GW source catalogues numbering in the thousands to tens of thousands of events will be constructed. These can then be cross-correlated with galaxy surveys such as the SKA1 \meddeeps~ to constrain cosmological models and determine properties of the BBH and NS host galaxy/halo populations.

GWs are lensed by intervening large-scale structure just as light is, and so
by cross-correlating foreground galaxies (that act as lenses) with a background of GWs, one can perform tests of general relativity and dark energy models in a way that is independent from current tests using galaxy surveys alone~\citep{Raccanelli:2017}. Forecasts are not currently available for an SKA1 HI galaxy survey, but $\mathcal{O}(10\%)$ constraints on $w_0$, $w_a$, $\mu_0$, and $\gamma_0$ are expected to be achievable with an SKA2 HI galaxy survey~\citep{Raccanelli:2017}.

The angular correlation of GW sources with different types of galaxies can also be used to understand if merging high-mass BBHs preferentially trace star-rich galaxies (as would be the case if they form from objects at the endpoint of stellar evolution), or the dark matter distribution (as would be the case if they are primordial black holes). The most star-rich galaxies are typically found in halos of mass $\sim10^{11-12}M_\odot$, while almost all mergers of primordial BBHs would happen in halos of $\lesssim 10^6 M_\odot$, as shown in~\cite{Bird2016}. Other models (e.g. where high-mass BBHs are the relics of Population III stars) also predict different host halo populations. The range of host halo masses determines the mean bias of the host population. This can be measured through the cross-correlation of galaxy populations of known bias with the GW source catalogue, therefore determining the nature of BBH progenitors~\citep{Raccanelli2016, Scelfo:2018}. HI galaxies are present across a wide range of halo masses, but there is expected to be a cut-off below $\sim 10^8 M_\odot$, where self-shielding of the HI from the ionising UV background fails \citep[e.g.][]{2010MNRAS.407..567B}. The cross-correlation between HI galaxies and GW sources can therefore be expected to strongly constrain the primordial black hole scenario.

\subsubsection{HI model uncertainties}
Cosmological constraints from the HI galaxy survey are also subject to uncertainties in the abundance and spatial distribution of neutral hydrogen. The combination of current astrophysical uncertainties on the neutral hydrogen density and bias parameters, $\Omega_{\rm HI}$ and $b_{\rm HI}$, can be shown to lead to about a $60-100\%$ uncertainty in current models of the HI power spectrum \citep{hptrcar2015}.

There have been numerous efforts to build accurate halo models of the HI distribution \citep[e.g.][]{2010MNRAS.407..567B, 2013MNRAS.434.2645D, Villaescusa-Navarro:2014cma, hpar2017, hparaa2017, 2018arXiv180410627P, 2018arXiv180409180V}, with free parameters typically constrained using some subset of currently available HI observables (galaxy number counts, intensity mapping observations, and Damped Lyman-$\alpha$ systems) across redshifts $0-5$ in the post-reionisation universe. HI galaxy redshift and HI intensity mapping surveys with SKA1 will greatly expand the amount of data available to constrain these models, leading to significantly enhanced precision in our knowledge of the relevant parameter values (e.g. see Table~\ref{tab:omHIbHI} below), and allowing the models themselves to be distinguished from one another. Recent forecasts by \cite{2018arXiv180410627P} also suggest that, once priors on HI model parameters from existing observations are applied, the cosmological parameter constraints from an SKA1 HI intensity mapping survey will be generally insensitive to remaining uncertainties in the astrophysical model. The same conclusion is also expected to hold for the HI galaxy redshift survey, at least if we restrict our attention to linear scales, $k \lesssim 0.14 {\rm Mpc}^{-1}$.

\section{HI intensity mapping}

\label{HI_IM}

Intensity mapping of the neutral Hydrogen line (HI IM) has been proposed as an innovative technique to probe the large scale structure of the Universe and deliver precision constraints on cosmology \citep{2001JApA...22...21B, 2004MNRAS.355.1339B, 2006astro.ph..6104P, 2008PhRvL.100p1301L, 2017arXiv170909066K}. 
It relies on observations of the sky intensity from the integrated 21cm line emission over a wide sky area.
For a reasonably large 3D pixel in solid angle and frequency interval, we expect to have several HI galaxies in each pixel so that their combined emission will provide a strong signal. Fluctuations in the observed intensity of this redshifted HI emission will follow fluctuations in the underlying matter density as traced by the HI emitting galaxies, allowing the density field to be reconstructed on sufficiently large scales from intensity maps. Although with low angular resolution, it is well matched to the scales required for cosmology.
Moreover, as we are probing a specific emission line (21cm) we have immediately a one to correspondence between observed frequency and redshift, which delivers very high redshift resolution. Such survey is much less time consuming than a spectroscopic galaxy survey, which requires a high signal to noise detection of each individual galaxy. 

On the other hand, there will be several foregrounds that will contaminate the HI intensity mapping signal at the observed frequencies. Cleaning such contaminants is therefore a crucial process in using this technique for cosmology and its convolution with instrumental effects poses a serious challenge \citep{2015ApJ...814..145A, 2014MNRAS.441.3271W, 2015aska.confE..35W, Olivari:2017bfv}. Note however that this line has little contamination from other spectral lines, which is an important advantage over the use of other intensity mapping tracers \citep{2017MNRAS.464.1948F}. 

Several experiments have been proposed in order to measure this signal, using single dish telescopes or interferometers \citep{2013MNRAS.434.1239B, 2014SPIE.9145E..22B, 2015ApJ...798...40X, 2016SPIE.9906E..5XN}. A precursor survey to the SKA1 with MeerKAT has also been proposed \citep{2017arXiv170906099S}.
Measurements using the Green Bank Telescope (GBT) produced the first tentative) detection of the cosmological HI intensity signal by cross-correlating with the WiggleZ redshift survey \citep{Chang:2010jp,2013MNRAS.434L..46S,2013ApJ...763L..20M}. More recently, a survey using the Parkes telescope made a detection in cross-correlation with the 2dF survey \citep{2018MNRAS.tmp..340A}.

The large dish array of the SKA-MID can be exploited for HI intensity mapping measurements. However, SKA-MID in interferometric mode does not provide enough short baselines to map the scales of interest with sufficient signal-to-noise \citep{Bull_2015}. The alternative is to use the array in single-dish mode instead. The large number of dishes available with SKA1-MID will guarantee a high survey speed for probing the HI signal and have the potential to probe cosmology over a wide range of scales with high signal to noise \citep{2015aska.confE..19S}. Keeping the interferometer data will allow to create high resolution sky images which can be used for other science as well as calibration.
In the following we consider the \wideims~ ($0.35 < z < 3$) using the auto-correlation information from each dish, although the same technique can in principle be used for the \meddeeps~ ($0<z<0.4$).

We also present the prospects of cosmology with \deeplowims~ for HI intensity mapping at $3<z<6$. One of the prime purposes of the LOW instrument is the detection of the HI gas distribution during the Epoch of Reionisation (EoR), which has been constrained to conclude at $z>6$. The $200-350$ MHz range of LOW is not the focus of EoR observations, but the EoR pipeline can provide intensity maps at these frequencies offering unique opportunities for high redshift cosmology. The combination of the SKA1-MID and LOW surveys considered here will provide an unique picture of HI on cosmological scales over a wide redshift range ($0<z<6$).

\subsection{The HI signal and power spectrum} 
\subsubsection{Temperature and bias}

The total brightness temperature at a given redshift and in a unit direction ${\bf n}$ on the sky can be written as 
\begin{equation}\label{ps-rsd}
T_{b}(z,{\bf n}) \approx \overline{T}_{b}(z) \Big[1+b_{\rm HI}(z)\delta_m(z,{\bf n})-\frac{(1+z)}{H(z)}\,n^i\partial_i\big({\bf n}\cdot {\bf v} \big)\Big], 
\end{equation}
where $b_{\rm HI}$ is the HI galaxy bias, $\delta_m$ is the matter density contrast, ${\bf v}$ is the peculiar velocity of emitters and the average signal $\overline{T}_{b}$ is determined by the comoving HI density fraction $\Omega_{\rm HI}$. The last term in braces describes the effect of Redshift Space Distortions (RSD). 
The signal will be completely specified once we have a prescription for the $\Omega_{\rm HI}$ and $b_{\rm HI}$. This can be obtained by making use of the halo mass function, ${\rm d}n/{\rm d}M$ 
and halo bias, relying on a model for the amount of HI mass in a dark matter halo of mass $M$, i.e. $M_{\rm HI}(M)$ \citep[see][for details]{2015aska.confE..19S}. 
Simulations have found that almost all HI in the post-reionisation Universe resides within dark matter halos \citep{Villaescusa-Navarro:2014cma, FVN_2018}. This fact justifies the usage of halo models to study the spatial distribution of cosmic neutral hydrogen \citep{Hamsa_2017,EmaPaco,2018arXiv180302477W,FVN_2018}.

\subsubsection{Power Spectrum}
The first aim of the intensity mapping survey will be to measure the HI power spectrum (or its large sky equivalent, the angular power spectrum). In addition, we will take advantage of multi-wavelength coverage (e.g. BOSS, DES, Euclid, LSST see section~\ref{sec:synergysurveys}) to detect the signal in cross-correlation. The HI power spectrum signal (with RSDs) can be written as
\begin{equation}
P^{\rm HI}(z,k) = \bar{T}_b(z)^2b_{\rm HI}(z)^2[1+\beta_{\rm HI}(z)\mu^2]^2P(z,k) \, ,
\end{equation} which allows to break the  degeneracy between $\Omega_{\rm HI}$ and $b_{\rm HI}$\citep{Masui:2012zc}. The cross-correlation power spectrum will also depend on the galaxy bias, $b_{\rm g}$ and the cross-correlation coefficient $r$ of the two probes, and can be used, as mentioned, to mitigate systematic effects.

The following forecasts make use of the Fisher matrix formulation. Details of the noise calculation for both MID (single dish and interferometer) and LOW can be found in \citet{Bull_2015} and \citet{2015aska.confE..19S}. Details on SKA1-LOW EoR surveys can be found in \citet{2015aska.confE...1K}. Particular care must be taken when combining MeerKAT and SKA1-MID dishes due to different primary beams and bands. Note that, when considering measurements with the interferometer, we assume a strict non-linear cutoff to define the maximum wavevector in the Fisher matrix, $k_{\rm max} = 0.2h/{\rm Mpc}$ at all redshifts. This is a conservative choice, much smaller than the instrumental cutoff.

The finite number of HI samples in the intensity maps also results into a shot noise contribution on the power spectrum measurements. In hydrodynamic simulations, \cite{FVN_2018} found that the amplitude of the HI shot-noise is negligible at $z\leq5$ \citep[see also][]{EmaPaco} and therefore BAO measurements through HI intensity mapping will barely be affected by this. They also found values of the linear HI bias equal to 0.84, 1.49, 2.03, 2.56, 2.82 and 3.18 at redshifts 0, 1, 2, 3, 4 and 5, respectively. While the HI bias is essentially scale-independent down to $k\simeq1~h{\rm Mpc}^{-1}$ at $z=1$, at redshifts $z\geq3$ the HI bias is scale-dependent already at $k=0.3~h{\rm Mpc}^{-1}$. In the following, we forecast the constraints on the linear bias $b_{\rm HI}$ by the SKA1 IM surveys, however, they will also be the first surveys to investigate the scale-dependence of the HI clustering signal for all redshifts $0<z<6$.\\ 

\begin{figure}
\centering
\includegraphics[width=0.95\columnwidth]{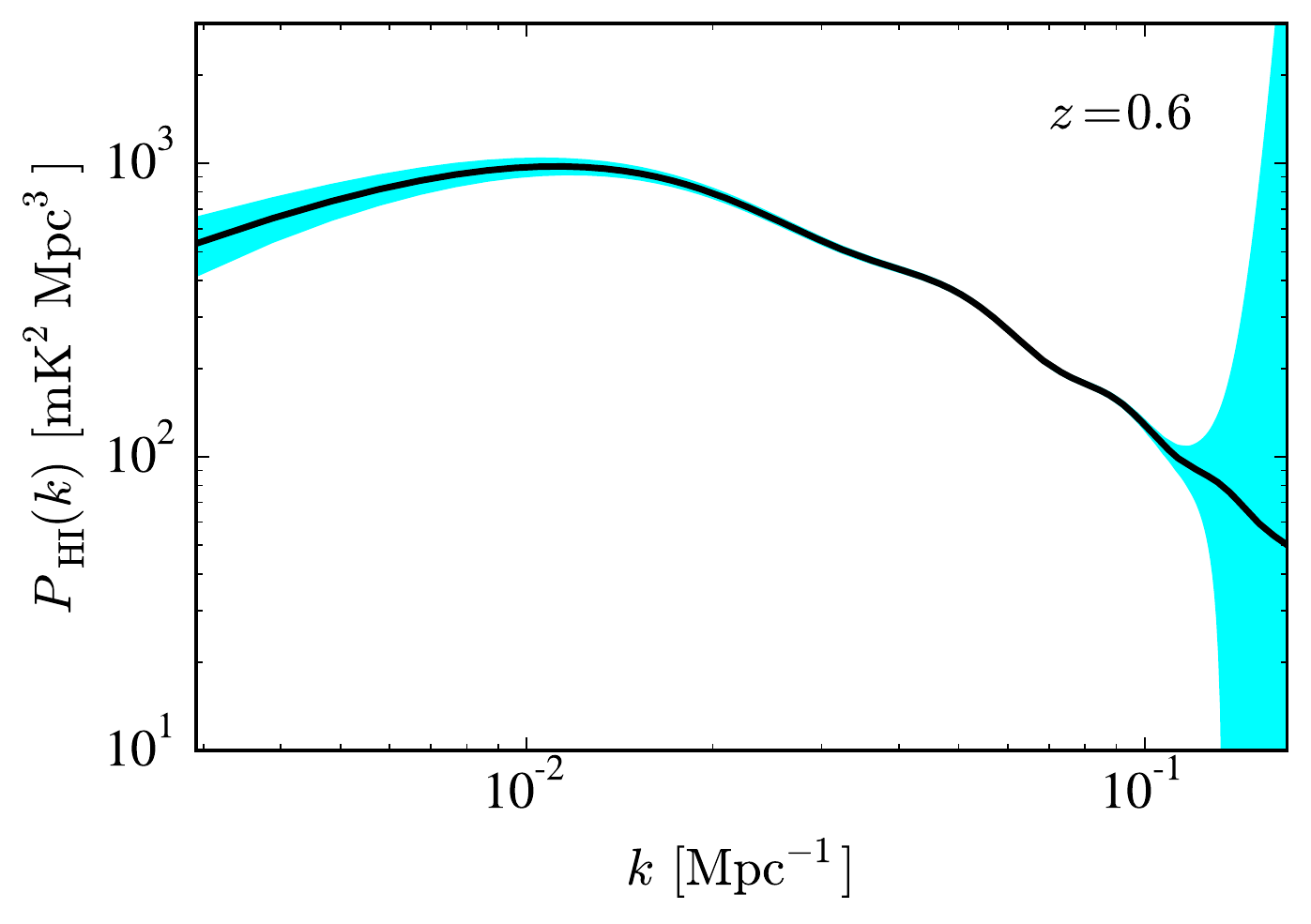}
\includegraphics[width=0.95\columnwidth]{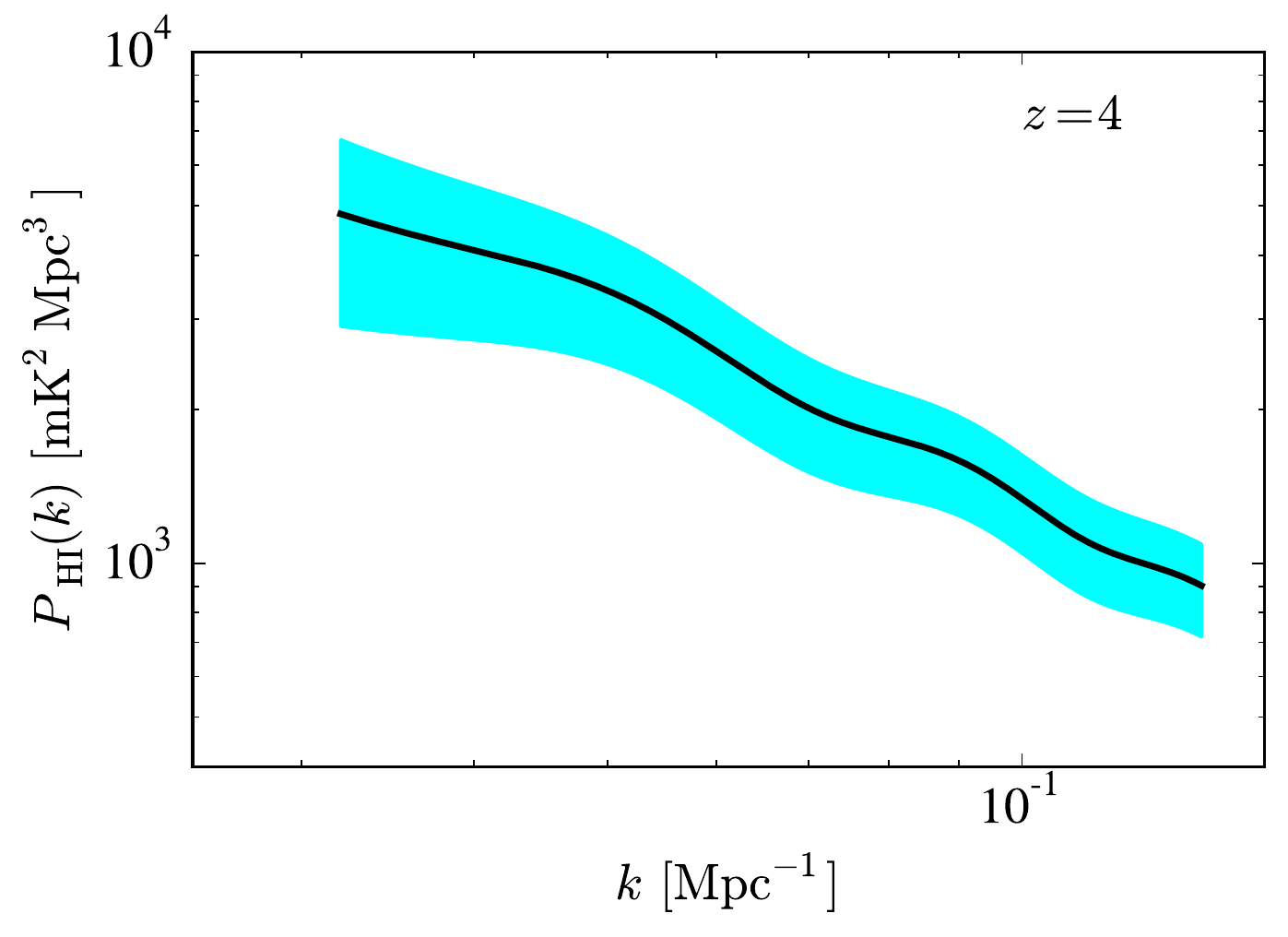}
\caption{{\it Upper panel}: HI detection with the SKA1-MID \wideims, showing the expected signal power spectrum (black solid) and measurement errors (cyan) from the HI auto-correlation power spectrum. The assumed $k$ binning is $\Delta k = 0.01 \, {\rm Mpc}^{-1}$.
{\it Lower panel}:  HI detection with the \deeplowims, signal power spectrum (solid black line) and measurement errors (cyan band) at $z=4$. We have used a $k$-binning $\Delta k = 0.01 \, {\rm Mpc}^{-1}$ and a redshift bin $\Delta z = 0.3$.}
\label{fig:PHI-z0p6}
\end{figure}

\noindent\textit{SKA1-MID} \\
The expected error on the measurement of the HI power spectrum from the \wideims~ is shown in Fig.~\ref{fig:PHI-z0p6} (top panel) for a redshift bin of width $\Delta z = 0.1$ centered at $z=0.6$. Keeping the cosmological parameters fixed to the Planck 2015 cosmology \citep{Ade:2015xua}, the only unknown in $P^{\rm HI}$ is ($\Omega_{\rm HI}b_{\rm HI}$). Employing a Fisher matrix analysis, we calculate the expected constraints on $\Omega_{\rm HI}b_{\rm HI}$ \citep{Pourtsidou:2016dzn}, which are summarized in the first column of Table~\ref{tab:omHIbHI-mid}. Using RSDs, the degeneracy between $\Omega_{\rm HI}$ and $b_{\rm HI}$ can be broken and the resulting constraints are presented in the second column of Table~\ref{tab:omHIbHI-mid}.\\

\begin{table}
\begin{center}
\caption{\label{tab:omHIbHI-mid} Forecasted fractional uncertainties on $\Omega_{\rm HI}b_{\rm HI}$, and $\Omega_{\rm HI}$ assuming the SKA1-MID \wideims~ and following the methodology in \citet{Pourtsidou:2016dzn}. For the $\Omega_{\rm HI}$ constraints we utilize the full HI power spectrum with RSDs. Note that the assumed redshift bin width is $\Delta z = 0.1$, but we show the results for half of the bins for brevity. The cosmological constraints are reported in Fig.~\ref{fig:hi-forecast} and Fig.~\ref{fig:hi-forecast-fs8}.}
\begin{tabular}{|c|c|c|}
\hline
$z$ & $\sigma(\Omega_{\rm HI}b_{\rm HI})/(\Omega_{\rm HI}b_{\rm HI})$ & 
$\sigma(\Omega_{\rm HI})/\Omega_{\rm HI}$ \\
\hline
0.4&0.002&0.009\\
0.6&0.003&0.011\\
0.8&0.004&0.013\\
1.0&0.005&0.017\\
1.2&0.006&0.022\\
1.4&0.008&0.029\\
1.6&0.010&0.036\\
1.8&0.013&0.046\\
2.0&0.016&0.058\\
2.2&0.020&0.072\\
2.4&0.025&0.091\\
2.6&0.030&0.115\\
2.8&0.038&0.145\\
3.0&0.046&0.183\\
\hline
\end{tabular}
\end{center}
\end{table}

\noindent\textit{SKA1-LOW}\\
Here we present predictions on the \deeplowims. Other possibilities (in terms of sky coverage and observation time) as well as an optimization study will be presented in an upcoming publication. 

In Fig.~\ref{fig:PHI-z0p6} (bottom panel) we show the predicted HI signal power spectrum neglecting the effect of RSDs, together with the predicted measurement errors at $z=4$ for the \deeplowims.
Performing a Fisher matrix analysis following the methodology in \citet{Pourtsidou:2016dzn} we can constrain $\Omega_{\rm HI}$ and $b_{\rm HI}$. Our derived constraints are quoted in Table~\ref{tab:omHIbHI-deep}. As we can see, intensity mapping with the \deeplowims~ probes the largely unexplored ``redshift desert'' era and can give us valuable information on the evolution of the HI abundance and bias across cosmic time. 

\begin{figure}[ht]
\centering
\includegraphics[width=0.95\columnwidth]{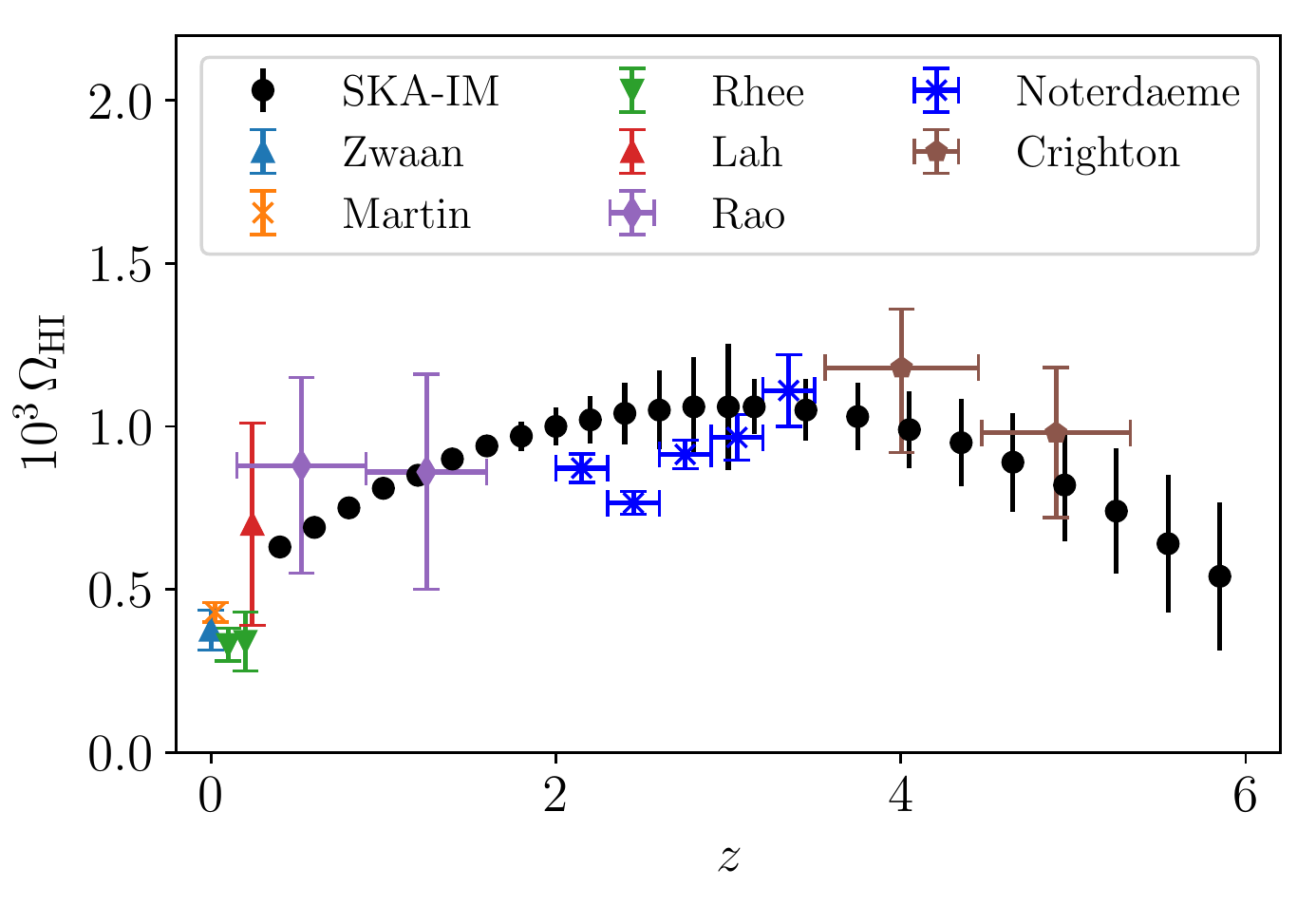}
\caption{Forecasts for the HI density, $\Omega_{\rm HI}$, using the \wideims~ and \deeplowims~ (black points), and comparison with measurements (see \citet{Crighton:2015pza} and references therein), following the methodology in \citet{Pourtsidou:2016dzn}. Note that we have used a very conservative non-linear $k_{\rm max}$ cutoff for these results.}
\label{fig:detection-IM}
\end{figure}

Finally, in Fig.~\ref{fig:detection-IM} we show the derived constraints for both SKA intensity mapping surveys (i.e. \wideims~ and \deeplowims~) on $\Omega_{\rm HI}$ compared to current measurements. 

At this point we note that our forecasts have ignored residual foreground contamination and other systematic effects. Assessing these effects using simulations and exploring the possibility of performing BAO measurements using this survey is the subject of ongoing work.

\begin{table}[ht]
\begin{center}
\caption{\label{tab:omHIbHI-deep} Forecast fractional uncertainties on HI parameters for intensity mapping with the \deeplowims, following the methodology in \citet{Pourtsidou:2016dzn}.}  
\begin{tabular}{|c|c|c|}
\hline
$z$ & $\sigma(\Omega_{\rm HI}b_{\rm HI})/(\Omega_{\rm HI}b_{\rm HI})$ & $\sigma(\Omega_{\rm HI})/\Omega_{\rm HI}$\\
\hline
3.15&0.010&0.08\\
3.45&0.011&0.09\\
3.75&0.012&0.10\\
4.05&0.014&0.12\\
4.35&0.015&0.14\\
4.65&0.018&0.17\\
4.95&0.021&0.21\\
5.25&0.024&0.26\\
5.55&0.029&0.33\\
5.85&0.035&0.42\\
\hline
\end{tabular}
\end{center}
\end{table}

\subsection{Cosmological probes using HI Intensity Mapping}

\subsubsection{Baryon Acoustic Oscillations and Redshift Space Distortions}
As already mentioned in section~\ref{HIGal_BAO}, Baryon Acoustic Oscillations (BAOs) can provide robust measurements on the angular diameter distance and Hubble rate as a function of redshift. Such measurements can in turn be used to constrain dark energy models and the curvature of the Universe \citep{Bull_2015, 2015aska.confE..24B, 2018MNRAS.477L.122W}. The same is true for RSDs, which can measure the growth rate, a crucial ingredient for instance in constraining modified gravity models. In this section we focus on what can be achieved with the \wideims. Exploring the same for \deeplowims~ is the subject of ongoing work.

\begin{table*}[!t]
\caption{\label{tab:sigfNLGR} Marginal errors on $f_{\rm NL}$, lensing ($\varepsilon_{\rm Lens}$) and GR effects ($\varepsilon_{\rm GR}$), which include the Doppler term ($\varepsilon_{\rm Doppler}$), Time Delay ($\varepsilon_{\rm TD}$), Sachs-Wolfe ($\varepsilon_{\rm SW}$) and Integrated Sachs-Wolfe $(\varepsilon_{\rm ISW})$, using the MT technique with HI intensity mapping with the SKA1 \wideims~ in conjugation with the Euclid and LSST surveys for three prior assumptions.}
\centering
\begin{tabular}{|l|c|c|c|c|c|c|c|}
\hline
Synergy &$\sigma(f_{\rm NL})$ & $\sigma(\varepsilon_{\rm Lens})$ & $\sigma(\varepsilon_{\rm GR})$ &$\sigma(\varepsilon_{\rm Doppler})$ & $\sigma(\varepsilon_{\rm TD})$ & $\sigma(\varepsilon_{\rm SW})$& $\sigma(\varepsilon_{\rm ISW})$\\
\hline
SKA1 HI\ IM $\times$ Euclid&1.1&-&-&-&-&-&-\\
&1.1&0.033&0.19&-&-&-&-\\
&1.3&0.033&-&0.19&5.3&5.5&16\\
\hline
SKA1 HI\ IM $\times$ LSST&0.67&-&-&-&-&-&-\\
&0.68&0.043&0.12&-&-&-&-\\
&0.96&0.043&-&0.13&5.7&4.0&7.5\\
\hline
\end{tabular}
\end{table*}

The relatively poor angular resolution of SKA1-MID in single-dish mode at high redshifts/low frequencies will partially smear out the shape of the BAO peak along the angular direction. Nevertheless, SKA1-MID can still provide competitive constraints on BAO measurements and its derived quantities using the HI intensity mapping technique. Following the Fisher matrix forecasting method described in \citet{Bull_2015,Bull2016}, Fig.~\ref{fig:hi-forecast} shows the expected constraints as a function of redshift on the angular diameter distance $D_A$ and Hubble rate $H$ while Fig.~\ref{fig:hi-forecast-fs8} shows the same for the growth rate $f\sigma_8$. We see that the constraints are still quite competitive when comparing to concurrent surveys (e.g. Euclid like). The high redshift resolution of the HI intensity mapping survey, makes it particularly fit for line of sight measurements, such as $H(z)$ and the growth rate. 

However, at frequencies $\nu\leqslant800$ MHz, the angular smoothing is so large that the BAO feature might be hard to extract from the angular direction. This depends on how well we can deconvolve the beam given the signal to noise. Even in this worst case scenario, the frequency resolution will be good enough to allow for a detection of the radial BAO. By means of numerical simulations incorporating the cosmological signal, instrumental effects and the presence of foregrounds, \cite{FVN_2016} demonstrated that the position of the radial BAO peak can be measured with percent precision accuracy through single-dish observations in the Band 1 of SKA1-MID.

\subsubsection{Ultra-large scale effects}

One of the "transformational" measurements expected from HI intensity mapping with the \wideims~ are the constraints on the power spectrum on ultra-large scales (past the equality peak). This is an area where a single dish survey with SKA1-MID can excel given its low resolution, but large survey speed \citep{2015ApJ...814..145A}. Such measurements can provide hints on new physics that only materialise on this ultra-large scales. 

One example of such an effect is Primordial non-Gaussianity (PNG). In particular, PNG of the local type $f_{\rm NL}$ introduces a  scale-dependent correction to clustering bias  \citep{Dalal:2007cu,Matarrese:2008nc} such that $b_{\rm HI}\propto f_{\rm NL}/k^2$. 
The $1/k^2$ term makes this effect particularly relevant on very large scales (small $k$) where statistical detectability is severely limited due to cosmic variance and large scale systematic effects. Using HI IM only we forecast $\sigma(f_{\rm NL})=2.8$, assuming Band 1 for SKA dishes and UHF band for the MeerKAT dishes. Note that our calculations take into account the telescope beams and marginalise over the biases. While it is not able to achieve $\sigma(f_{\rm NL})<1$, as opposed to more futuristic SKA upgrades \citep[see][]{2013PhRvL.111q1302C}, thus distinguishing between single-field and multi-field inflation, it will be an improvement on $\sigma^{\rm Planck}(f_{\rm NL})=5.0$. 

Another type of very large scale signatures are the so called General Relativistic (GR) effects. These GR effects introduce corrections to the tracers' transfer function as leading to a set of terms which are usually gathered together as a single  contribution. They are an important prediction of GR over the very largest distances that it is possible to probe observationally, and so constitute a valuable test of alternative gravitational theories \citep{Hall:2012wd, Lombriser:2013aj, 2015ApJ...811..116B}. \citet{Alonso:2015uua} has shown that these effects are not detectable in the single tracer case due to cosmic variance. However, it will be crucial to correctly model these relativistic corrections in future large-scale structure surveys, in order not to bias the estimation of other ultra-large scale effects such as primordial non-Gaussianity \citep{2015MNRAS.451L..80C}.

\begin{figure}[!t]
\centering
\includegraphics[width=\columnwidth]{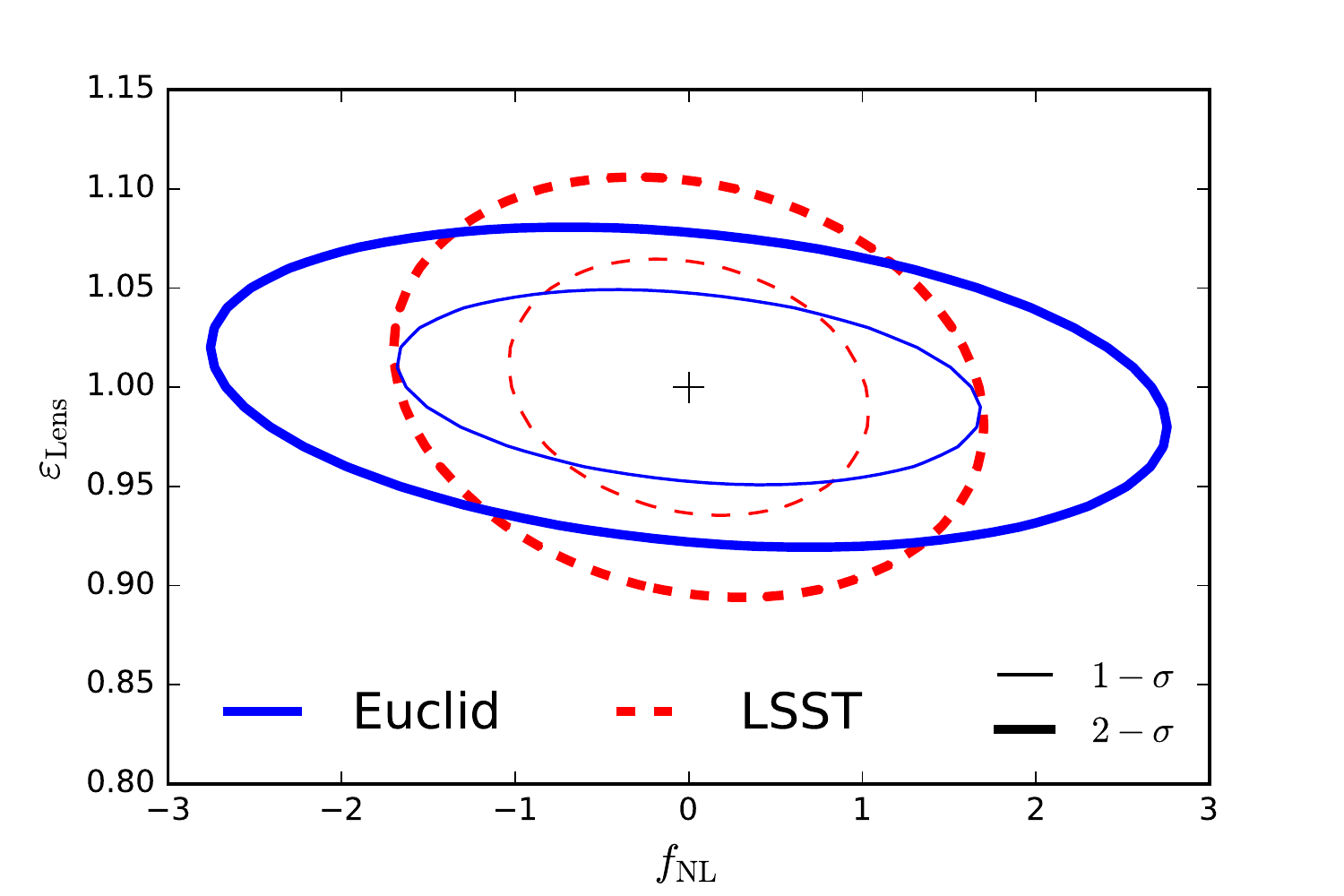}
\includegraphics[width=\columnwidth]{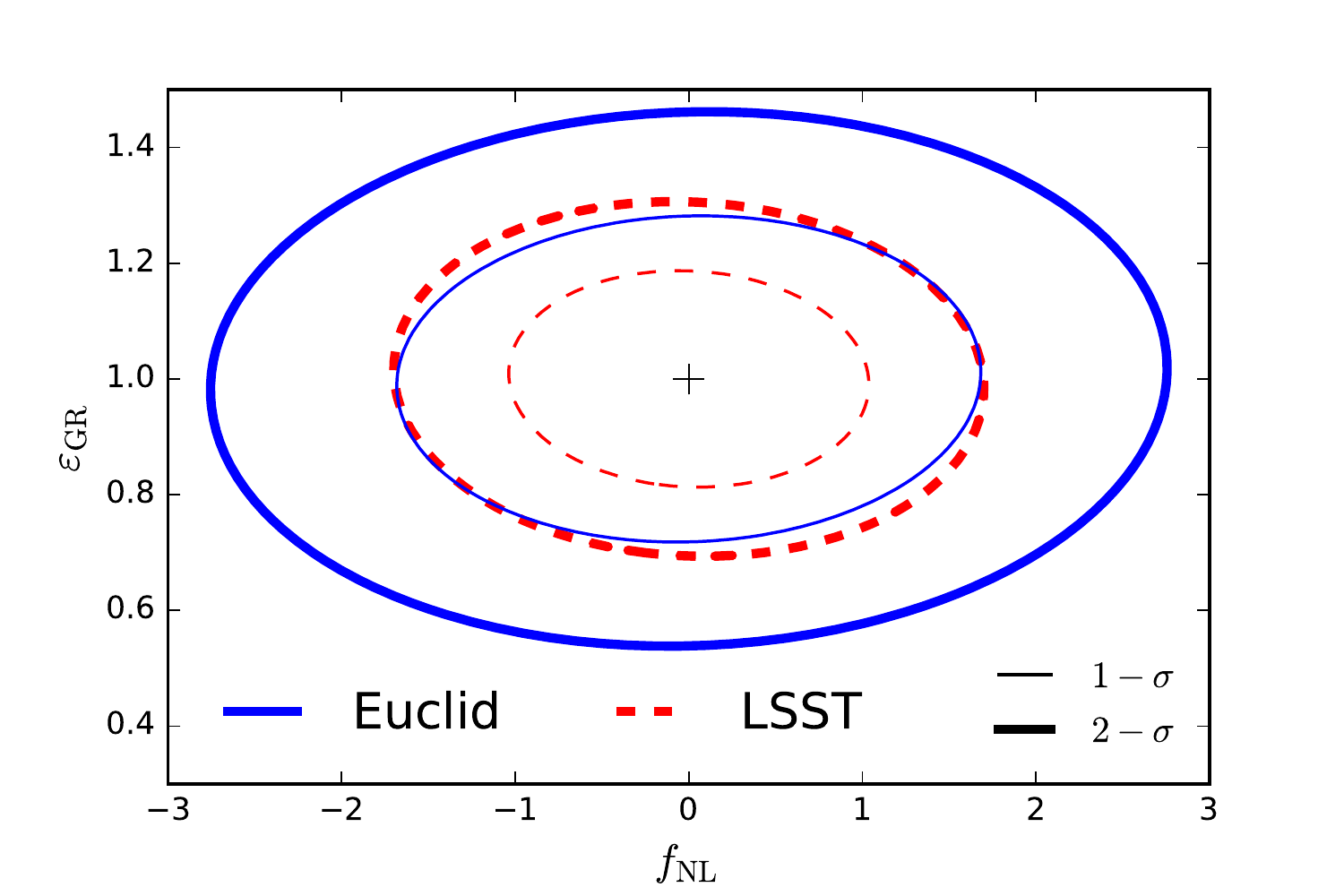}
\caption{The $1\sigma$ (thin) and $2\sigma$ (thick) contours for the forecasted marginal errors on $f_{\rm NL}$ and Lensing (\emph{top}), and GR effects (\emph{bottom}) using the multi-tracer technique from HI intensity mapping with the SKA1 \wideims~ in combination with Euclid data (solid blue line) and LSST data (dashed red line). These forecasts assume Case 2 as presented in Table~\ref{tab:sigfNLGR}. Combination with LSST will allow to probe $f_{\rm NL}\sim 1$ as well as detect large scale GR effects.}
\label{fig:MT:contour_fnl_eps}
\end{figure}

It is possible to overcome cosmic variance with the multi-tracer (MT) technique \citep{Seljak:2008xr}, where one combines two differently biased dark matter tracers in such a way that the fundamental statistical uncertainty coming from cosmic variance can be bypassed. We updated the forecasts of \citet{Alonso:2015sfa} and \citet{Fonseca:2015laa} for $f_{\rm NL}$ and GR effects using the multi-tracer technique with HI IM with SKA1 in combination with an overlapping $10,000\deg^2$ Euclid-like survey and $14,000\deg^2$ LSST-like photometric surveys.
In Table~\ref{tab:sigfNLGR} we show the forecast marginal errors on $f_{\rm NL}$ and GR effects for 3 different sets of cosmological parameters: Case 1 -- marginal errors on $f_{\rm NL}$ without including GR effects; Case 2 -- marginal errors on $f_{\rm NL}$ including Lensing and GR effects all together; Case 3 -- marginal errors on $f_{\rm NL}$ including Lensing and each GR effect individually. Note that all of the $\epsilon$ parameters have a fiducial value of $\epsilon=1$ (see \citealt{2018MNRAS.479.3490F} for the definitions).
In Fig.~\ref{fig:MT:contour_fnl_eps} we show the degeneracy between $f_{\rm NL}$ and lensing (\emph{top}) and GR effects (\emph{bottom}) for the two synergy surveys considered assuming Case 2. It can be seen that using the multi-tracer technique, we will be able to break the barrier $\sigma(f_{\rm NL})<1$ and make a detection of some GR effects such as the Doppler term.

\subsubsection{HI detection via synergies with optical surveys}
Cross correlations between HI intensity mapping and optical galaxy surveys can also provide precise and robust cosmological measurements, as they have the advantage of mitigating major issues like systematics and foreground contaminants that are relevant for one type of survey but not for the other. For example, in \citet{Masui:2012zc} the intensity maps acquired at the Green Bank telescope were combined with the WiggleZ galaxy survey to constrain the quantity $\Omega_{\rm HI}b_{\rm HI}r$ at $z\sim 0.8$ with a statistical fractional error $\sim 16\%$. $r$ is the cross-correlation efficient of the two observables ranging $0<r<1$.

\begin{figure}
\centering
{\includegraphics[width=\columnwidth]{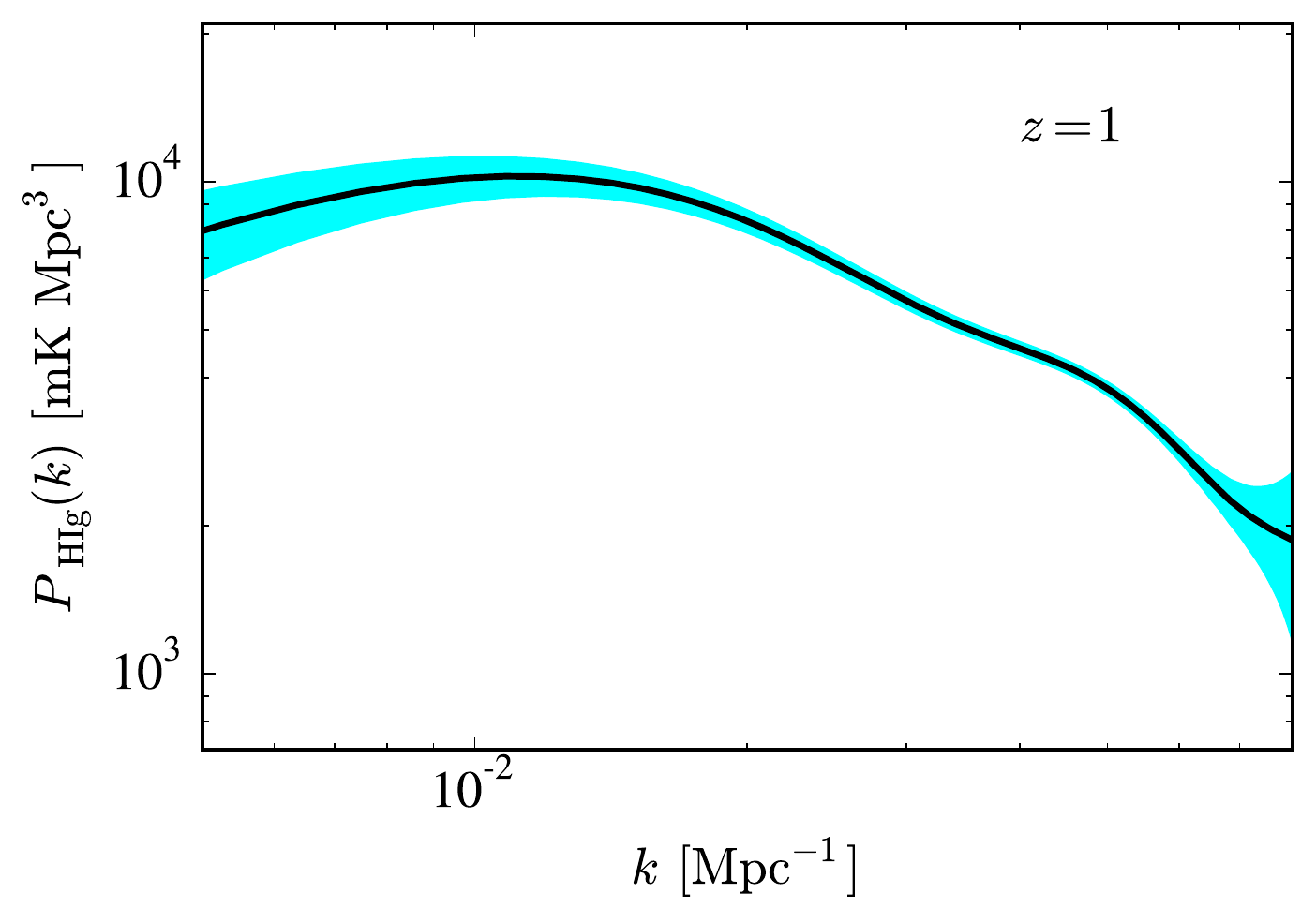}}
{\includegraphics[width=\columnwidth]{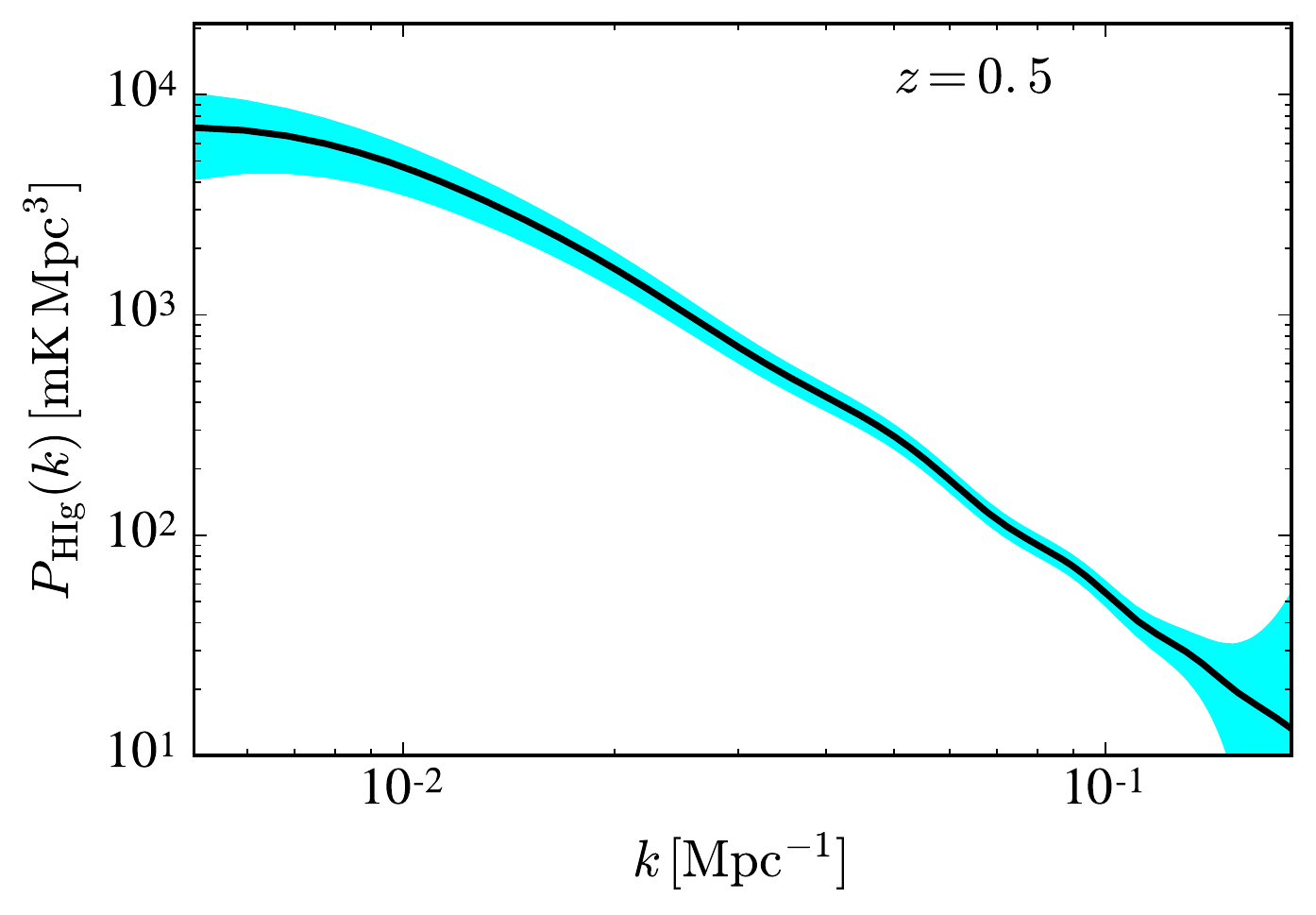}}
\caption{HI intensity mapping with SKA1 \wideims~ in cross-correlation with optical surveys, showing the expected signal power spectrum (black solid) and measurement errors (cyan). {\it Top}: Cross-correlation with a Euclid-like spectroscopic optical galaxy survey with 10,000\,deg$^2$ overlap{\it Bottom}: Cross-correlation with a DES-like photometric optical galaxy survey with 5,000\,deg$^2$ overlap. The assumed $k$ binning is $\Delta k = 0.01$.}
\label{fig:PHI}
\end{figure}

We start by looking at the intensity mapping cross-correlations with a spectroscopic optical galaxy survey, following \citet{Pourtsidou:2016dzn}. Fig.~\ref{fig:PHI}, top panel, shows the expected signal and  errors for a Euclid-like spectroscopic sample \citep{Majerotto:2012mf} for a redshift bin of width $\Delta z = 0.1$ centered at $z=1$. The assumed sky overlap  is 10,000\,deg$^2$ with corresponding 5,800\,hr total observing time for the IM survey which can be approximately achieved with the suggested SKA1 \wideims. The resulting constraints on $\Omega_{\rm HI}b_{\rm HI}r$ (keeping the cosmological parameters fixed to the Planck 2015 cosmology \citep{Ade:2015xua}) are summarised in Table~\ref{tab:omHIbHI}. This table also shows constraints on $f\sigma_8$, $D_A$ and $H$ from cross-correlations with Euclid, considering $\bar{T}_b$ is known. 

Cross-correlations with photometric optical galaxy surveys can also be used to constrain HI properties, and perform joint probes studies \citep{Pourtsidou:2015mia}. Fig.~\ref{fig:PHI}, bottom panel, shows the expected signal and  errors for Stage III DES-like photometric sample for a redshift bin of width $\Delta z = 0.1$ centered at $z=0.5$. The assumed sky overlap  is 5,000\,deg$^2$. We can also combine probes such as HI clustering and optical lensing, or HI clustering and the CMB, to constrain gravity \citep{Pourtsidou:2015ksn} and inflation \citep{Pourtsidou:2016ctq}. 

\begin{table}
\begin{center}
\caption{\label{tab:omHIbHI} Forecast fractional uncertainties on HI and cosmological parameters assuming HI intensity mapping with the SKA1 \wideims~ and Euclid-like cross-correlation described in the main text, following the methodology in \citet{Pourtsidou:2016dzn}. Note that the assumed redshift bin width is $\Delta z = 0.1$, but we show the results for half of the bins for brevity.}  
\begin{tabular}{|c|c|c|c|c|}
\hline
$z$ & $\frac{\sigma(\Omega_{\rm HI}b_{\rm HI}r)}{(\Omega_{\rm HI}b_{\rm HI}r)}$
& $\frac{\sigma(f\sigma_8)}{(f\sigma_8)}$
& $\frac{\sigma(D_A)}{D_A}$
& $\frac{\sigma(H)}{H}$\\
\hline
1.0&0.014&0.04&0.02&0.02\\
1.2&0.018&0.06&0.03&0.02\\
1.4&0.024&0.08&0.05&0.02\\
1.6&0.030&0.10&0.06&0.02\\
1.8&0.038&0.12&0.08&0.03\\
2.0&0.047&0.15&0.09&0.03\\
\hline
\end{tabular}
\end{center}
\end{table}

\subsubsection{Neutrino masses}
\begin{figure}
 \centering
 \hspace{-0.6cm}\includegraphics[width=1.03\columnwidth]{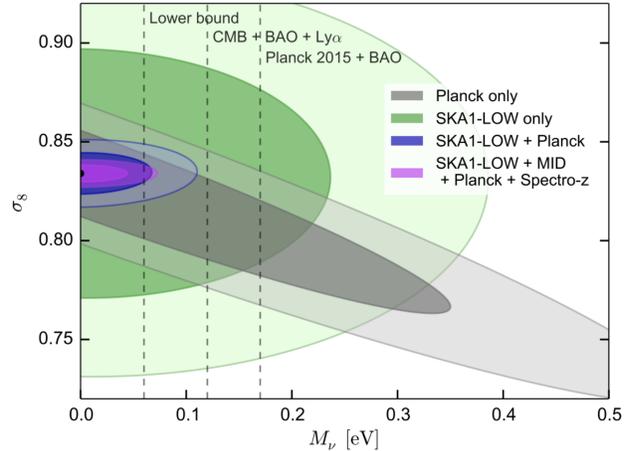}
 \caption{This figure shows 1$\sigma$ and 2$\sigma$ constraints on the $M_\nu-\sigma_8$ plane from {\it Planck} CMB 2015 alone (grey), SKA1-LOW (green), SKA1-LOW plus {\it Planck} CMB 2015 (blue) and SKA1-LOW plus SKA1-MID plus {\it Planck} CMB 2015 plus a spectroscopic galaxy survey (magenta). The lower limit from neutrino oscillations, together with recent cosmological upper bounds are shown with dashed vertical lines.}
 \vspace{-0.4cm}
 \label{fig:neutrinos} 
\end{figure}

The impact of massive neutrinos on the abundance and clustering of cosmic neutral hydrogen has been studied in \cite{FVN_2015} through hydrodynamic simulations. It was found that neutrino masses do not affect much the halo HI mass function\footnote{This function represents the average HI mass inside a dark matter halo of mass $M$ at redshift $z$.}, $M_{\rm HI}(M,z)$. Therefore, neutrino effects on HI properties can easily be explained through simple HI halo models. \cite{FVN_2015} used those ingredients to forecast the sensitivity of the phase 1 of SKA to neutrino masses, finding that observations by SKA1-MID plus SKA1-LOW alone can place a constrain of $\sigma(M_\nu)=0.18$eV ($2\sigma$), where $M_\nu=\sum_i m_{\nu_i}$. By adding information from {\it Planck} CMB 2015 data alone that limit can shrink to $\sigma(M_\nu)=0.067$eV ($2\sigma$), while a combination of data from SKA1-MID, SKA1-LOW, {\it Planck} and a spectroscopic galaxy survey like Euclid can yield a very competitive constraint of $\sigma(M_\nu)=0.057$eV ($2\sigma$). Those constraints have been derived with the \wideims~ assuming observations in Band 1 and 2, and 10,000 hours of interferometry observations by SKA1-LOW over 20 ${\rm deg}^2$ at frequencies $\nu\in[200,355]$ MHz. Fig.~ \ref{fig:neutrinos} show those constraints projected in the $M_\nu-\sigma_8$ plane.

\subsubsection{Probing inflationary features}
\begin{figure}
 \centering
 \includegraphics[width=0.95\columnwidth]{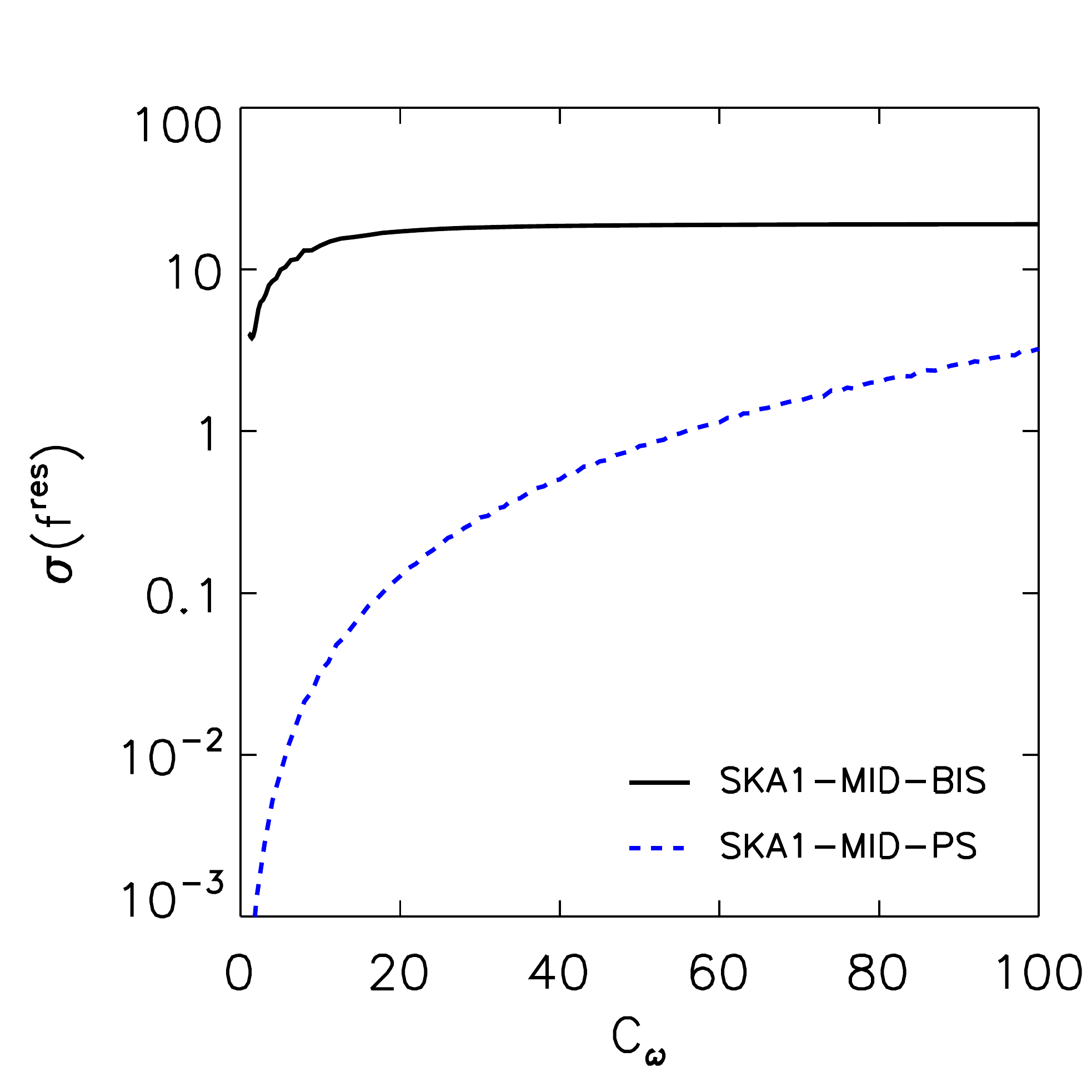}
 \caption{The marginalized $1-\sigma$ error on the resonance amplitude as a function of frequency in the resonant inflationary model, using HI intensity mapping power spectrum measurements {\it (blue dashed line)} and bispectrum measurements {\it(black solid line)} from  the SKA1-MID \wideims~ (fiducial value is $f^{\rm res}=0$).}
\label{fig:infl_res} 
\end{figure}

\begin{figure}[!t]
\centering
\includegraphics[width=0.95\columnwidth]{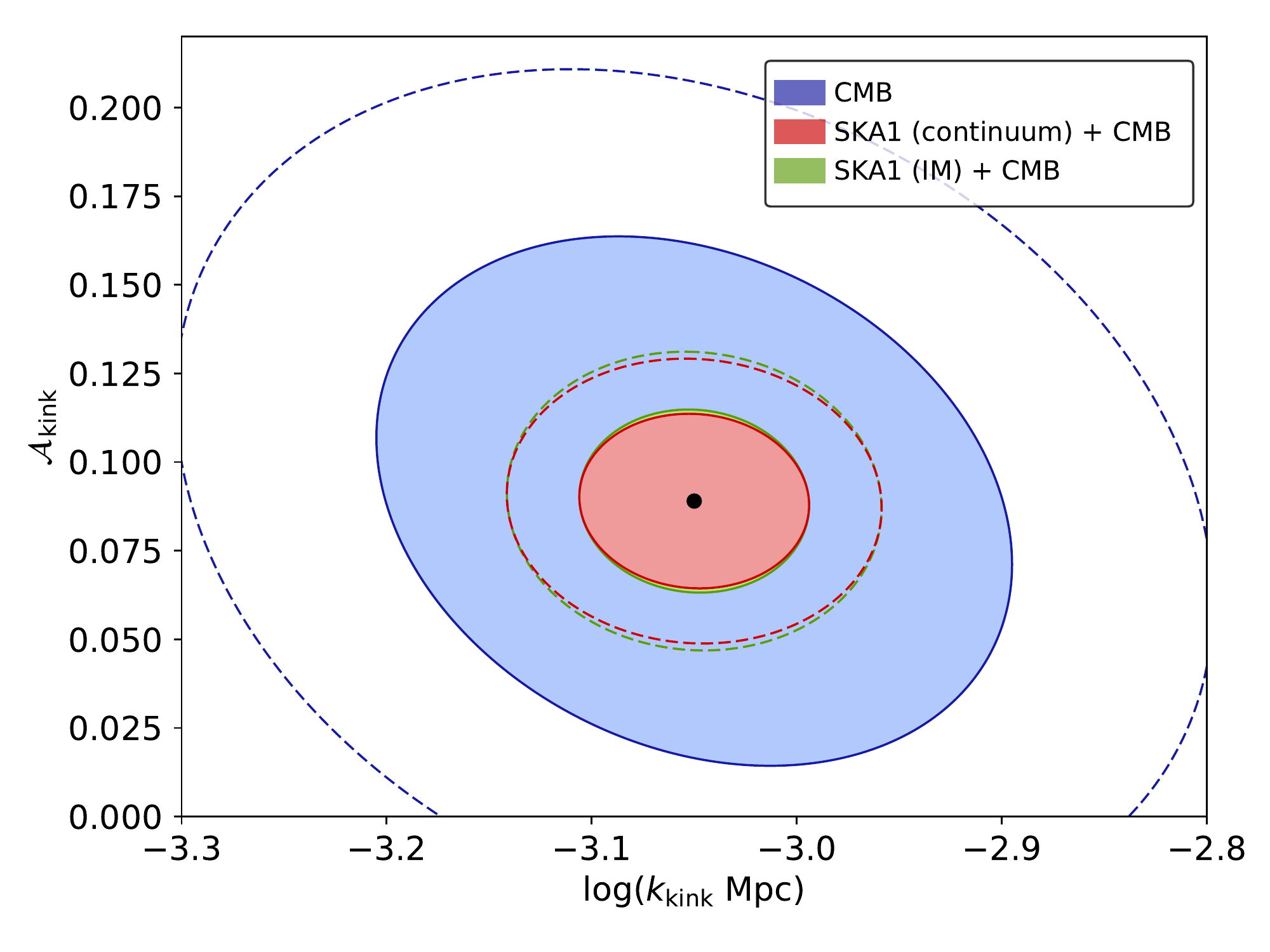}
\includegraphics[width=0.95\columnwidth]{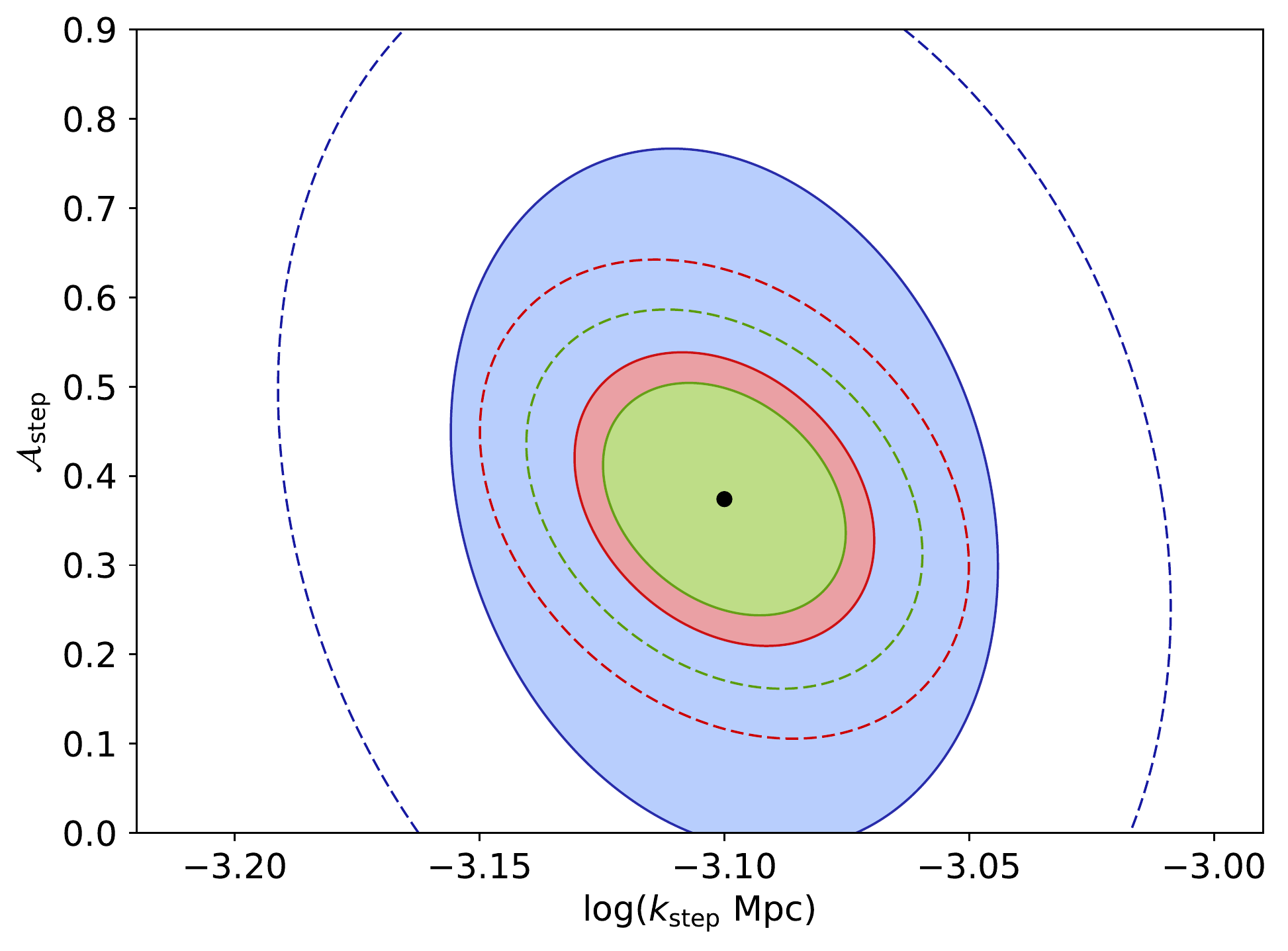}
\includegraphics[width=0.95\columnwidth]{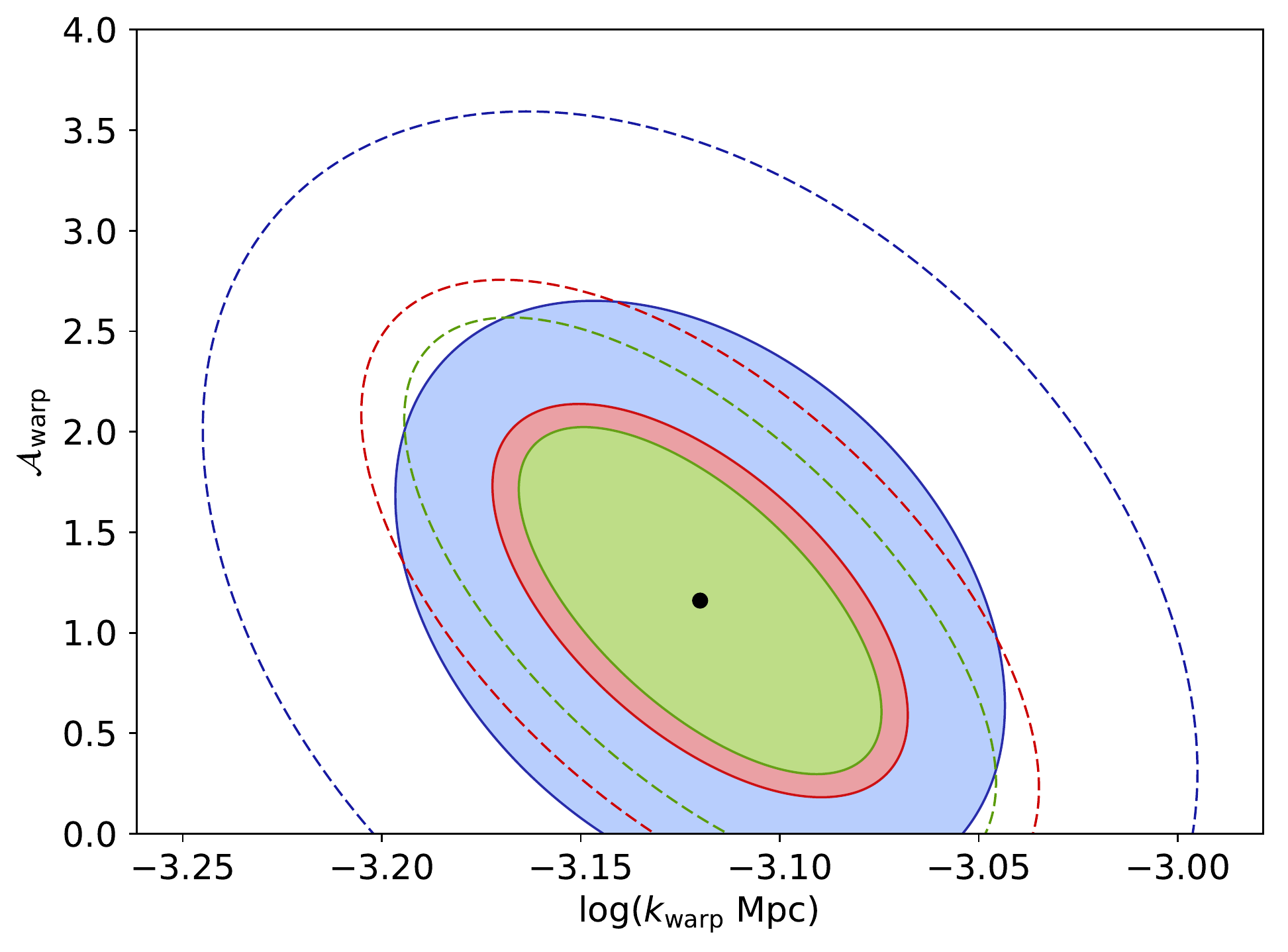}
\caption{Marginalized 68\% (shaded areas) and 95\% (dashed lines) confidence level contours for the feature wave-number in the kink (top), step (middle) 
and warp (bottom) inflationary models, using the \textit{Planck} CMB 2015 alone and combining intensity mapping and continuum data from the SKA1 \wideims~ with the CMB.}
\label{prim-feat}
\end{figure}

Possible anomalies observed in the CMB by WMAP \citep{Peiris:2003ff} and \planck\ \citep{Planck:2013jfk,Ade:2015lrj,Akrami:2018odb} may be connected to features on ultra-large scales ($10^{-3}<k\, {\rm Mpc}/h<10^{-2}$) in the primordial power spectrum, that are generated by a violation of slow-roll. Constraints on such primordial features from inflation are shown in \cite{Xu2016,2018JCAP...04..044B} to be significantly improved by using the ultra-large scale HI intensity mapping and continuum surveys of SKA1-MID.
The potential of such surveys for constraining the ``resonant'' \citep{2008JCAP...04..010C}, 
``kink'' \citep{1992ZhPmR..55..477S}, ``step'' \citep{2001PhRvD..64l3514A,2012PhRvD..85b3531A} and ``warp'' \citep{2012PhRvD..86f3529M} 
inflation models is illustrated in Figs.~\ref{fig:infl_res} and \ref{prim-feat}.

Fig.~\ref{fig:infl_res} shows constraints on the amplitude of the resonant
non-Gaussianity, $f^{\rm res}$, as a function of the resonance
frequency $C_\omega$, using either the scale-dependent bias of the power spectrum or the bispectrum, with the \wideims~ of SKA1-MID (adding Band 2 intensity mapping observations for $z<0.4$), combining the single-dish observation mode with the interferometric mode.
The results show that even in the presence of foreground contamination, the upcoming HI intensity mapping observations of the large-scale structure with the SKA1-MID alone could put extremely tight constraints on the feature models, potentially achieving orders-of-magnitude improvements over the two-dimensional CMB measurements. For details on the parameterization and forecasts see \cite{Xu2016}.

Fig.~\ref{prim-feat} shows the Fisher forecast constraints on the amplitude of the feature versus the scale of the feature in Fourier space, using \wideims~ in both intensity mapping and continuum on SKA1-MID. SKA1 can constrain parameters of the feature models at $>3\sigma$ (for details, see \citealt{2018JCAP...04..044B}). 

HI intensity mapping surveys could be also used in combination with CMB experiments to constrain the scalar spectral index ($n_{\rm s}$) and its runnings ($\alpha_{\rm s}, \beta_{\rm s}$) and test the predictions of popular single-field slow-roll inflation models. Current constraints from Planck are $\sigma(n_{\rm s})=0.006$ and $\sigma(\alpha_{\rm s})=0.007$. A Stage IV CORE-like CMB survey \citep{Finelli:2016cyd} combined with an HI intensity mapping survey with SKA1 \wideims~ could reach $\sigma(n_{\rm s})=0.0011$ and $\sigma(\alpha_{\rm s})=0.0019$ \citep{Pourtsidou:2016ctq}.

\subsubsection{Unveiling the nature of dark matter}

\begin{figure}
  \includegraphics[width=\columnwidth]{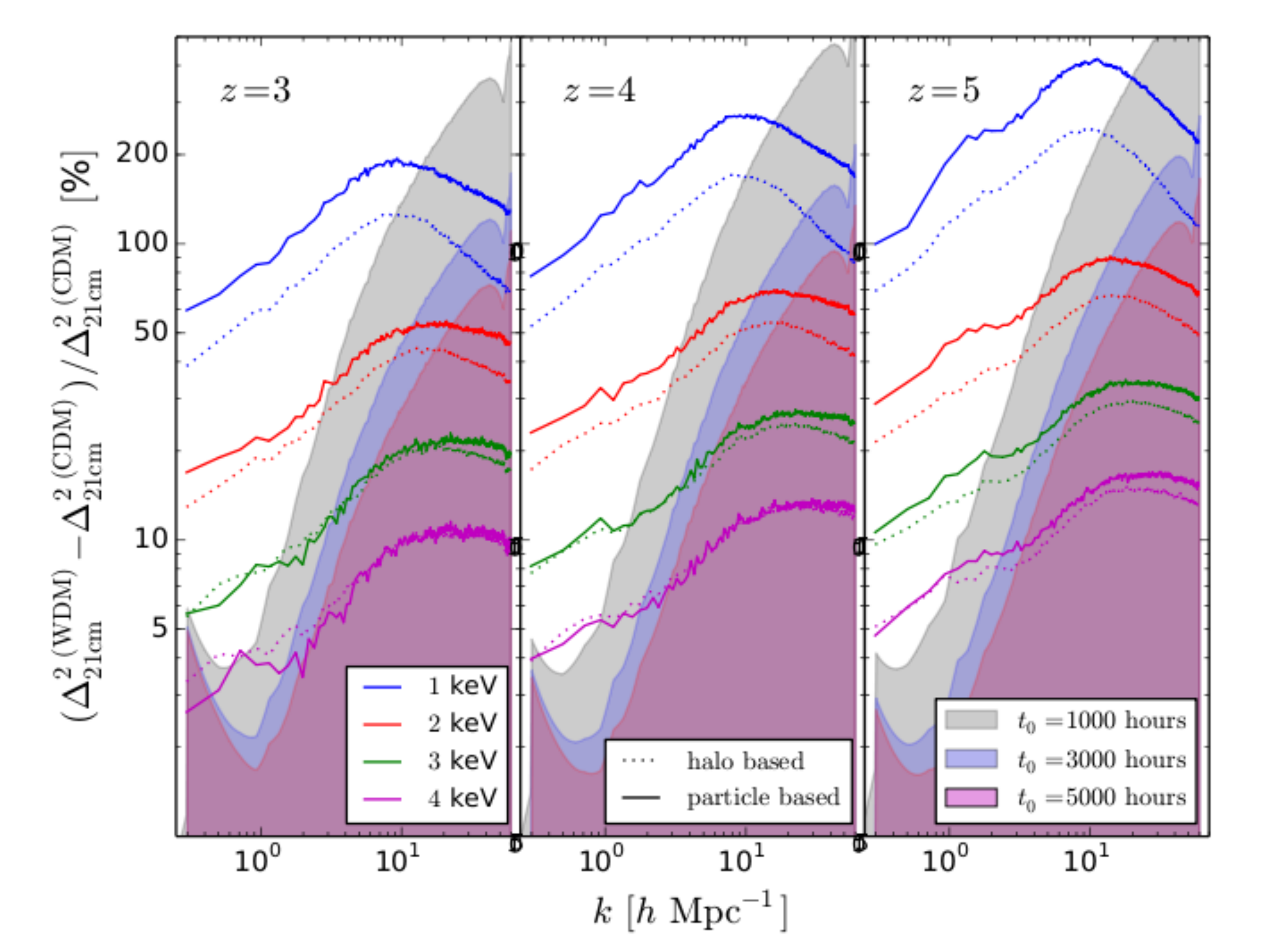}
  \caption{Percentage difference for the HI intensity mapping power
    spectrum when the HI distribution is modeled using two different methods:
    the halo based method (dotted lines) and the particle based method
    (solid lines). Results are shown at z = 3 (left), z = 4 (middle)
    and z = 5 (right). The error on the HI power spectrum of the
    model with CDM, normalized to the amplitude of the 21 cm 
    (CDM) power spectrum is shown in a shaded region
    for three different observation times: $t_0 = 1,000$ hours (grey), $t_0
    = 3,000$ hours (blue) and $t_0 = 5,000$ hours (fuchsia). The field-of-view for this example corresponds to an area of between 2.7 and 6 deg$^2$, at $z=3-5$. For clarity, we show the error from one HI-assignment method only
    because both are very similar and overlap at the scale of the plot.
  }
  \label{fig1} 
  \vspace{-0.3cm}
\end{figure}

\begin{figure}
\vspace{-0.4cm}
  \includegraphics[width=\columnwidth]{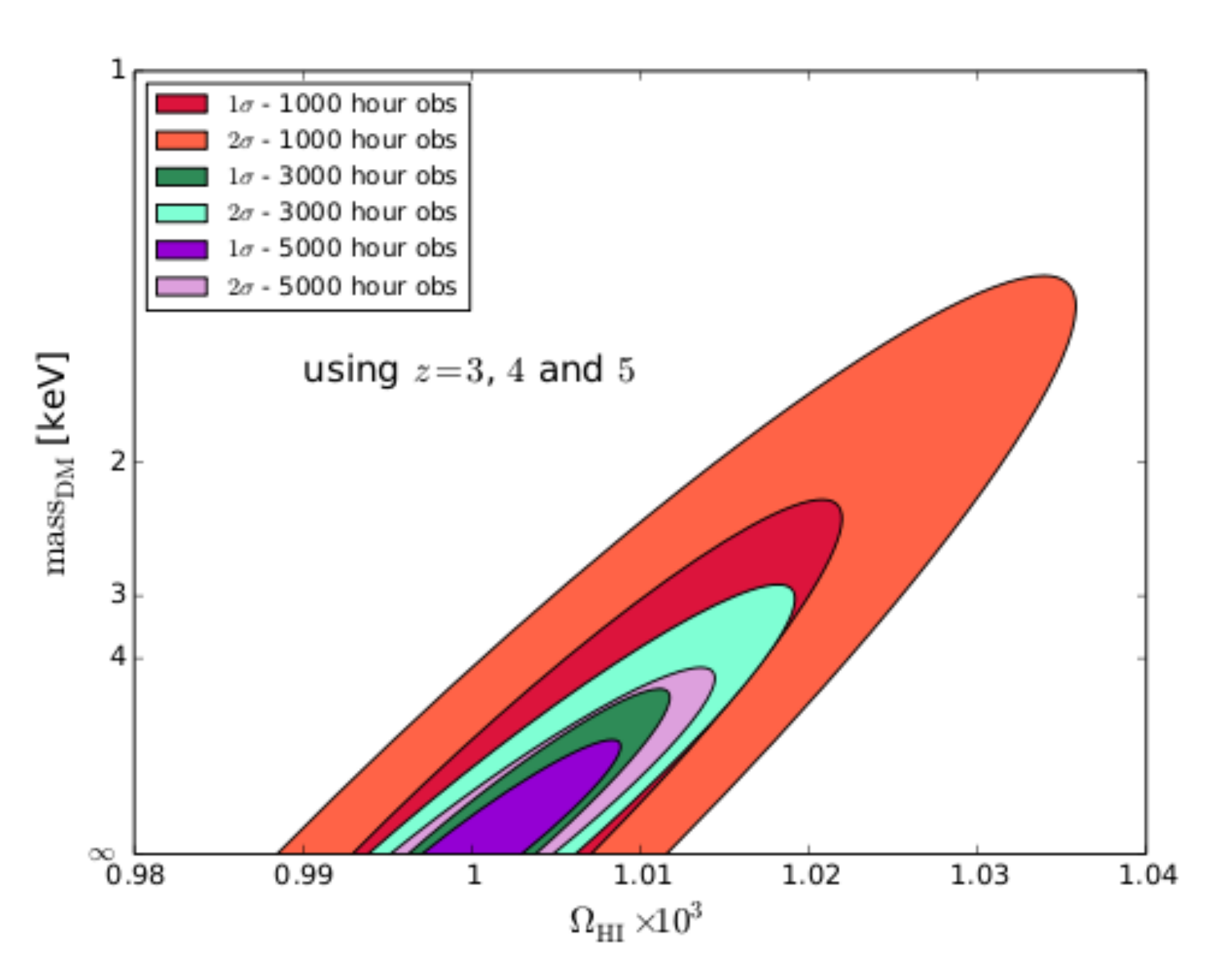}
  \caption{$1\sigma$ and $2\sigma$ contours (dark and light areas) of the values of $\Omega_{\rm HI}$ and m$_{\rm WDM}$ determined
using the HI power spectrum measured by SKA1-LOW with three different observation times:
1,000, 3,000 and 5,000 hours (red, green and violet) and using a field-of-view between 2.7-6 deg$^2$. The Fisher matrix analysis is performed using
information coming from redshift $z = 3, 4 \text{ and } 5$.}
  \label{fig2}
\end{figure}

Intensity mapping offers the opportunity to measure the matter power
spectrum also at intermediate and small scales. At such scales there
could be a signature of the so-called free streaming of dark matter
particles (as in the case of Warm Dark Matter - WDM), which produces a suppression of power \citep{Smith:2011ev}. 
It is thus natural to explore what could be the constraints achieved by looking at neutral
hydrogen in emission as probed by intensity mapping surveys. In \citep{carucci15} the
impact of WDM thermal relics is investigated on the 21 cm intensity
mapping signal focusing on the high redshift, where structure
formation is closer to the linear regime \citep{vielbaldi};   the authors find that there is no suppression of power but there is an
increase of power in a redshift and scale dependent way at mildly
non-linear scales. 
In Fig.~\ref{fig1} we show the difference for the HI intensity mapping power spectrum which is expected between the WDM model and a corresponding CDM model
with the same cosmological parameters, assuming a deep and narrow intensity mapping survey with SKA1-LOW with an area of $\sim 3-6$ deg$^2$ at $z=3-5$ with a range of observation times as described in the caption. It will be quite important to obtain independent constraints on
$\Omega_{\rm HI}$ since, as it is shown in Fig.~\ref{fig2}, it is evident that
that there exists a relatively strong degeneracy between the HI cosmic density
and the WDM mass.


The results indicate that we will be able to rule out a 4 keV WDM
model with 5,000 hours of observations at $z > 3$, with a statistical
significance larger than 3$\sigma$, while a smaller mass of 3 keV,
comparable to present day constraints, can be ruled out at more than
2$\sigma$ confidence level with 1,000 hours of observations at $z > 5$.

\subsubsection{Photometric redshift calibration}
\label{sec:photo-calib}

\begin{figure*}
\centering
	\includegraphics[width=0.8\textwidth]{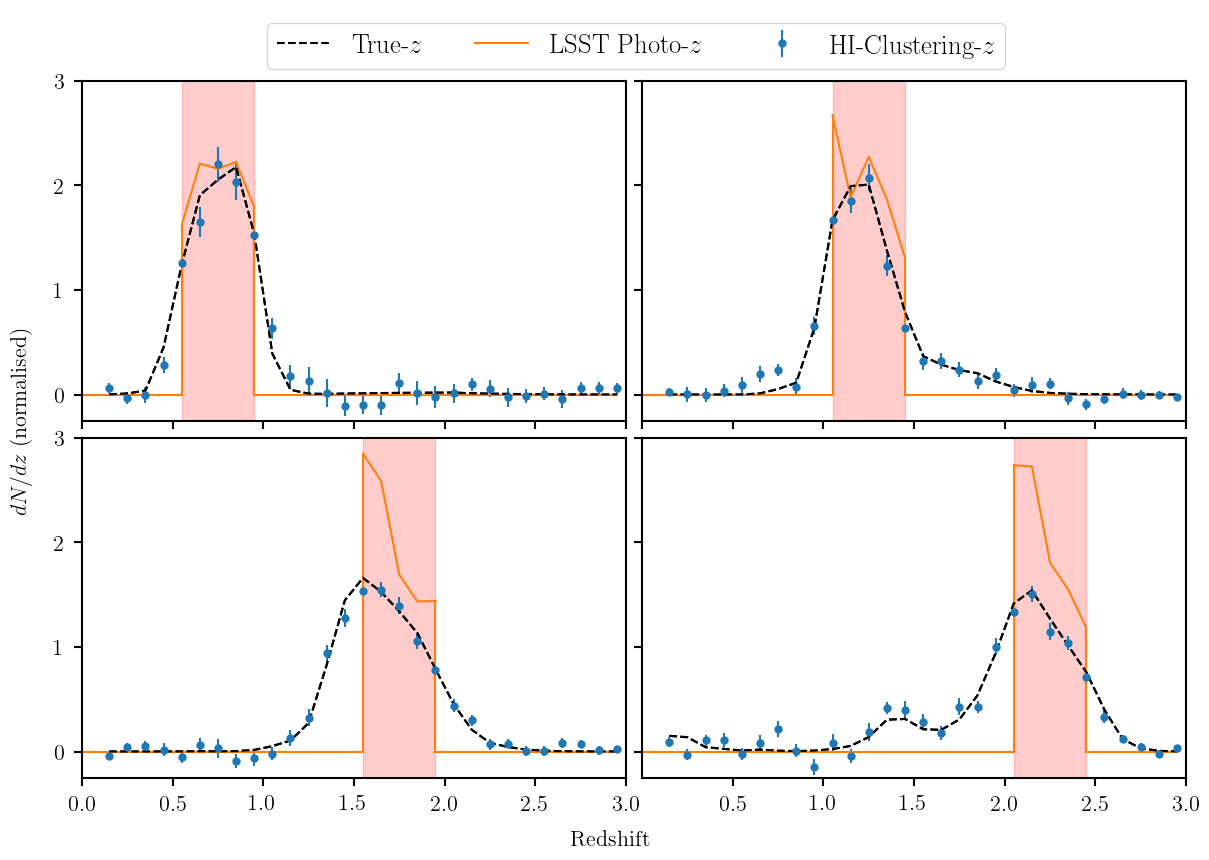}
    \caption{Photometric redshift estimation using HI intensity maps as calibrator for an optical survey such as the LSST. The pink shaded regions show the range in photometric redshift which galaxies are selected from. The orange line shows the distribution of these chosen galaxies according to their LSST-like photometric redshift \citep{AscasoA2A}. The black-dashed line shows the true distribution, and the blue points show the HI clustering redshift estimate \citep{2018arXiv180504498C}.}
\label{PhotozError}
\end{figure*}

With next generation optical surveys such as \textit{Euclid} and LSST promising to deliver unprecedented numbers of resolved galaxies, immense strain will be placed on the amount of spectroscopic follow-up required. Through a clustering-based redshift estimation method which utilities HI intensity maps with excellent redshift resolution, a well-constrained prediction can be made on the redshift distribution for an arbitrarily large optical population.

SKA1 HI intensity mapping would be capable of reducing uncertainties in photometric redshift measurements below the requirements of DES and LSST \citep{2017PhRvD..96d3515A} assuming adequate foreground cleaning could be achieved. Tests have been carried out whereby attempts were made to recover the redshift distribution for a simulated optical galaxy catalogue by cross-correlating its clustering with HI intensity maps \citep{2018arXiv180504498C}. 

This method relies on an estimate of $b_{\rm g}$, $b_{\rm HI}$ (the bias for the optical galaxies and HI intensity maps respectively) and a model for the measurement of the mean HI brightness temperature $\bar{T}_{\rm HI}$. Assuming these are in hand,  Fig.~\ref{PhotozError} presents a proof of concept example using a small survey (25 deg$^2$) with 1' beam size, neglecting noise, where HI emission is estimated using a HI-halo mass relation. \citet{2018arXiv180504498C} shows that this result is comparable to that for the proposed intensity mapping experiment with the SKA1 \wideims. We use the \citet{AscasoA2A} catalogue in which accurate LSST photometric redshifts are simulated, with  2 pixels per arcminute resolution and 30 redshift bins over a range of 0 < z < 3. This gives the optical catalogue a low number density of 0.27 galaxies per voxel, but despite this, a clustering redshift recovery is still possible. This is at the expense of the simulations' halo mass resolution for the galaxies which must increase for larger skies due to computational cost. This means that the wide results are on the conservative side, since including lower mass galaxies would mean a more complete representation of the underlying mass density, potentially improving the cross-correlation.

\subsection{Systematics}

In this section, we discuss some of the main systematics that can affect the signal. These are usually a convolution of strong sky contaminants with imperfections in the telescope. In principle, with a high fidelity model of the instrument it should be possible to model many of these out, however, many systematics can appear highly degenerate without knowledge of the origin of the contamination.
The success of intensity mapping with the SKA1 will rely heavily on our ability to suppress residual uncertainties below the level of the thermal noise. In this section we present some recent results based on simulations, indicating where appropriate, where further work will be necessary. There is a long list of important systematics which still need to be considered including primary beam effects (sidelobes), polarization leakage, and standing waves. Combining simulations with actual observations and data analysis will be crucial in defining which of these will be the limiting factor. Upcoming experiments such as the recently built MeerKAT telescope, which will become part of SKA1-Mid, and the bespoke BINGO telescope \citep{2013MNRAS.434.1239B} will play an important role in this process in the context of single-dish observations.

\subsubsection{Foregrounds}
Intensity mapping observations suffer from contamination from Galactic and extra-Galactic foregrounds. The main components of the Galactic foregrounds in intensity mapping are synchrotron and free-free emission with amplitudes up to 4 to 5 orders of magnitudes higher than the redshifted HI signal. Current all-sky observations of the foregrounds are sparse in frequency \citep{1988A&AS...74....7R} and suffer under high systematic contaminations \citep{2015MNRAS.451.4311R} which limit the possibility of template-fitting. However, the spectral smoothness of the foregrounds - each component approximately following a power law in frequency - allows one to  separate the spectrally varying HI signal from the foregrounds. Results of Green Bank Telescope data analysis show that blind component separation techniques like Singular-Value Decomposition (SVD, \citealt{2015ApJ...815...51S}) and independent component analysis (fastICA, \citealt{2015arXiv151005453W}) had some success in separating signal and foregrounds. Studies of the performance of these methods on large sky areas (see \citealt{2014MNRAS.441.3271W,2015aska.confE..35W,2015MNRAS.447..400A}) show that large angular scales $\ell <30$ suffer the most  from foreground contamination. In addition, \cite{Alonso:2014dhk} demonstrates how the leakage of polarised foregrounds can affect the cosmological analysis. Alternative, promising separation techniques have been proposed by \cite{2013MNRAS.429..165C,2016MNRAS.456.2749O,2016ApJS..222....3Z,2018arXiv180104082Z}, which provide a diverse collection of techniques to tackle the foreground subtraction of the SKA data. Moreover, the overall effects of foreground residuals on the cosmological interpretation is dramatically reduced by combining intensity mapping data with optical galaxy surveys. \cite{2015aska.confE..35W} presents a comparison of foreground removal in the context of intensity mapping with the SKA.

\subsubsection{Red Noise}
Red noise, also termed $1/f$ noise, is a form of noise inherent to radio receivers which is correlated in time and manifests itself as gain fluctuations (see \citealt{2018MNRAS.478.2416H} for a detailed exposition of the subject in the context of intensity mapping).

On time scales larger than $1/f_{\rm k}$, where is $f_{\rm k}$ is called the knee frequency, the noise no longer behaves as "white noise" and does not integrate down as square root of time. This behavior typically leads to scan synchronous "stripes" in maps. Techniques have been proposed to clean such effects directly on the time ordered data \citep{1996astro.ph..2009J,2002A&A...387..356M}, which should provide unbiased results even for time scales longer that the knee frequency, but this is traded for an overall increase of the noise variance which could ultimately prevent single dish observations from being useful for cosmology.

When scanning the sky with the telescope at a particular scan speed, the timescale $1/f_{\rm k}$ will translate into an angular scale, which should be larger than the scale of the feature (e.g. the BAO scale) one is trying to measure in the thermal noise dominated regime. Hence, scanning as fast possible can help some or all the effects of the red noise. This may not be sufficient with the SKA for the BAO scale and a knee frequency is $\sim 1\,{\rm Hz}$ since the maximum scan speed will be $\sim 3\,{\rm deg}\,{\sec}^{-1}$. However, one would expect that the red  noise is strongly correlated along the frequency direction. If that is the case, it might be possible to remove its effects as part of the foreground cleaning process. In \citet{2018MNRAS.478.2416H}, such frequency correlations were injected directly in the noise power spectrum density in a simulated spectroscopic receiver. For levels of correlation expected for a typically SKA receiver, it was found that the effects could be removed to a level where the noise is within a factor of two of the thermal noise.

An alternative would be to try to calibrate such fluctuations using a noise diode or the sky itself. To \textit{calibrate out} $1/f$ noise using a noise diode signal, the uncertainties on the calibration measurement will need to be significantly better than the r.m.s. fluctuations of the $1/f$ noise ($\sigma(1/f)$). This can not be done on short timescales over which $\sigma(1/f)$ itself is very small (and therefore, not expected to be a problem), but it might be possible to calibrate the SKA receiver on 100 seconds or longer timescales, for feasible diode brightnesses. On 100 second timescales, for a bandwidth of $\Delta \nu~=~$50 MHz and diode brightness of $25\,{\rm K}$, the diode signal stability needs to be better than 1 part in $10^{4}$. It might be possible to use the noise diode in conjunction with component separation techniques described above to relax this requirement.

\subsubsection{Bandpass Calibration}

Bandpass calibration errors are multiplicative with the total system temperature of the receiver. As the system temperature is typically many orders-of-magnitude of greater than the HI intensity signal, even very small bandpass calibration errors can have a big impact on signal recovery. For the SKA receivers, the system temperature is approximately $T_\mathrm{sys} = 22$\,K (at 1200\,MHz), while the expected HI fluctuation scale will be approximately $\sigma_\mathrm{HI} = 0.1$\,mK in a 10\,MHz channel bandwidth. Assuming that at a minimum, the r.m.s. of the HI signal and bandpass calibration errors ($\delta$) should be equal, then

\begin{equation}
	\delta = \frac{\sigma_\mathrm{HI}}{T_\mathrm{sys}} \approx 5 \times 10^{-6}\,,
\end{equation}
at the scale of interest for the HI signal (e.g. $\ell \approx 100$ corresponding to angular scales $\sim 2^{\circ}$). This is the calibration error that should be aimed for in the final SKA HI IM survey per voxel with $4\,{\rm deg}^2\times 10\,{\rm MHz}$. 

Assuming that calibration will be performed $N$ times throughout a survey, and assuming that the bandpass calibration uncertainties are Gaussian then the bandpass uncertainty per calibration should not exceed
\begin{equation}
	\Delta = 5 \times 10^{-6} \sqrt{N_{\rm c}N_\mathrm{dishes}},
\end{equation}
where $N_c$ is the number of bandpass calibrations per dish, and $N_\mathrm{dishes}$ is the number of dishes ($N = N_c N_\mathrm{dishes}$). No particular calibration procedure has been assumed here (e.g. calibration can be performed from a noise diode or astronomical source), and it assumed that there is no uncertainty in the calibration procedure being performed. This calculation also neglects many of the complexities expected of real calibration errors, such as possible non-Gaussianity, and correlations in frequency. On the other hand, there could also be the possibility of dealing with these uncalibrated uncertainties at the power spectrum level, depending on the behavior of such fluctuations.

\subsubsection{RFI from Navigation Satellites}

\begin{figure}
\centering
\includegraphics[width=0.5\textwidth]{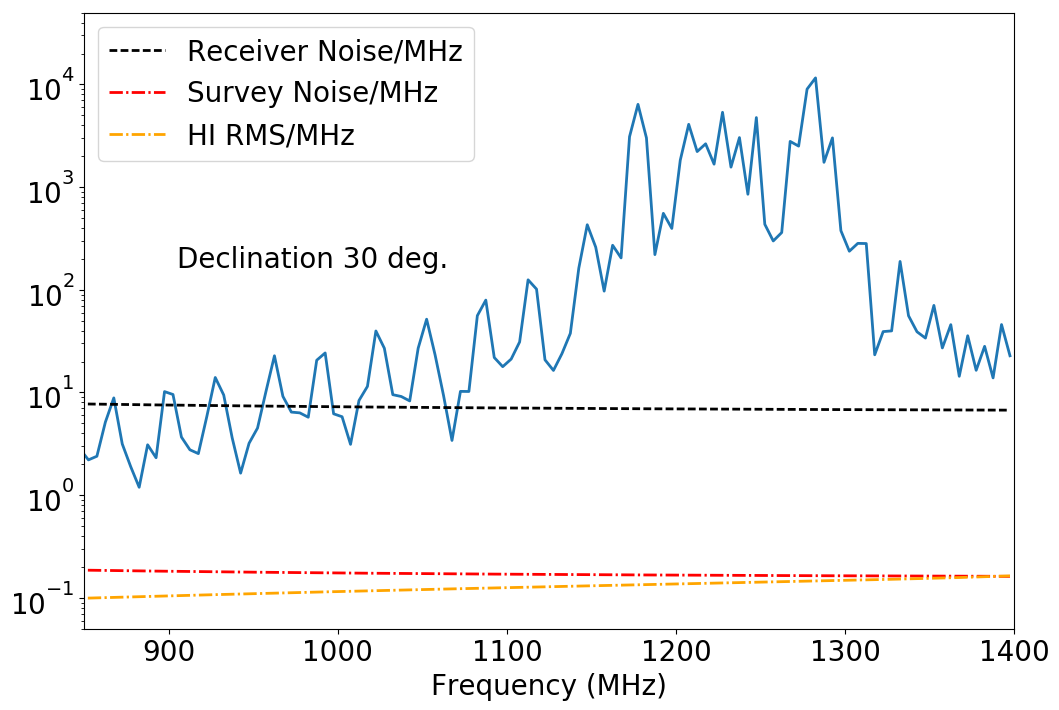}
\caption{Expected r.m.s. of emission from global navigation satellite services within SKA1 Band 2 compared with the expected instantaneous receiver noise (black \textit{dashed} line) for 1\,MHz channel widths. The red \textit{dot-dashed} line shows the sensitivity for an SKA HI IM survey in Band 2 with $30,000 {\rm deg}^2$, 200\,dishes and 30\,days while the orange \textit{dot-dashed} line shows the expected HI signal. Note however that the proposed wide HI intensity mapping survey is in Band 1.}
\label{fig:gnss}
\end{figure}

Residual contamination from satellites can also pose a problem for HI intensity mapping measurements. Although the proposed IM survey is in band 1 where such contamination is expected to be smaller than band 2. Here, we review the recent study of the effect for band 2 \citep{Harper2018} which will indicate some aspects of the problem.

Fig.~\ref{fig:gnss} shows the expected r.m.s. fluctuations of L-band emission from satellites when filtering all satellite within 5\,degrees of the main beam. The figure shows that the satellite signal is comparable to the expected instantaneous sensitivity of the SKA receivers, and  greatly exceeds the SKA survey sensitivity if we consider a large survey in band 2 ($20,000 {\rm deg}^2$, 200\,dishes, 30\,days, 1\,MHz channel widths). In \citet{Harper2018} it is shown that in general the satellite emission does not integrate down on the sky and the residual structure exceeds the expected HI signal fluctuations at all frequencies within SKA Band 2. However, some regions of the sky are clearer than others such as a $8,700 {\rm deg}^2$  patch around the South Celestial Pole ($\delta < -65^{\circ}$) that might make for a good SKA HI IM survey location if done in band 2.

\section{Discussion and conclusions}

In this paper we have brought together the present state of the science case for cosmology using SKA1. A brief summary of the main conclusions are listed below.

\paragraph{Continuum galaxy surveys:} 
\begin{itemize}

\item A continuum survey with SKA1-MID Band 2 of $5,000\,\mathrm{deg}^2$ (the \meddeeps) is expected to yield a number density of resolved star-forming galaxies of $2.7\,\mathrm{arcmin}^{-2}$ usable for a weak lensing shear analysis.
\item By separating these galaxies into tomographic bins and measuring their ellipticities to $\mathcal{O}(10^{-4})$ accuracy it will be possible to measure the dark energy equation of state to a DETF Figure of Merit of $\sim 1.5$ alone, along with measurement of matter parameters to $\sim 5\%$ and modified gravity parameters to $\sim 20\%$.
\item Cross-correlating the weak lensing shear maps made with SKA1 with optical weak lensing experiments will retain nearly all of the statistical power of each survey, while gaining significant robustness to both additive and multiplicative systematics on the cosmological parameter measurements.
\item A continuum survey with SKA1-MID Band 1, covering $20,000 \mathrm{deg}^2$ (the \wideims) will provide a high quality data set for angular clustering analysis of large scale structure, and cross-correlation with the CMB.
\item Large-scale clustering data will provide measurements of primordial non-Gaussianity with statistical errors around half the best current constraints from the CMB, as well as measurements of the dark energy, modified gravity and homogeneous curvature that are independent from, but complementary to, other cosmological probes.
\item The large-area, high redshift radio continuum galaxy sample will allow measurements of the cosmic dipole, providing an accurate and independent test of the origin of the dipole that is impossible with current infrared and optical data.

\end{itemize}

\paragraph{HI galaxy redshift survey:}
\begin{itemize}
\item A SKA1 HI galaxy sample from the 5,000 deg$^2$ \meddeeps~ will provide new independent measurements of the cosmic expansion rate, distance-redshift relation, and linear growth rate from $0 < z \lesssim 0.4$. This will cover a significant additional fraction of the southern sky compared to existing optical surveys, improving constraints on dark energy and modified gravity theories in the important late-time (low redshift) regime.
\item The angular sizes and linewidths of a subset of the detected HI galaxies can be used to infer line-of-sight peculiar velocities through the Doppler magnification and Tully-Fisher methods respectively. The statistics of the measured cosmic velocity field can provide unique constraints on modified gravity theories.
\item The HI galaxy sample will reach extremely high number densities at $z  \lesssim 0.2$, making it possible to reliably identify even small cosmic voids, and obtain high-SNR cross-correlations with $\gamma$-ray maps. The resulting void sample can be used as a complementary probe of matter clustering that is particularly sensitive to modified gravity effects, while the $\gamma$-ray cross-correlations can be used to detect dark matter annihilation.
\end{itemize}

\paragraph{HI intensity mapping:}
\begin{itemize}
\item The SKA1 \wideims~ in combination with the \deeplowims~ will provide a legacy dataset of the large-scale matter distribution measuring the cosmic HI abundance through cosmic time ($0<z<6$) with unprecedented precision.
\item The excellent redshift precision of the intensity maps covering large areas allows one to constrain the expansion history and growth of structure in the Universe, providing constraints on dark energy with SKA1-MID comparable to concurrent surveys at other wavelengths. 
\item The HI intensity mapping surveys will also allow to measure neutrino masses, test Warm Dark Matter models, and inflationary physics. 
\item Synergies of the intensity mapping surveys with optical surveys such as LSST and \textit{Euclid} are crucial for multi-wavelength cosmology and systematics mitigation (see more detailed discussion below). In particular, they will provide ground breaking constraints on ultra-large scale effects such as primordial non-Gaussianity, potentially a factor of 10 better than current measurements.
\end{itemize}

\paragraph{Synergies with other surveys:}
\begin{itemize}
\item We have noted the improved systematic control likely to be possible using the combination of radio data from the SKA1 with optical/NIR data from {\it Euclid} and LSST. Cross-correlations are in principle able to remove all additive residual systematics.

\item Using the SKA with other telescopes can provide complementary physical constraints, e.g. from the combination of optical weak lensing with radio intensity mapping, and vital cross-checks of results by comparing dark energy constraints from optical surveys to those from the SKA. Cross-correlations of probes can measure signatures which would otherwise be buried in noise. 

\item Different radio and optical populations of galaxies afford a multi-tracer approach to large-scale structure measurements, removing sample variance.

\item In addition, optical and radio surveys mutually support one another through the provision of redshifts; intensity mapping can provide calibration of optical photometric redshifts, while optical surveys can provide photometric redshift information for the SKA continuum survey. 

\end{itemize}

\noindent The prospects for observational cosmology in the next decade are particularly promising, with the SKA playing an important part in concert with the Stage IV optical surveys. 

\section*{Acknowledgments}

All authors of the paper contributed to the work presented and the writing of the paper. There was an editorial board who were responsible for the planning and bringing together all the material. This comprised of the working group chairs, Richard Battye and Laura Wolz who are the corresponding authors for the paper, and members from each of the Focus Groups: David Bacon and Stefano Camera for Synergies, Phil Bull for HI galaxies, Ian Harrison for Weak Lensing, David Parkinson for Continuum Science, Alkistis Pourtsidou and M\'ario Santos for HI intensity mapping (Mario Santos was previously working group chair and started the Red Book project), as well as Pedro Ferreira in editorial capacity. This Editorial Board is listed as the first few authors alphabetically and then all other contributors as another alphabetical list.

Some of the results in this work have been derived using the HEALPix~\citep{2005ApJ...622..759G} package. We have used the public Boltzmann code CLASS~\citep{Lesgourgues:2011re} to compute some theoretical observables.

RB,CD and SH acknowledge support from an STFC Consolidated Grant (ST/P000649/1). SC is supported by the Italian Ministry of Education, University and Research (MIUR) through Rita Levi Montalcini project `\textsc{prometheus} -- Probing and Relating Observables with Multi-wavelength Experiments To Help Enlightening the Universe's Structure', and by the `Departments of Excellence 2018-2022' Grant awarded by MIUR (L.\ 232/2016). 
PGF acknowledges support from ERC Grant No: 693024, the Beecroft Trust and STFC.
RM and MGS acknowledge support from the South African Square Kilometre Array Project and National Research Foundation (Grant Nos. 75415 and 84156).
LW is supported by an ARC Discovery Early Career Researcher Award (DE170100356).
YA acknowledges support from the Netherlands Organization for Scientific Research (NWO) and the Dutch Ministry of Education, Culture and Science (OCW), and also from the D-ITP consortium, a program of the NWO that is funded by the OCW. YA is also supported by LabEx ENS-ICFP: ANR-10-LABX-0010/ANR-10-IDEX-0001-02 PSL*.
HP's research is supported by the Tomalla Foundation.
AR has received funding from the People Programme (Marie Curie Actions) of the European Union H2020 Programme under REA grant agreement number 706896 (COSMOFLAGS).
TS and DJS gratefully acknowledge support from the Deutsche Forschungsgemeinschaft (DFG)
within the Research Training Group 1620 `Models of Gravity`.


\bibliographystyle{pasa-mnras}
\bibliography{SKA_cosm}

\begin{thebibliography}{}
\makeatletter
\relax
\def\mn@urlcharsother{\let\do\@makeother \do\$\do\&\do\#\do\^\do\_\do\%\do\~}
\definecolor{darkblue}{rgb}{0,0,0.597656}
\def\mndoi{\begingroup\mn@urlcharsother \@ifnextchar [ {\mndoi@} {\mndoi@[]}}
\def\mndoi@[#1]#2{\def\@tempa{#1}\ifx\@tempa\@empty \href
  {http://dx.doi.org/#2} {\textcolor{darkblue}{doi:#2}}\else \href
  {http://dx.doi.org/#2} {\textcolor{darkblue}{#1}}\fi \endgroup}
\def\mn@eprint#1#2{\mn@eprint@#1:#2::\@nil}
\def\mn@eprint@arXiv#1{\href {http://arxiv.org/abs/#1} {{\tt arXiv:#1}}}
\def\mn@eprint@dblp#1{\href {http://dblp.uni-trier.de/rec/bibtex/#1.xml}
  {dblp:#1}}
\def\mn@eprint@#1:#2:#3:#4\@nil{\def\@tempa {#1}\def\@tempb {#2}\def\@tempc
  {#3}\ifx \@tempc \@empty \let \@tempc \@tempb \let \@tempb \@tempa \fi \ifx
  \@tempb \@empty \def\@tempb {arXiv}\fi \@ifundefined
  {mn@eprint@\@tempb}{\@tempb:\@tempc}{\expandafter \expandafter \csname
  mn@eprint@\@tempb\endcsname \expandafter{\@tempc}}}

\bibitem[\protect\citeauthoryear{{Abdalla} et~al.,}{{Abdalla}
  et~al.}{2015}]{2015aska.confE..17A}
{Abdalla} F.~B.,  et~al., 2015, Advancing Astrophysics with the Square
  Kilometre Array (AASKA14), \href
  {http://adsabs.harvard.edu/abs/2015aska.confE..17A} {p.~17}

\bibitem[\protect\citeauthoryear{{Adams}, {Cresswell}  \& {Easther}}{{Adams}
  et~al.}{2001}]{2001PhRvD..64l3514A}
{Adams} J.,  {Cresswell} B.,   {Easther} R.,  2001, \mndoi [\prd]
  {10.1103/PhysRevD.64.123514}, \href
  {https://ui.adsabs.harvard.edu/#abs/2001PhRvD..64l3514A} {64, 123514}

\bibitem[\protect\citeauthoryear{Ade et~al.}{Ade et~al.}{2014}]{Planck:2013jfk}
Ade P. A.~R.,  et~al., 2014, \mndoi [\aap] {10.1051/0004-6361/201321569}, 571,
  A22

\bibitem[\protect\citeauthoryear{Ade et~al.}{Ade et~al.}{2016a}]{Ade:2015xua}
Ade P. A.~R.,  et~al., 2016a, \mndoi [\aap] {10.1051/0004-6361/201525830}, 594,
  A13

\bibitem[\protect\citeauthoryear{Ade et~al.}{Ade et~al.}{2016b}]{Ade:2015lrj}
Ade P. A.~R.,  et~al., 2016b, \mndoi [\aap] {10.1051/0004-6361/201525898}, 594,
  A20

\bibitem[\protect\citeauthoryear{{Adshead}, {Dvorkin}, {Hu}  \&
  {Lim}}{{Adshead} et~al.}{2012}]{2012PhRvD..85b3531A}
{Adshead} P.,  {Dvorkin} C.,  {Hu} W.,   {Lim} E.~A.,  2012, \mndoi [\prd]
  {10.1103/PhysRevD.85.023531}, \href
  {http://adsabs.harvard.edu/abs/2012PhRvD..85b3531A} {85, 023531}

\bibitem[\protect\citeauthoryear{Aghamousa et~al.}{Aghamousa
  et~al.}{2016}]{Aghamousa:2016zmz}
Aghamousa A.,  et~al., 2016, preprint (\mn@eprint {arXiv} {1611.00036})

\bibitem[\protect\citeauthoryear{Akrami et~al.}{Akrami
  et~al.}{2018}]{Akrami:2018odb}
Akrami Y.,  et~al., 2018, preprint (\mn@eprint {arXiv} {1807.06211})

\bibitem[\protect\citeauthoryear{{Alam} et~al.,}{{Alam}
  et~al.}{2017}]{2017MNRAS.470.2617A}
{Alam} S.,  et~al., 2017, \mndoi [\mnras] {10.1093/mnras/stx721}, \href
  {http://adsabs.harvard.edu/abs/2017MNRAS.470.2617A} {470, 2617}

\bibitem[\protect\citeauthoryear{Albrecht et~al.}{Albrecht
  et~al.}{2009}]{Albrecht:2009ct}
Albrecht A.,  et~al., 2009, preprint (\mn@eprint {arXiv} {0901.0721})

\bibitem[\protect\citeauthoryear{Alonso \& Ferreira}{Alonso \&
  Ferreira}{2015}]{Alonso:2015sfa}
Alonso D.,  Ferreira P.~G.,  2015, \mndoi [\prd] {10.1103/PhysRevD.92.063525},
  D92, 063525

\bibitem[\protect\citeauthoryear{{Alonso}, {Bull}, {Ferreira}  \&
  {Santos}}{{Alonso} et~al.}{2015a}]{2015MNRAS.447..400A}
{Alonso} D.,  {Bull} P.,  {Ferreira} P.~G.,   {Santos} M.~G.,  2015a, \mndoi
  [\mnras] {10.1093/mnras/stu2474}, \href
  {https://ui.adsabs.harvard.edu/#abs/2015MNRAS.447..400A} {447, 400}

\bibitem[\protect\citeauthoryear{Alonso, Bull, Ferreira  \& Santos}{Alonso
  et~al.}{2015b}]{Alonso:2014dhk}
Alonso D.,  Bull P.,  Ferreira P.~G.,   Santos M.~G.,  2015b, \mndoi [\mnras]
  {10.1093/mnras/stu2474}, 447, 400

\bibitem[\protect\citeauthoryear{{Alonso}, {Bull}, {Ferreira}, {Maartens}  \&
  {Santos}}{{Alonso} et~al.}{2015c}]{2015ApJ...814..145A}
{Alonso} D.,  {Bull} P.,  {Ferreira} P.~G.,  {Maartens} R.,   {Santos} M.~G.,
  2015c, \mndoi [\apj] {10.1088/0004-637X/814/2/145}, \href
  {http://adsabs.harvard.edu/abs/2015ApJ...814..145A} {814, 145}

\bibitem[\protect\citeauthoryear{Alonso, Bull, Ferreira, Maartens  \&
  Santos}{Alonso et~al.}{2015d}]{Alonso:2015uua}
Alonso D.,  Bull P.,  Ferreira P.~G.,  Maartens R.,   Santos M.,  2015d, \mndoi
  [\apj] {10.1088/0004-637X/814/2/145}, 814, 145

\bibitem[\protect\citeauthoryear{{Alonso}, {Ferreira}, {Jarvis}  \&
  {Moodley}}{{Alonso} et~al.}{2017}]{2017PhRvD..96d3515A}
{Alonso} D.,  {Ferreira} P.~G.,  {Jarvis} M.~J.,   {Moodley} K.,  2017, \mndoi
  [\prd] {10.1103/PhysRevD.96.043515}, \href
  {https://ui.adsabs.harvard.edu/#abs/2017PhRvD..96d3515A} {96, 043515}

\bibitem[\protect\citeauthoryear{Alonso et~al.}{Alonso
  et~al.}{2018}]{Mandelbaum:2018ouv}
Alonso D.,  et~al., 2018, preprint (\mn@eprint {arXiv} {1809.01669})

\bibitem[\protect\citeauthoryear{{Amara} \& {R{\'e}fr{\'e}gier}}{{Amara} \&
  {R{\'e}fr{\'e}gier}}{2008}]{2008MNRAS.391..228A}
{Amara} A.,  {R{\'e}fr{\'e}gier} A.,  2008, \mndoi [\mnras]
  {10.1111/j.1365-2966.2008.13880.x}, \href
  {http://adsabs.harvard.edu/abs/2008MNRAS.391..228A} {391, 228}

\bibitem[\protect\citeauthoryear{Amendola et~al.}{Amendola
  et~al.}{2013}]{Amendola:2012ys}
Amendola L.,  et~al., 2013, \mndoi [Living Rev. Rel.] {10.12942/lrr-2013-6},
  16, 6

\bibitem[\protect\citeauthoryear{Amendola et~al.}{Amendola
  et~al.}{2018}]{Amendola:2016saw}
Amendola L.,  et~al., 2018, \mndoi [Living Rev. Rel.]
  {10.1007/s41114-017-0010-3}, 21, 2

\bibitem[\protect\citeauthoryear{{Anderson} et~al.,}{{Anderson}
  et~al.}{2018}]{2018MNRAS.tmp..340A}
{Anderson} C.~J.,  et~al., 2018, \mndoi [\mnras] {10.1093/mnras/sty346}, \href
  {https://ui.adsabs.harvard.edu/#abs/2018MNRAS.476.3382A} {476, 3382}

\bibitem[\protect\citeauthoryear{{Andrianomena}, {Bonvin}, {Bacon}, {Bull},
  {Clarkson}, {Maartens}  \& {Moloi}}{{Andrianomena}
  et~al.}{2018}]{2018arXiv181012793A}
{Andrianomena} S.,  {Bonvin} C.,  {Bacon} D.,  {Bull} P.,  {Clarkson} C.,
  {Maartens} R.,   {Moloi} T.,  2018, preprint, \href
  {https://ui.adsabs.harvard.edu/#abs/2018arXiv181012793A} {p.
  arXiv:1810.12793} (\mn@eprint {arXiv} {1810.12793})

\bibitem[\protect\citeauthoryear{{Ascaso}, {Mei}  \& {Ben{\'{\i}}tez}}{{Ascaso}
  et~al.}{2015}]{AscasoA2A}
{Ascaso} B.,  {Mei} S.,   {Ben{\'{\i}}tez} N.,  2015, \mndoi [\mnras]
  {10.1093/mnras/stv1597}, \href
  {http://adsabs.harvard.edu/abs/2015MNRAS.453.2515A} {453, 2515}

\bibitem[\protect\citeauthoryear{{Astier} et~al.,}{{Astier}
  et~al.}{2006}]{2006A&A...447...31A}
{Astier} P.,  et~al., 2006, \mndoi [\aap] {10.1051/0004-6361:20054185}, \href
  {http://adsabs.harvard.edu/abs/2006A%26A...447...31A} {447, 31}

\bibitem[\protect\citeauthoryear{{Atwood} et~al.,}{{Atwood}
  et~al.}{2009}]{2009ApJ...697.1071A}
{Atwood} W.~B.,  et~al., 2009, \mndoi [\apj] {10.1088/0004-637X/697/2/1071},
  \href {http://adsabs.harvard.edu/abs/2009ApJ...697.1071A} {697, 1071}

\bibitem[\protect\citeauthoryear{{Bacon}, {Andrianomena}, {Clarkson}, {Bolejko}
   \& {Maartens}}{{Bacon} et~al.}{2014}]{2014MNRAS.443.1900B}
{Bacon} D.~J.,  {Andrianomena} S.,  {Clarkson} C.,  {Bolejko} K.,   {Maartens}
  R.,  2014, \mndoi [\mnras] {10.1093/mnras/stu1270}, \href
  {http://adsabs.harvard.edu/abs/2014MNRAS.443.1900B} {443, 1900}

\bibitem[\protect\citeauthoryear{{Bagla}, {Khandai}  \& {Datta}}{{Bagla}
  et~al.}{2010}]{2010MNRAS.407..567B}
{Bagla} J.~S.,  {Khandai} N.,   {Datta} K.~K.,  2010, \mndoi [\mnras]
  {10.1111/j.1365-2966.2010.16933.x}, \href
  {http://adsabs.harvard.edu/abs/2010MNRAS.407..567B} {407, 567}

\bibitem[\protect\citeauthoryear{{Baker} \& {Bull}}{{Baker} \&
  {Bull}}{2015}]{2015ApJ...811..116B}
{Baker} T.,  {Bull} P.,  2015, \mndoi [\apj] {10.1088/0004-637X/811/2/116},
  \href {http://adsabs.harvard.edu/abs/2015ApJ...811..116B} {811, 116}

\bibitem[\protect\citeauthoryear{Baker, Ferreira  \& Skordis}{Baker
  et~al.}{2014}]{Baker:2013hia}
Baker T.,  Ferreira P.~G.,   Skordis C.,  2014, \mndoi [\prd]
  {10.1103/PhysRevD.89.024026}, D89, 024026

\bibitem[\protect\citeauthoryear{{Ballardini}, {Finelli}, {Maartens}  \&
  {Moscardini}}{{Ballardini} et~al.}{2018}]{2018JCAP...04..044B}
{Ballardini} M.,  {Finelli} F.,  {Maartens} R.,   {Moscardini} L.,  2018,
  \mndoi [Journal of Cosmology and Astro-Particle Physics]
  {10.1088/1475-7516/2018/04/044}, \href
  {https://ui.adsabs.harvard.edu/#abs/2018JCAP...04..044B} {2018, 044}

\bibitem[\protect\citeauthoryear{{Bandura} et~al.,}{{Bandura}
  et~al.}{2014}]{2014SPIE.9145E..22B}
{Bandura} K.,  et~al., 2014, in Ground-based and Airborne Telescopes V. p.
  914522 (\mn@eprint {arXiv} {1406.2288}), \mndoi{10.1117/12.2054950}

\bibitem[\protect\citeauthoryear{{Bartelmann} \& {Schneider}}{{Bartelmann} \&
  {Schneider}}{2001}]{Bartelmann:2001}
{Bartelmann} M.,  {Schneider} P.,  2001, \mndoi [\physrep]
  {10.1016/S0370-1573(00)00082-X}, \href
  {https://ui.adsabs.harvard.edu/#abs/2001PhR...340..291B} {340, 291}

\bibitem[\protect\citeauthoryear{{Battye}, {Davies}  \& {Weller}}{{Battye}
  et~al.}{2004}]{2004MNRAS.355.1339B}
{Battye} R.~A.,  {Davies} R.~D.,   {Weller} J.,  2004, \mndoi [\mnras]
  {10.1111/j.1365-2966.2004.08416.x}, \href
  {http://adsabs.harvard.edu/abs/2004MNRAS.355.1339B} {355, 1339}

\bibitem[\protect\citeauthoryear{{Battye}, {Browne}, {Dickinson}, {Heron},
  {Maffei}  \& {Pourtsidou}}{{Battye} et~al.}{2013}]{2013MNRAS.434.1239B}
{Battye} R.~A.,  {Browne} I.~W.~A.,  {Dickinson} C.,  {Heron} G.,  {Maffei} B.,
    {Pourtsidou} A.,  2013, \mndoi [\mnras] {10.1093/mnras/stt1082}, \href
  {http://adsabs.harvard.edu/abs/2013MNRAS.434.1239B} {434, 1239}

\bibitem[\protect\citeauthoryear{Bengaly, Siewert, Schwarz  \&
  Maartens}{Bengaly et~al.}{2018a}]{Bengaly:2018ykb}
Bengaly C. A.~P.,  Siewert T.~M.,  Schwarz D.~J.,   Maartens R.,  2018a,
  preprint (\mn@eprint {arXiv} {1810.04960})

\bibitem[\protect\citeauthoryear{{Bengaly}, {Maartens}  \& {Santos}}{{Bengaly}
  et~al.}{2018b}]{2017arXiv171008804B}
{Bengaly} C. A.~P.,  {Maartens} R.,   {Santos} M.~G.,  2018b, \mndoi [\jcap]
  {10.1088/1475-7516/2018/04/031}, \href
  {https://ui.adsabs.harvard.edu/#abs/2018JCAP...04..031B} {2018, 031}

\bibitem[\protect\citeauthoryear{{Bharadwaj}, {Nath}  \& {Sethi}}{{Bharadwaj}
  et~al.}{2001}]{2001JApA...22...21B}
{Bharadwaj} S.,  {Nath} B.~B.,   {Sethi} S.~K.,  2001, \mndoi [\jcap]
  {10.1007/BF02933588}, \href
  {http://adsabs.harvard.edu/abs/2001JApA...22...21B} {22, 21}

\bibitem[\protect\citeauthoryear{{Bird}, {Cholis}, {Mu{\~n}oz}, {Ali-
  Ha{\"\i}moud}, {Kamionkowski}, {Kovetz}, {Raccanelli}  \& {Riess}}{{Bird}
  et~al.}{2016}]{Bird2016}
{Bird} S.,  {Cholis} I.,  {Mu{\~n}oz} J.~B.,  {Ali- Ha{\"\i}moud} Y.,
  {Kamionkowski} M.,  {Kovetz} E.~D.,  {Raccanelli} A.,   {Riess} A.~G.,  2016,
  \mndoi [\prl] {10.1103/PhysRevLett.116.201301}, \href
  {https://ui.adsabs.harvard.edu/#abs/2016PhRvL.116t1301B} {116, 201301}

\bibitem[\protect\citeauthoryear{{Blake} \& {Wall}}{{Blake} \&
  {Wall}}{2002}]{2002Natur.416..150B}
{Blake} C.,  {Wall} J.,  2002, \mndoi [\nat] {10.1038/416150a}, \href
  {http://adsabs.harvard.edu/abs/2002Natur.416..150B} {416, 150}

\bibitem[\protect\citeauthoryear{{Bonaldi}, {Harrison}, {Camera}  \&
  {Brown}}{{Bonaldi} et~al.}{2016}]{2016MNRAS.463.3686B}
{Bonaldi} A.,  {Harrison} I.,  {Camera} S.,   {Brown} M.~L.,  2016, \mndoi
  [\mnras] {10.1093/mnras/stw2104}, \href
  {http://adsabs.harvard.edu/abs/2016MNRAS.463.3686B} {463, 3686}

\bibitem[\protect\citeauthoryear{{Bonaldi}, {Bonato}, {Galluzzi}, {Harrison},
  {Massardi}, {Kay}, {De Zotti}  \& {Brown}}{{Bonaldi}
  et~al.}{2018}]{2018arXiv180505222B}
{Bonaldi} A.,  {Bonato} M.,  {Galluzzi} V.,  {Harrison} I.,  {Massardi} M.,
  {Kay} S.,  {De Zotti} G.,   {Brown} M.~L.,  2018, preprint, \href
  {http://adsabs.harvard.edu/abs/2018arXiv180505222B} {} (\mn@eprint {arXiv}
  {1805.05222})

\bibitem[\protect\citeauthoryear{{Bonvin}}{{Bonvin}}{2008}]{2008PhRvD..78l3530B}
{Bonvin} C.,  2008, \prd, 78, 123530

\bibitem[\protect\citeauthoryear{{Bonvin}, {Andrianomena}, {Bacon}, {Clarkson},
  {Maartens}, {Moloi}  \& {Bull}}{{Bonvin} et~al.}{2017}]{2017MNRAS.472.3936B}
{Bonvin} C.,  {Andrianomena} S.,  {Bacon} D.,  {Clarkson} C.,  {Maartens} R.,
  {Moloi} T.,   {Bull} P.,  2017, \mndoi [\mnras] {10.1093/mnras/stx2049},
  \href {https://ui.adsabs.harvard.edu/#abs/2017MNRAS.472.3936B} {472, 3936}

\bibitem[\protect\citeauthoryear{{Borzyszkowski}, {Bertacca}  \&
  {Porciani}}{{Borzyszkowski} et~al.}{2017}]{2017MNRAS.471.3899B}
{Borzyszkowski} M.,  {Bertacca} D.,   {Porciani} C.,  2017, \mndoi [\mnras]
  {10.1093/mnras/stx1423}, \href
  {http://adsabs.harvard.edu/abs/2017MNRAS.471.3899B} {471, 3899}

\bibitem[\protect\citeauthoryear{Branchini, Camera, Cuoco, Fornengo, Regis,
  Viel  \& Xia}{Branchini et~al.}{2017}]{Branchini:2016glc}
Branchini E.,  Camera S.,  Cuoco A.,  Fornengo N.,  Regis M.,  Viel M.,   Xia
  J.-Q.,  2017, \mndoi [\apjs] {10.3847/1538-4365/228/1/8}, 228, 8

\bibitem[\protect\citeauthoryear{{Braun}, {Bourke}, {Green}, {Keane}  \&
  {Wagg}}{{Braun} et~al.}{2015}]{2015aska.confE.174B}
{Braun} R.,  {Bourke} T.,  {Green} J.~A.,  {Keane} E.,   {Wagg} J.,  2015,
  Advancing Astrophysics with the Square Kilometre Array (AASKA14), \href
  {http://ukads.nottingham.ac.uk/abs/2015aska.confE.174B} {p.~174}

\bibitem[\protect\citeauthoryear{Brown \& Battye}{Brown \&
  Battye}{2010}]{Brown:2010rr}
Brown M.~L.,  Battye R.~A.,  2010, \mndoi [\mnras]
  {10.1111/j.1365-2966.2010.17583.x}, 410

\bibitem[\protect\citeauthoryear{Brown \& Battye}{Brown \&
  Battye}{2011}]{Brown:2011db}
Brown M.~L.,  Battye R.~A.,  2011, \mndoi [\apj] {10.1088/2041-8205/735/1/L23},
  735

\bibitem[\protect\citeauthoryear{{Bull}}{{Bull}}{2016}]{Bull2016}
{Bull} P.,  2016, \mndoi [\apj] {10.3847/0004-637X/817/1/26}, \href
  {http://adsabs.harvard.edu/abs/2016ApJ...817...26B} {817, 26}

\bibitem[\protect\citeauthoryear{{Bull}, {Camera}, {Raccanelli}, {Blake},
  {Ferreira}, {Santos}  \& {Schwarz}}{{Bull}
  et~al.}{2015a}]{2015aska.confE..24B}
{Bull} P.,  {Camera} S.,  {Raccanelli} A.,  {Blake} C.,  {Ferreira} P.,
  {Santos} M.,   {Schwarz} D.~J.,  2015a, Advancing Astrophysics with the
  Square Kilometre Array (AASKA14), \href
  {http://adsabs.harvard.edu/abs/2015aska.confE..24B} {p.~24}

\bibitem[\protect\citeauthoryear{{Bull}, {Ferreira}, {Patel}  \&
  {Santos}}{{Bull} et~al.}{2015b}]{Bull_2015}
{Bull} P.,  {Ferreira} P.~G.,  {Patel} P.,   {Santos} M.~G.,  2015b, \mndoi
  [\apj] {10.1088/0004-637X/803/1/21}, \href
  {http://adsabs.harvard.edu/abs/2015ApJ...803...21B} {803, 21}

\bibitem[\protect\citeauthoryear{Camera, Santos, Bacon, Jarvis, McAlpine,
  Norris, Raccanelli  \& Rottgering}{Camera et~al.}{2012}]{Camera:2012ez}
Camera S.,  Santos M.~G.,  Bacon D.~J.,  Jarvis M.~J.,  McAlpine K.,  Norris
  R.~P.,  Raccanelli A.,   Rottgering H.,  2012, \mndoi [\mnras]
  {10.1111/j.1365-2966.2012.22073.x}, 427, 2079

\bibitem[\protect\citeauthoryear{{Camera}, {Santos}, {Ferreira}  \&
  {Ferramacho}}{{Camera} et~al.}{2013a}]{2013PhRvL.111q1302C}
{Camera} S.,  {Santos} M.~G.,  {Ferreira} P.~G.,   {Ferramacho} L.,  2013a,
  \mndoi [\prl] {10.1103/PhysRevLett.111.171302}, \href
  {http://adsabs.harvard.edu/abs/2013PhRvL.111q1302C} {111, 171302}

\bibitem[\protect\citeauthoryear{Camera, Fornasa, Fornengo  \& Regis}{Camera
  et~al.}{2013b}]{Camera:2012cj}
Camera S.,  Fornasa M.,  Fornengo N.,   Regis M.,  2013b, \mndoi [\apj]
  {10.1088/2041-8205/771/1/L5}, 771, L5

\bibitem[\protect\citeauthoryear{Camera, Santos  \& Maartens}{Camera
  et~al.}{2015a}]{Camera:2014bwa}
Camera S.,  Santos M.~G.,   Maartens R.,  2015a, \mndoi [\mnras]
  {10.1093/mnras/stv040, 10.1093/mnras/stx159}, 448, 1035

\bibitem[\protect\citeauthoryear{{Camera}, {Maartens}  \& {Santos}}{{Camera}
  et~al.}{2015b}]{2015MNRAS.451L..80C}
{Camera} S.,  {Maartens} R.,   {Santos} M.~G.,  2015b, \mndoi [\mnras]
  {10.1093/mnrasl/slv069}, \href
  {http://adsabs.harvard.edu/abs/2015MNRAS.451L..80C} {451, L80}

\bibitem[\protect\citeauthoryear{Camera, Fornasa, Fornengo  \& Regis}{Camera
  et~al.}{2015c}]{Camera:2014rja}
Camera S.,  Fornasa M.,  Fornengo N.,   Regis M.,  2015c, \mndoi [\jcap]
  {10.1088/1475-7516/2015/06/029}, 1506, 029

\bibitem[\protect\citeauthoryear{Camera, Harrison, Bonaldi  \& Brown}{Camera
  et~al.}{2017}]{Camera:2016owj}
Camera S.,  Harrison I.,  Bonaldi A.,   Brown M.~L.,  2017, \mndoi [\mnras]
  {10.1093/mnras/stw2688}, 464, 4747

\bibitem[\protect\citeauthoryear{{Carucci}, {Villaescusa-Navarro}, {Viel}  \&
  {Lapi}}{{Carucci} et~al.}{2015}]{carucci15}
{Carucci} I.~P.,  {Villaescusa-Navarro} F.,  {Viel} M.,   {Lapi} A.,  2015,
  \mndoi [\jcap] {10.1088/1475-7516/2015/07/047}, \href
  {http://adsabs.harvard.edu/abs/2015JCAP...07..047C} {7, 047}

\bibitem[\protect\citeauthoryear{{Castorina} \&
  {Villaescusa-Navarro}}{{Castorina} \& {Villaescusa-Navarro}}{2017}]{EmaPaco}
{Castorina} E.,  {Villaescusa-Navarro} F.,  2017, \mndoi [\mnras]
  {10.1093/mnras/stx1599}, \href
  {http://adsabs.harvard.edu/abs/2017MNRAS.471.1788C} {471, 1788}

\bibitem[\protect\citeauthoryear{{Challinor} \& {Lewis}}{{Challinor} \&
  {Lewis}}{2011}]{2011PhRvD..84d3516C}
{Challinor} A.,  {Lewis} A.,  2011, \mndoi [\prd] {10.1103/PhysRevD.84.043516},
  \href {http://adsabs.harvard.edu/abs/2011PhRvD..84d3516C} {84, 043516}

\bibitem[\protect\citeauthoryear{{Chang}, {Refregier}  \& {Helfand}}{{Chang}
  et~al.}{2004}]{2004ApJ...617..794C}
{Chang} T.-C.,  {Refregier} A.,   {Helfand} D.~J.,  2004, \mndoi [\apj]
  {10.1086/425491}, \href {http://adsabs.harvard.edu/abs/2004ApJ...617..794C}
  {617, 794}

\bibitem[\protect\citeauthoryear{Chang, Pen, Bandura  \& Peterson}{Chang
  et~al.}{2010}]{Chang:2010jp}
Chang T.-C.,  Pen U.-L.,  Bandura K.,   Peterson J.~B.,  2010, \mndoi [\nat]
  {10.1038/nature09187}, 466, 463

\bibitem[\protect\citeauthoryear{{Chapman} et~al.,}{{Chapman}
  et~al.}{2013}]{2013MNRAS.429..165C}
{Chapman} E.,  et~al., 2013, \mndoi [\mnras] {10.1093/mnras/sts333}, \href
  {https://ui.adsabs.harvard.edu/#abs/2013MNRAS.429..165C} {429, 165}

\bibitem[\protect\citeauthoryear{{Chen}, {Easther}  \& {Lim}}{{Chen}
  et~al.}{2008}]{2008JCAP...04..010C}
{Chen} X.,  {Easther} R.,   {Lim} E.~A.,  2008, \mndoi [\jcap]
  {10.1088/1475-7516/2008/04/010}, \href
  {http://adsabs.harvard.edu/abs/2008JCAP...04..010C} {4, 010}

\bibitem[\protect\citeauthoryear{{Chevallier} \& {Polarski}}{{Chevallier} \&
  {Polarski}}{2001}]{CPL_w0wa}
{Chevallier} M.,  {Polarski} D.,  2001, \mndoi [International Journal of Modern
  Physics D] {10.1142/S0218271801000822}, \href
  {http://adsabs.harvard.edu/abs/2001IJMPD..10..213C} {10, 213}

\bibitem[\protect\citeauthoryear{{Cole} et~al.,}{{Cole}
  et~al.}{2005}]{2005MNRAS.362..505C}
{Cole} S.,  et~al., 2005, \mndoi [\mnras] {10.1111/j.1365-2966.2005.09318.x},
  \href {http://adsabs.harvard.edu/abs/2005MNRAS.362..505C} {362, 505}

\bibitem[\protect\citeauthoryear{{Colin}, {Mohayaee}, {Rameez}  \&
  {Sarkar}}{{Colin} et~al.}{2017}]{2017MNRAS.471.1045C}
{Colin} J.,  {Mohayaee} R.,  {Rameez} M.,   {Sarkar} S.,  2017, \mndoi [\mnras]
  {10.1093/mnras/stx1631}, \href
  {http://adsabs.harvard.edu/abs/2017MNRAS.471.1045C} {471, 1045}

\bibitem[\protect\citeauthoryear{Crighton et~al.}{Crighton
  et~al.}{2015}]{Crighton:2015pza}
Crighton N. H.~M.,  et~al., 2015, \mndoi [\mnras] {10.1093/mnras/stv1182}, 452,
  217

\bibitem[\protect\citeauthoryear{{Cunnington}, {Harrison}, {Pourtsidou}  \&
  {Bacon}}{{Cunnington} et~al.}{2018}]{2018arXiv180504498C}
{Cunnington} S.,  {Harrison} I.,  {Pourtsidou} A.,   {Bacon} D.,  2018,
  preprint, \href {http://adsabs.harvard.edu/abs/2018arXiv180504498C} {}
  (\mn@eprint {arXiv} {1805.04498})

\bibitem[\protect\citeauthoryear{Cuoco, Xia, Regis, Branchini, Fornengo  \&
  Viel}{Cuoco et~al.}{2015}]{Cuoco:2015rfa}
Cuoco A.,  Xia J.-Q.,  Regis M.,  Branchini E.,  Fornengo N.,   Viel M.,  2015,
  \mndoi [\apjs] {10.1088/0067-0049/221/2/29}, 221, 29

\bibitem[\protect\citeauthoryear{{DESI Collaboration} et~al.,}{{DESI
  Collaboration} et~al.}{2016}]{2016arXiv161100036D}
{DESI Collaboration} et~al., 2016, preprint, \href
  {http://adsabs.harvard.edu/abs/2016arXiv161100036D} {} (\mn@eprint {arXiv}
  {1611.00036})

\bibitem[\protect\citeauthoryear{Dalal, Dore, Huterer  \& Shirokov}{Dalal
  et~al.}{2008}]{Dalal:2007cu}
Dalal N.,  Dore O.,  Huterer D.,   Shirokov A.,  2008, \mndoi [\prd]
  {10.1103/PhysRevD.77.123514}, D77, 123514

\bibitem[\protect\citeauthoryear{{Dark Energy Survey Collaboration}
  et~al.,}{{Dark Energy Survey Collaboration}
  et~al.}{2016}]{2016MNRAS.460.1270D}
{Dark Energy Survey Collaboration} et~al., 2016, \mndoi [\mnras]
  {10.1093/mnras/stw641}, \href
  {http://adsabs.harvard.edu/abs/2016MNRAS.460.1270D} {460, 1270}

\bibitem[\protect\citeauthoryear{{Dav{\'e}}, {Katz}, {Oppenheimer}, {Kollmeier}
   \& {Weinberg}}{{Dav{\'e}} et~al.}{2013}]{2013MNRAS.434.2645D}
{Dav{\'e}} R.,  {Katz} N.,  {Oppenheimer} B.~D.,  {Kollmeier} J.~A.,
  {Weinberg} D.~H.,  2013, \mndoi [\mnras] {10.1093/mnras/stt1274}, \href
  {http://adsabs.harvard.edu/abs/2013MNRAS.434.2645D} {434, 2645}

\bibitem[\protect\citeauthoryear{{Demetroullas} \& {Brown}}{{Demetroullas} \&
  {Brown}}{2016}]{2016MNRAS.456.3100D}
{Demetroullas} C.,  {Brown} M.~L.,  2016, \mndoi [\mnras]
  {10.1093/mnras/stv2876}, \href
  {http://adsabs.harvard.edu/abs/2016MNRAS.456.3100D} {456, 3100}

\bibitem[\protect\citeauthoryear{{Demetroullas} \& {Brown}}{{Demetroullas} \&
  {Brown}}{2018}]{2018MNRAS.473..937D}
{Demetroullas} C.,  {Brown} M.~L.,  2018, \mndoi [\mnras]
  {10.1093/mnras/stx2366}, \href
  {http://adsabs.harvard.edu/abs/2018MNRAS.473..937D} {473, 937}

\bibitem[\protect\citeauthoryear{{Efstathiou}, {Sutherland}  \&
  {Maddox}}{{Efstathiou} et~al.}{1990}]{1990Natur.348..705E}
{Efstathiou} G.,  {Sutherland} W.~J.,   {Maddox} S.~J.,  1990, \mndoi [\nat]
  {10.1038/348705a0}, \href {http://adsabs.harvard.edu/abs/1990Natur.348..705E}
  {348, 705}

\bibitem[\protect\citeauthoryear{{Ellis} \& {Baldwin}}{{Ellis} \&
  {Baldwin}}{1984}]{1984MNRAS.206..377E}
{Ellis} G.~F.~R.,  {Baldwin} J.~E.,  1984, \mndoi [\mnras]
  {10.1093/mnras/206.2.377}, \href
  {http://adsabs.harvard.edu/abs/1984MNRAS.206..377E} {206, 377}

\bibitem[\protect\citeauthoryear{{Ferramacho}, {Santos}, {Jarvis}  \&
  {Camera}}{{Ferramacho} et~al.}{2014}]{2014MNRAS.442.2511F}
{Ferramacho} L.~D.,  {Santos} M.~G.,  {Jarvis} M.~J.,   {Camera} S.,  2014,
  \mndoi [\mnras] {10.1093/mnras/stu1015}, \href
  {https://ui.adsabs.harvard.edu/#abs/2014MNRAS.442.2511F} {442, 2511}

\bibitem[\protect\citeauthoryear{Finelli et~al.}{Finelli
  et~al.}{2018}]{Finelli:2016cyd}
Finelli F.,  et~al., 2018, \mndoi [\jcap] {10.1088/1475-7516/2018/04/016},
  1804, 016

\bibitem[\protect\citeauthoryear{Fonseca, Camera, Santos  \& Maartens}{Fonseca
  et~al.}{2015}]{Fonseca:2015laa}
Fonseca J.,  Camera S.,  Santos M.,   Maartens R.,  2015, \mndoi [\apj]
  {10.1088/2041-8205/812/2/L22}, 812, L22

\bibitem[\protect\citeauthoryear{{Fonseca}, {Silva}, {Santos}  \&
  {Cooray}}{{Fonseca} et~al.}{2017}]{2017MNRAS.464.1948F}
{Fonseca} J.,  {Silva} M.~B.,  {Santos} M.~G.,   {Cooray} A.,  2017, \mndoi
  [\mnras] {10.1093/mnras/stw2470}, \href
  {http://adsabs.harvard.edu/abs/2017MNRAS.464.1948F} {464, 1948}

\bibitem[\protect\citeauthoryear{{Fonseca}, {Maartens}  \& {Santos}}{{Fonseca}
  et~al.}{2018}]{2018MNRAS.479.3490F}
{Fonseca} J.,  {Maartens} R.,   {Santos} M.~G.,  2018, \mndoi [\mnras]
  {10.1093/mnras/sty1702}, \href
  {http://adsabs.harvard.edu/abs/2018MNRAS.479.3490F} {479, 3490}

\bibitem[\protect\citeauthoryear{Fornengo \& Regis}{Fornengo \&
  Regis}{2014}]{Fornengo:2013rga}
Fornengo N.,  Regis M.,  2014, \mndoi [Front. Physics]
  {10.3389/fphy.2014.00006}, 2, 6

\bibitem[\protect\citeauthoryear{Fornengo, Perotto, Regis  \& Camera}{Fornengo
  et~al.}{2015}]{Fornengo:2014cya}
Fornengo N.,  Perotto L.,  Regis M.,   Camera S.,  2015, \mndoi [\apj]
  {10.1088/2041-8205/802/1/L1}, 802, L1

\bibitem[\protect\citeauthoryear{{G{\'o}rski}, {Hivon}, {Banday}, {Wandelt},
  {Hansen}, {Reinecke}  \& {Bartelmann}}{{G{\'o}rski}
  et~al.}{2005}]{2005ApJ...622..759G}
{G{\'o}rski} K.~M.,  {Hivon} E.,  {Banday} A.~J.,  {Wandelt} B.~D.,  {Hansen}
  F.~K.,  {Reinecke} M.,   {Bartelmann} M.,  2005, \mndoi [\apj]
  {10.1086/427976}, \href {http://adsabs.harvard.edu/abs/2005ApJ...622..759G}
  {622, 759}

\bibitem[\protect\citeauthoryear{Hall, Bonvin  \& Challinor}{Hall
  et~al.}{2013}]{Hall:2012wd}
Hall A.,  Bonvin C.,   Challinor A.,  2013, \mndoi [\prd]
  {10.1103/PhysRevD.87.064026}, D87, 064026

\bibitem[\protect\citeauthoryear{{Harper} \& {Dickinson}}{{Harper} \&
  {Dickinson}}{2018}]{Harper2018}
{Harper} S.,  {Dickinson} C.,  2018, submitted: \mnras, \href
  {http://adsabs.harvard.edu/abs/2018arXiv180306314H} {}

\bibitem[\protect\citeauthoryear{{Harper}, {Dickinson}, {Battye},
  {Roychowdhury}, {Browne}, {Ma}, {Olivari}  \& {Chen}}{{Harper}
  et~al.}{2018}]{2018MNRAS.478.2416H}
{Harper} S.~E.,  {Dickinson} C.,  {Battye} R.~A.,  {Roychowdhury} S.,  {Browne}
  I.~W.~A.,  {Ma} Y.~Z.,  {Olivari} L.~C.,   {Chen} T.,  2018, \mndoi [\mnras]
  {10.1093/mnras/sty1238}, \href
  {https://ui.adsabs.harvard.edu/#abs/2018MNRAS.478.2416H} {478, 2416}

\bibitem[\protect\citeauthoryear{{Harrison} \& {Brown}}{{Harrison} \&
  {Brown}}{2015}]{2015arXiv150706639H}
{Harrison} I.,  {Brown} M.~L.,  2015, preprint, \href
  {http://adsabs.harvard.edu/abs/2015arXiv150706639H} {} (\mn@eprint {arXiv}
  {1507.06639})

\bibitem[\protect\citeauthoryear{{Harrison}, {Camera}, {Zuntz}  \&
  {Brown}}{{Harrison} et~al.}{2016}]{2016MNRAS.463.3674H}
{Harrison} I.,  {Camera} S.,  {Zuntz} J.,   {Brown} M.~L.,  2016, \mndoi
  [\mnras] {10.1093/mnras/stw2082}, \href
  {http://adsabs.harvard.edu/abs/2016MNRAS.463.3674H} {463, 3674}

\bibitem[\protect\citeauthoryear{{Harrison}, {Lochner}  \& {Brown}}{{Harrison}
  et~al.}{2017}]{2017arXiv170408278H}
{Harrison} I.,  {Lochner} M.,   {Brown} M.~L.,  2017, preprint, \href
  {https://ui.adsabs.harvard.edu/#abs/2017arXiv170408278H} {p.
  arXiv:1704.08278} (\mn@eprint {arXiv} {1704.08278})

\bibitem[\protect\citeauthoryear{Hellwing, Barreira, Frenk, Li  \&
  Cole}{Hellwing et~al.}{2014}]{Hellwing:2014nma}
Hellwing W.~A.,  Barreira A.,  Frenk C.~S.,  Li B.,   Cole S.,  2014, \mndoi
  [\prl] {10.1103/PhysRevLett.112.221102}, 112, 221102

\bibitem[\protect\citeauthoryear{{Hillier}, {Brown}, {Harrison}  \&
  {Whittaker}}{{Hillier} et~al.}{2018}]{2018arXiv181001220H}
{Hillier} T.,  {Brown} M.~L.,  {Harrison} I.,   {Whittaker} L.,  2018,
  preprint, \href {http://adsabs.harvard.edu/abs/2018arXiv181001220H} {}
  (\mn@eprint {arXiv} {1810.01220})

\bibitem[\protect\citeauthoryear{{Hinshaw} et~al.,}{{Hinshaw}
  et~al.}{2013}]{2013ApJS..208...19H}
{Hinshaw} G.,  et~al., 2013, \mndoi [\apjs] {10.1088/0067-0049/208/2/19}, \href
  {http://adsabs.harvard.edu/abs/2013ApJS..208...19H} {208, 19}

\bibitem[\protect\citeauthoryear{{H{\"o}gbom}}{{H{\"o}gbom}}{1974}]{1974A&AS...15..417H}
{H{\"o}gbom} J.~A.,  1974, \aaps, \href
  {http://cdsads.u-strasbg.fr/abs/1974A%26AS...15..417H} {15, 417}

\bibitem[\protect\citeauthoryear{Ivarsen, Bull, Llinares  \& Mota}{Ivarsen
  et~al.}{2016}]{Ivarsen:2016xre}
Ivarsen M.~F.,  Bull P.,  Llinares C.,   Mota D.~F.,  2016, \mndoi [\aap]
  {10.1051/0004-6361/201628604}, 595, A40

\bibitem[\protect\citeauthoryear{{Jain} \& {Zhang}}{{Jain} \&
  {Zhang}}{2008}]{2008PhRvD..78f3503J}
{Jain} B.,  {Zhang} P.,  2008, \mndoi [\prd] {10.1103/PhysRevD.78.063503},
  \href {http://adsabs.harvard.edu/abs/2008PhRvD..78f3503J} {78, 063503}

\bibitem[\protect\citeauthoryear{{Janssen} et~al.,}{{Janssen}
  et~al.}{1996}]{1996astro.ph..2009J}
{Janssen} M.~A.,  et~al., 1996, preprint, \href
  {https://ui.adsabs.harvard.edu/#abs/1996astro.ph..2009J} {pp
  astro--ph/9602009} (\mn@eprint {arXiv} {astro-ph/9602009})

\bibitem[\protect\citeauthoryear{{Kaiser}}{{Kaiser}}{1987}]{1987MNRAS.227....1K}
{Kaiser} N.,  1987, \mndoi [\mnras] {10.1093/mnras/227.1.1}, \href
  {http://adsabs.harvard.edu/abs/1987MNRAS.227....1K} {227, 1}

\bibitem[\protect\citeauthoryear{{Kazin} et~al.,}{{Kazin}
  et~al.}{2014}]{2014MNRAS.441.3524K}
{Kazin} E.~A.,  et~al., 2014, \mndoi [\mnras] {10.1093/mnras/stu778}, \href
  {http://adsabs.harvard.edu/abs/2014MNRAS.441.3524K} {441, 3524}

\bibitem[\protect\citeauthoryear{{Koda} et~al.,}{{Koda}
  et~al.}{2014}]{2014MNRAS.445.4267K}
{Koda} J.,  et~al., 2014, \mndoi [\mnras] {10.1093/mnras/stu1610}, \href
  {http://adsabs.harvard.edu/abs/2014MNRAS.445.4267K} {445, 4267}

\bibitem[\protect\citeauthoryear{{Koopmans} et~al.,}{{Koopmans}
  et~al.}{2015}]{2015aska.confE...1K}
{Koopmans} L.,  et~al., 2015, Advancing Astrophysics with the Square Kilometre
  Array (AASKA14), \href {http://adsabs.harvard.edu/abs/2015aska.confE...1K}
  {p.~1}

\bibitem[\protect\citeauthoryear{{Kovetz} et~al.,}{{Kovetz}
  et~al.}{2017a}]{2017arXiv170909066K}
{Kovetz} E.~D.,  et~al., 2017a, preprint, \href
  {http://adsabs.harvard.edu/abs/2017arXiv170909066K} {} (\mn@eprint {arXiv}
  {1709.09066})

\bibitem[\protect\citeauthoryear{{Kovetz}, {Raccanelli}  \& {Rahman}}{{Kovetz}
  et~al.}{2017b}]{Kovetz2017}
{Kovetz} E.~D.,  {Raccanelli} A.,   {Rahman} M.,  2017b, \mndoi [\mnras]
  {10.1093/mnras/stx691}, \href
  {http://adsabs.harvard.edu/abs/2017MNRAS.468.3650K} {468, 3650}

\bibitem[\protect\citeauthoryear{{LSST Dark Energy Science
  Collaboration}}{{LSST Dark Energy Science
  Collaboration}}{2012}]{2012arXiv1211.0310L}
{LSST Dark Energy Science Collaboration} 2012, preprint, \href
  {http://adsabs.harvard.edu/abs/2012arXiv1211.0310L} {} (\mn@eprint {arXiv}
  {1211.0310})

\bibitem[\protect\citeauthoryear{{LSST Science Collaboration} \& et al.}{{LSST
  Science Collaboration} \& et~al.}{2009}]{2009arXiv0912.0201L}
{LSST Science Collaboration} et al. 2009, preprint, \href
  {http://adsabs.harvard.edu/abs/2009arXiv0912.0201L} {} (\mn@eprint {arXiv}
  {0912.0201})

\bibitem[\protect\citeauthoryear{Lahav, Kiakotou, Abdalla  \& Blake}{Lahav
  et~al.}{2010}]{Lahav:2009zr}
Lahav O.,  Kiakotou A.,  Abdalla F.~B.,   Blake C.,  2010, \mndoi [\mnras]
  {10.1111/j.1365-2966.2010.16472.x}, 405, 168

\bibitem[\protect\citeauthoryear{{Laureijs} et~al.,}{{Laureijs}
  et~al.}{2011a}]{2011arXiv1110.3193L}
{Laureijs} R.,  et~al., 2011a, preprint, \href
  {http://ukads.nottingham.ac.uk/abs/2011arXiv1110.3193L} {} (\mn@eprint
  {arXiv} {1110.3193})

\bibitem[\protect\citeauthoryear{Laureijs et~al.}{Laureijs
  et~al.}{2011b}]{Laureijs:2011gra}
Laureijs R.,  et~al., 2011b, preprint (\mn@eprint {arXiv} {1110.3193})

\bibitem[\protect\citeauthoryear{Lesgourgues}{Lesgourgues}{2011}]{Lesgourgues:2011re}
Lesgourgues J.,  2011, preprint (\mn@eprint {arXiv} {1104.2932})

\bibitem[\protect\citeauthoryear{{Loeb} \& {Wyithe}}{{Loeb} \&
  {Wyithe}}{2008}]{2008PhRvL.100p1301L}
{Loeb} A.,  {Wyithe} J.~S.~B.,  2008, \mndoi [\prl]
  {10.1103/PhysRevLett.100.161301}, \href
  {http://adsabs.harvard.edu/abs/2008PhRvL.100p1301L} {100, 161301}

\bibitem[\protect\citeauthoryear{Lombriser, Yoo  \& Koyama}{Lombriser
  et~al.}{2013}]{Lombriser:2013aj}
Lombriser L.,  Yoo J.,   Koyama K.,  2013, \mndoi [\prd]
  {10.1103/PhysRevD.87.104019}, D87, 104019

\bibitem[\protect\citeauthoryear{{Maartens}, {Abdalla}, {Jarvis}  \&
  {Santos}}{{Maartens} et~al.}{2015}]{2015aska.confE..16M}
{Maartens} R.,  {Abdalla} F.~B.,  {Jarvis} M.,   {Santos} M.~G.,  2015,
  Advancing Astrophysics with the Square Kilometre Array (AASKA14), \href
  {http://adsabs.harvard.edu/abs/2015aska.confE..16M} {p.~16}

\bibitem[\protect\citeauthoryear{Macaulay, Wehus  \& Eriksen}{Macaulay
  et~al.}{2013}]{Macaulay:2013swa}
Macaulay E.,  Wehus I.~K.,   Eriksen H.~K.,  2013, \mndoi [\prl]
  {10.1103/PhysRevLett.111.161301}, 111, 161301

\bibitem[\protect\citeauthoryear{{Maino}, {Burigana}, {G{\'o}rski}, {Mandolesi}
   \& {Bersanelli}}{{Maino} et~al.}{2002}]{2002A&A...387..356M}
{Maino} D.,  {Burigana} C.,  {G{\'o}rski} K.~M.,  {Mandolesi} N.,
  {Bersanelli} M.,  2002, \mndoi [\aap] {10.1051/0004-6361:20020242}, \href
  {http://adsabs.harvard.edu/abs/2002A%26A...387..356M} {387, 356}

\bibitem[\protect\citeauthoryear{Majerotto et~al.}{Majerotto
  et~al.}{2012}]{Majerotto:2012mf}
Majerotto E.,  et~al., 2012, \mndoi [\mnras]
  {10.1111/j.1365-2966.2012.21323.x}, 424, 1392

\bibitem[\protect\citeauthoryear{Markovic et~al.,}{Markovic
  et~al.}{2017}]{Markovic:2016qzf}
Markovic K.,  et~al., 2017, \mndoi [\mnras] {10.1093/mnras/stx283}, 467, 3677

\bibitem[\protect\citeauthoryear{Massara, Villaescusa-Navarro, Viel  \&
  Sutter}{Massara et~al.}{2015}]{Massara:2015msa}
Massara E.,  Villaescusa-Navarro F.,  Viel M.,   Sutter P.~M.,  2015, \mndoi
  [\jcap] {10.1088/1475-7516/2015/11/018}, 1511, 018

\bibitem[\protect\citeauthoryear{{Masui} et~al.}{{Masui}
  et~al.}{2013a}]{2013ApJ...763L..20M}
{Masui} K.~W.,  et~al., 2013a, \mndoi [\apjl] {10.1088/2041-8205/763/1/L20},
  \href {http://adsabs.harvard.edu/abs/2013ApJ...763L..20M} {763, L20}

\bibitem[\protect\citeauthoryear{Masui et~al.}{Masui
  et~al.}{2013b}]{Masui:2012zc}
Masui K.~W.,  et~al., 2013b, \mndoi [\apj] {10.1088/2041-8205/763/1/L20}, 763,
  L20

\bibitem[\protect\citeauthoryear{Matarrese \& Verde}{Matarrese \&
  Verde}{2008}]{Matarrese:2008nc}
Matarrese S.,  Verde L.,  2008, \mndoi [\apj] {10.1086/587840}, 677, L77

\bibitem[\protect\citeauthoryear{{Miranda}, {Hu}  \& {Adshead}}{{Miranda}
  et~al.}{2012}]{2012PhRvD..86f3529M}
{Miranda} V.,  {Hu} W.,   {Adshead} P.,  2012, \mndoi [\prd]
  {10.1103/PhysRevD.86.063529}, \href
  {https://ui.adsabs.harvard.edu/#abs/2012PhRvD..86f3529M} {86, 063529}

\bibitem[\protect\citeauthoryear{Morales}{Morales}{2006}]{Morales:2006fq}
Morales M.~F.,  2006, \mndoi [\apj] {10.1086/508614}, 650, L21

\bibitem[\protect\citeauthoryear{{Newburgh} et~al.,}{{Newburgh}
  et~al.}{2016}]{2016SPIE.9906E..5XN}
{Newburgh} L.~B.,  et~al., 2016, in Ground-based and Airborne Telescopes VI. p.
  99065X (\mn@eprint {arXiv} {1607.02059}), \mndoi{10.1117/12.2234286}

\bibitem[\protect\citeauthoryear{{Ng}, {Wong}, {Broadhurst}  \& {Li}}{{Ng}
  et~al.}{2018}]{2018PhRvD..97b3012N}
{Ng} K.~K.~Y.,  {Wong} K.~W.~K.,  {Broadhurst} T.,   {Li} T.~G.~F.,  2018,
  \mndoi [\prd] {10.1103/PhysRevD.97.023012}, \href
  {http://adsabs.harvard.edu/abs/2018PhRvD..97b3012N} {97, 023012}

\bibitem[\protect\citeauthoryear{Obreschkow \& Rawlings}{Obreschkow \&
  Rawlings}{2009}]{Obreschkow:2009ha}
Obreschkow D.,  Rawlings S.,  2009, \mndoi [\apj]
  {10.1088/0004-637X/703/2/1890}, 703, 1890

\bibitem[\protect\citeauthoryear{{Olivari}, {Remazeilles}  \&
  {Dickinson}}{{Olivari} et~al.}{2016}]{2016MNRAS.456.2749O}
{Olivari} L.~C.,  {Remazeilles} M.,   {Dickinson} C.,  2016, \mndoi [\mnras]
  {10.1093/mnras/stv2884}, \href
  {https://ui.adsabs.harvard.edu/#abs/2016MNRAS.456.2749O} {456, 2749}

\bibitem[\protect\citeauthoryear{Olivari, Dickinson, Battye, Ma, Costa,
  Remazeilles  \& Harper}{Olivari et~al.}{2018}]{Olivari:2017bfv}
Olivari L.~C.,  Dickinson C.,  Battye R.~A.,  Ma Y.-Z.,  Costa A.~A.,
  Remazeilles M.,   Harper S.,  2018, \mndoi [\mnras] {10.1093/mnras/stx2621},
  473, 4242

\bibitem[\protect\citeauthoryear{{Padmanabhan} \& {Refregier}}{{Padmanabhan} \&
  {Refregier}}{2017a}]{hpar2017}
{Padmanabhan} H.,  {Refregier} A.,  2017a, \mndoi [\mnras]
  {10.1093/mnras/stw2706}, \href
  {http://adsabs.harvard.edu/abs/2017MNRAS.464.4008P} {464, 4008}

\bibitem[\protect\citeauthoryear{{Padmanabhan} \& {Refregier}}{{Padmanabhan} \&
  {Refregier}}{2017b}]{Hamsa_2017}
{Padmanabhan} H.,  {Refregier} A.,  2017b, \mndoi [\mnras]
  {10.1093/mnras/stw2706}, \href
  {http://adsabs.harvard.edu/abs/2017MNRAS.464.4008P} {464, 4008}

\bibitem[\protect\citeauthoryear{{Padmanabhan} et~al.,}{{Padmanabhan}
  et~al.}{2007}]{Padmanabhan2007}
{Padmanabhan} N.,  et~al., 2007, \mndoi [\mnras]
  {10.1111/j.1365-2966.2007.11593.x}, \href
  {http://adsabs.harvard.edu/abs/2007MNRAS.378..852P} {378, 852}

\bibitem[\protect\citeauthoryear{{Padmanabhan}, {Choudhury}  \&
  {Refregier}}{{Padmanabhan} et~al.}{2015}]{hptrcar2015}
{Padmanabhan} H.,  {Choudhury} T.~R.,   {Refregier} A.,  2015, \mndoi [\mnras]
  {10.1093/mnras/stu2702}, \href
  {http://adsabs.harvard.edu/abs/2015MNRAS.447.3745P} {447, 3745}

\bibitem[\protect\citeauthoryear{{Padmanabhan}, {Refregier}  \&
  {Amara}}{{Padmanabhan} et~al.}{2017}]{hparaa2017}
{Padmanabhan} H.,  {Refregier} A.,   {Amara} A.,  2017, \mndoi [\mnras]
  {10.1093/mnras/stx979}, \href
  {http://adsabs.harvard.edu/abs/2017MNRAS.469.2323P} {469, 2323}

\bibitem[\protect\citeauthoryear{{Padmanabhan}, {Refregier}  \&
  {Amara}}{{Padmanabhan} et~al.}{2018}]{2018arXiv180410627P}
{Padmanabhan} H.,  {Refregier} A.,   {Amara} A.,  2018, preprint, \href
  {http://adsabs.harvard.edu/abs/2018arXiv180410627P} {} (\mn@eprint {arXiv}
  {1804.10627})

\bibitem[\protect\citeauthoryear{{Patel}, {Bacon}, {Beswick}, {Muxlow}  \&
  {Hoyle}}{{Patel} et~al.}{2010}]{2010MNRAS.401.2572P}
{Patel} P.,  {Bacon} D.~J.,  {Beswick} R.~J.,  {Muxlow} T.~W.~B.,   {Hoyle} B.,
   2010, \mndoi [\mnras] {10.1111/j.1365-2966.2009.15836.x}, \href
  {http://adsabs.harvard.edu/abs/2010MNRAS.401.2572P} {401, 2572}

\bibitem[\protect\citeauthoryear{{Patel} et~al.,}{{Patel}
  et~al.}{2015}]{2015aska.confE..30P}
{Patel} P.,  et~al., 2015, Advancing Astrophysics with the Square Kilometre
  Array (AASKA14), \href {http://adsabs.harvard.edu/abs/2015aska.confE..30P}
  {p.~30}

\bibitem[\protect\citeauthoryear{Peiris et~al.}{Peiris
  et~al.}{2003}]{Peiris:2003ff}
Peiris H.~V.,  et~al., 2003, \mndoi [\apjs] {10.1086/377228}, 148, 213

\bibitem[\protect\citeauthoryear{{Percival}, {Samushia}, {Ross}, {Shapiro}  \&
  {Raccanelli}}{{Percival} et~al.}{2011}]{2011RSPTA.369.5058P}
{Percival} W.~J.,  {Samushia} L.,  {Ross} A.~J.,  {Shapiro} C.,   {Raccanelli}
  A.,  2011, \mndoi [\PhilTrans] {10.1098/rsta.2011.0370}, \href
  {http://adsabs.harvard.edu/abs/2011RSPTA.369.5058P} {369, 5058}

\bibitem[\protect\citeauthoryear{{Peterson}, {Bandura}  \& {Pen}}{{Peterson}
  et~al.}{2006}]{2006astro.ph..6104P}
{Peterson} J.~B.,  {Bandura} K.,   {Pen} U.~L.,  2006, preprint, \href
  {http://adsabs.harvard.edu/abs/2006astro.ph..6104P} {} (\mn@eprint {}
  {astro-ph/0606104})

\bibitem[\protect\citeauthoryear{Pisani, Sutter, Hamaus, Alizadeh, Biswas,
  Wandelt  \& Hirata}{Pisani et~al.}{2015}]{Pisani:2015jha}
Pisani A.,  Sutter P.~M.,  Hamaus N.,  Alizadeh E.,  Biswas R.,  Wandelt B.~D.,
    Hirata C.~M.,  2015, \mndoi [\prd] {10.1103/PhysRevD.92.083531}, D92,
  083531

\bibitem[\protect\citeauthoryear{{Planck Collaboration} et~al.,}{{Planck
  Collaboration} et~al.}{2016a}]{2016A&A...594A..13P}
{Planck Collaboration} et~al., 2016a, \mndoi [\aap]
  {10.1051/0004-6361/201525830}, \href
  {http://adsabs.harvard.edu/abs/2016A%26A...594A..13P} {594, A13}

\bibitem[\protect\citeauthoryear{{Planck Collaboration} et~al.,}{{Planck
  Collaboration} et~al.}{2016b}]{PlanckNG2016}
{Planck Collaboration} et~al., 2016b, \mndoi [\aap]
  {10.1051/0004-6361/201525836}, \href
  {http://adsabs.harvard.edu/abs/2016A%26A...594A..17P} {594, A17}

\bibitem[\protect\citeauthoryear{{Planck Collaboration} et~al.,}{{Planck
  Collaboration} et~al.}{2016c}]{2016A&A...594A..24P}
{Planck Collaboration} et~al., 2016c, \mndoi [\aap]
  {10.1051/0004-6361/201525833}, \href
  {http://ukads.nottingham.ac.uk/abs/2016A%26A...594A..24P} {594, A24}

\bibitem[\protect\citeauthoryear{{Planck Collaboration} et~al.,}{{Planck
  Collaboration} et~al.}{2018}]{2018arXiv180706209P}
{Planck Collaboration} et~al., 2018, preprint, \href
  {http://adsabs.harvard.edu/abs/2018arXiv180706209P} {} (\mn@eprint {arXiv}
  {1807.06209})

\bibitem[\protect\citeauthoryear{Pourtsidou}{Pourtsidou}{2016a}]{Pourtsidou:2016ctq}
Pourtsidou A.,  2016a, preprint (\mn@eprint {arXiv} {1612.05138})

\bibitem[\protect\citeauthoryear{Pourtsidou}{Pourtsidou}{2016b}]{Pourtsidou:2015ksn}
Pourtsidou A.,  2016b, \mndoi [\mnras] {10.1093/mnras/stw1406}, 461, 1457

\bibitem[\protect\citeauthoryear{Pourtsidou, Bacon, Crittenden  \&
  Metcalf}{Pourtsidou et~al.}{2016}]{Pourtsidou:2015mia}
Pourtsidou A.,  Bacon D.,  Crittenden R.,   Metcalf R.~B.,  2016, \mndoi
  [\mnras] {10.1093/mnras/stw658}, 459, 863

\bibitem[\protect\citeauthoryear{Pourtsidou, Bacon  \& Crittenden}{Pourtsidou
  et~al.}{2017}]{Pourtsidou:2016dzn}
Pourtsidou A.,  Bacon D.,   Crittenden R.,  2017, \mndoi [\mnras]
  {10.1093/mnras/stx1479}, 470, 4251

\bibitem[\protect\citeauthoryear{{Raccanelli}}{{Raccanelli}}{2017}]{Raccanelli:2017}
{Raccanelli} A.,  2017, \mnras, 469

\bibitem[\protect\citeauthoryear{Raccanelli et~al.,}{Raccanelli
  et~al.}{2011}]{Raccanelli2011}
Raccanelli A.,  et~al., 2011, \mnras, 424, 19

\bibitem[\protect\citeauthoryear{{Raccanelli} et~al.,}{{Raccanelli}
  et~al.}{2015}]{Raccanelli2014}
{Raccanelli} A.,  et~al., 2015, \mndoi [\jcap] {10.1088/1475-7516/2015/01/042},
  \href {http://adsabs.harvard.edu/abs/2015JCAP...01..042R} {1, 042}

\bibitem[\protect\citeauthoryear{{Raccanelli}, {Kovetz}, {Bird}, {Cholis}  \&
  {Mu{\~n}oz}}{{Raccanelli} et~al.}{2016}]{Raccanelli2016}
{Raccanelli} A.,  {Kovetz} E.~D.,  {Bird} S.,  {Cholis} I.,   {Mu{\~n}oz}
  J.~B.,  2016, \mndoi [\prd] {10.1103/PhysRevD.94.023516}, \href
  {https://ui.adsabs.harvard.edu/#abs/2016PhRvD..94b3516R} {94, 023516}

\bibitem[\protect\citeauthoryear{{Raccanelli}, {Shiraishi}, {Bartolo},
  {Bertacca}, {Liguori}, {Matarrese}, {Norris}  \& {Parkinson}}{{Raccanelli}
  et~al.}{2017}]{Raccanelli2017}
{Raccanelli} A.,  {Shiraishi} M.,  {Bartolo} N.,  {Bertacca} D.,  {Liguori} M.,
   {Matarrese} S.,  {Norris} R.~P.,   {Parkinson} D.,  2017, \mndoi [Physics of
  the Dark Universe] {10.1016/j.dark.2016.10.006}, \href
  {http://adsabs.harvard.edu/abs/2017PDU....15...35R} {15, 35}

\bibitem[\protect\citeauthoryear{{Raveri}, {Martinelli}, {Zhao}  \&
  {Wang}}{{Raveri} et~al.}{2016a}]{2016arXiv160606268R}
{Raveri} M.,  {Martinelli} M.,  {Zhao} G.,   {Wang} Y.,  2016a, preprint, \href
  {http://adsabs.harvard.edu/abs/2016arXiv160606268R} {} (\mn@eprint {arXiv}
  {1606.06268})

\bibitem[\protect\citeauthoryear{{Raveri}, {Martinelli}, {Zhao}  \&
  {Wang}}{{Raveri} et~al.}{2016b}]{2016arXiv160606273R}
{Raveri} M.,  {Martinelli} M.,  {Zhao} G.,   {Wang} Y.,  2016b, preprint, \href
  {http://adsabs.harvard.edu/abs/2016arXiv160606273R} {} (\mn@eprint {arXiv}
  {1606.06273})

\bibitem[\protect\citeauthoryear{{Reich} \& {Reich}}{{Reich} \&
  {Reich}}{1988}]{1988A&AS...74....7R}
{Reich} P.,  {Reich} W.,  1988, \aaps, \href
  {https://ui.adsabs.harvard.edu/#abs/1988A&AS...74....7R} {74, 7}

\bibitem[\protect\citeauthoryear{{Remazeilles}, {Dickinson}, {Banday}, {Bigot-
  Sazy}  \& {Ghosh}}{{Remazeilles} et~al.}{2015}]{2015MNRAS.451.4311R}
{Remazeilles} M.,  {Dickinson} C.,  {Banday} A.~J.,  {Bigot- Sazy} M.~A.,
  {Ghosh} T.,  2015, \mndoi [\mnras] {10.1093/mnras/stv1274}, \href
  {https://ui.adsabs.harvard.edu/#abs/2015MNRAS.451.4311R} {451, 4311}

\bibitem[\protect\citeauthoryear{{Rivi} \& {Miller}}{{Rivi} \&
  {Miller}}{2018}]{2018MNRAS.tmp..362R}
{Rivi} M.,  {Miller} L.,  2018, \mndoi [\mnras] {10.1093/mnras/sty371}, \href
  {https://ui.adsabs.harvard.edu/#abs/2018MNRAS.476.2053R} {476, 2053}

\bibitem[\protect\citeauthoryear{{Rivi}, {Lochner}, {Balan}, {Harrison}  \&
  {Abdalla}}{{Rivi} et~al.}{2018}]{2018arXiv180506799R}
{Rivi} M.,  {Lochner} M.,  {Balan} S.~T.,  {Harrison} I.,   {Abdalla} F.~B.,
  2018, \mndoi [\mnras] {10.1093/mnras/sty2700}, \href
  {http://adsabs.harvard.edu/abs/2018MNRAS.tmp.2581R} {}

\bibitem[\protect\citeauthoryear{{Rubart} \& {Schwarz}}{{Rubart} \&
  {Schwarz}}{2013}]{2013A&A...555A.117R}
{Rubart} M.,  {Schwarz} D.~J.,  2013, \mndoi [\aap]
  {10.1051/0004-6361/201321215}, \href
  {http://adsabs.harvard.edu/abs/2013A%26A...555A.117R} {555, A117}

\bibitem[\protect\citeauthoryear{{Sachs} \& {Wolfe}}{{Sachs} \&
  {Wolfe}}{1967}]{ISW_paper}
{Sachs} R.~K.,  {Wolfe} A.~M.,  1967, \mndoi [\apj] {10.1086/148982}, \href
  {http://adsabs.harvard.edu/abs/1967ApJ...147...73S} {147, 73}

\bibitem[\protect\citeauthoryear{{Sahl{\'e}n}}{{Sahl{\'e}n}}{2018}]{2018arXiv180702470S}
{Sahl{\'e}n} M.,  2018, preprint, \href
  {http://adsabs.harvard.edu/abs/2018arXiv180702470S} {} (\mn@eprint {arXiv}
  {1807.02470})

\bibitem[\protect\citeauthoryear{Sahl\'{e}n \& Silk}{Sahl\'{e}n \&
  Silk}{2016}]{Sahlen:2016kzx}
Sahl\'{e}n M.,  Silk J.,  2016, preprint (\mn@eprint {arXiv} {1612.06595})

\bibitem[\protect\citeauthoryear{Sahl\'{e}n, Zubeldia  \& Silk}{Sahl\'{e}n
  et~al.}{2016}]{Sahlen:2015wpc}
Sahl\'{e}n M.,  Zubeldia I.,   Silk J.,  2016, \mndoi [\apj]
  {10.3847/2041-8205/820/1/L7}, 820, L7

\bibitem[\protect\citeauthoryear{{Santos}, {Bull}, {Alonso}  \& et
  al.}{{Santos} et~al.}{2015}]{2015aska.confE..19S}
{Santos} M.,  {Bull} P.,  {Alonso} D.,   et al. 2015, Advancing Astrophysics
  with the Square Kilometre Array (AASKA14), \href
  {http://adsabs.harvard.edu/abs/2015aska.confE..19S} {p.~19}

\bibitem[\protect\citeauthoryear{{Santos} et~al.,}{{Santos}
  et~al.}{2017}]{2017arXiv170906099S}
{Santos} M.~G.,  et~al., 2017, preprint, \href
  {http://adsabs.harvard.edu/abs/2017arXiv170906099S} {} (\mn@eprint {arXiv}
  {1709.06099})

\bibitem[\protect\citeauthoryear{{Scelfo}, {Bellomo}, {Raccanelli}, {Verde}  \&
  {Matarrese}}{{Scelfo} et~al.}{2018}]{Scelfo:2018}
{Scelfo} G.,  {Bellomo} N.,  {Raccanelli} A.,  {Verde} L.,   {Matarrese} S.,
  2018, \jcap, 09

\bibitem[\protect\citeauthoryear{{Schwarz} et~al.,}{{Schwarz}
  et~al.}{2015}]{2015aska.confE..32S}
{Schwarz} D.~J.,  et~al., 2015, Advancing Astrophysics with the Square
  Kilometre Array (AASKA14), \href
  {http://adsabs.harvard.edu/abs/2015aska.confE..32S} {p.~32}

\bibitem[\protect\citeauthoryear{{Scoccimarro}}{{Scoccimarro}}{2004}]{2004PhRvD..70h3007S}
{Scoccimarro} R.,  2004, \mndoi [\prd] {10.1103/PhysRevD.70.083007}, \href
  {http://adsabs.harvard.edu/abs/2004PhRvD..70h3007S} {70, 083007}

\bibitem[\protect\citeauthoryear{Seljak}{Seljak}{2009}]{Seljak:2008xr}
Seljak U.,  2009, \mndoi [\prl] {10.1103/PhysRevLett.102.021302}, 102, 021302

\bibitem[\protect\citeauthoryear{{Singal}}{{Singal}}{2011}]{2011ApJ...742L..23S}
{Singal} A.~K.,  2011, \mndoi [\apjl] {10.1088/2041-8205/742/2/L23}, \href
  {http://adsabs.harvard.edu/abs/2011ApJ...742L..23S} {742, L23}

\bibitem[\protect\citeauthoryear{Smith \& Markovic}{Smith \&
  Markovic}{2011}]{Smith:2011ev}
Smith R.~E.,  Markovic K.,  2011, \mndoi [\prd] {10.1103/PhysRevD.84.063507},
  D84, 063507

\bibitem[\protect\citeauthoryear{Spolyar, Sahl\'{e}n  \& Silk}{Spolyar
  et~al.}{2013}]{Spolyar:2013maa}
Spolyar D.,  Sahl\'{e}n M.,   Silk J.,  2013, \mndoi [\prl]
  {10.1103/PhysRevLett.111.241103}, 111, 241103

\bibitem[\protect\citeauthoryear{{Starobinskii}}{{Starobinskii}}{1992}]{1992ZhPmR..55..477S}
{Starobinskii} A.~A.,  1992, ZhETF Pisma Redaktsiiu, \href
  {https://ui.adsabs.harvard.edu/#abs/1992ZhPmR..55..477S} {55, 477}

\bibitem[\protect\citeauthoryear{Sutter, Pisani, Wandelt  \& Weinberg}{Sutter
  et~al.}{2014}]{Sutter:2014oca}
Sutter P.~M.,  Pisani A.,  Wandelt B.~D.,   Weinberg D.~H.,  2014, \mndoi
  [\mnras] {10.1093/mnras/stu1392}, 443, 2983

\bibitem[\protect\citeauthoryear{{Switzer} et~al.,}{{Switzer}
  et~al.}{2013}]{2013MNRAS.434L..46S}
{Switzer} E.~R.,  et~al., 2013, \mndoi [\mnras] {10.1093/mnrasl/slt074}, \href
  {http://adsabs.harvard.edu/abs/2013MNRAS.434L..46S} {434, L46}

\bibitem[\protect\citeauthoryear{{Switzer}, {Chang}, {Masui}, {Pen}  \&
  {Voytek}}{{Switzer} et~al.}{2015}]{2015ApJ...815...51S}
{Switzer} E.~R.,  {Chang} T.~C.,  {Masui} K.~W.,  {Pen} U.~L.,   {Voytek}
  T.~C.,  2015, \mndoi [\apj] {10.1088/0004-637X/815/1/51}, \href
  {https://ui.adsabs.harvard.edu/#abs/2015ApJ...815...51S} {815}

\bibitem[\protect\citeauthoryear{Thomas, Whittaker, Camera  \& Brown}{Thomas
  et~al.}{2017}]{Thomas:2016xhb}
Thomas D.~B.,  Whittaker L.,  Camera S.,   Brown M.~L.,  2017, \mndoi [\mnras]
  {10.1093/mnras/stx1468}, 470, 3131

\bibitem[\protect\citeauthoryear{{Tiwari} \& {Nusser}}{{Tiwari} \&
  {Nusser}}{2016}]{2016JCAP...03..062T}
{Tiwari} P.,  {Nusser} A.,  2016, \mndoi [\jcap]
  {10.1088/1475-7516/2016/03/062}, \href
  {http://adsabs.harvard.edu/abs/2016JCAP...03..062T} {3, 062}

\bibitem[\protect\citeauthoryear{{Tr{\"o}ster} et~al.,}{{Tr{\"o}ster}
  et~al.}{2017}]{Troster:2016sgf}
{Tr{\"o}ster} T.,  et~al., 2017, \mndoi [\mnras] {10.1093/mnras/stx365}, \href
  {http://adsabs.harvard.edu/abs/2017MNRAS.467.2706T} {467, 2706}

\bibitem[\protect\citeauthoryear{{Tully} \& {Fisher}}{{Tully} \&
  {Fisher}}{1977}]{1977A&A....54..661T}
{Tully} R.~B.,  {Fisher} J.~R.,  1977, \aap, \href
  {http://adsabs.harvard.edu/abs/1977A%26A....54..661T} {54, 661}

\bibitem[\protect\citeauthoryear{{Tunbridge}, {Harrison}  \&
  {Brown}}{{Tunbridge} et~al.}{2016}]{2016MNRAS.463.3339T}
{Tunbridge} B.,  {Harrison} I.,   {Brown} M.~L.,  2016, \mndoi [\mnras]
  {10.1093/mnras/stw2224}, \href
  {http://adsabs.harvard.edu/abs/2016MNRAS.463.3339T} {463, 3339}

\bibitem[\protect\citeauthoryear{{Viel}, {Markovi{\v c}}, {Baldi}  \&
  {Weller}}{{Viel} et~al.}{2012}]{vielbaldi}
{Viel} M.,  {Markovi{\v c}} K.,  {Baldi} M.,   {Weller} J.,  2012, \mndoi
  [\mnras] {10.1111/j.1365-2966.2011.19910.x}, \href
  {http://adsabs.harvard.edu/abs/2012MNRAS.421...50V} {421, 50}

\bibitem[\protect\citeauthoryear{Villaescusa-Navarro, Viel, Datta  \&
  Choudhury}{Villaescusa-Navarro et~al.}{2014}]{Villaescusa-Navarro:2014cma}
Villaescusa-Navarro F.,  Viel M.,  Datta K.~K.,   Choudhury T.~R.,  2014,
  \mndoi [\jcap] {10.1088/1475-7516/2014/09/050}, 1409, 050

\bibitem[\protect\citeauthoryear{{Villaescusa-Navarro}, {Bull}  \&
  {Viel}}{{Villaescusa-Navarro} et~al.}{2015}]{FVN_2015}
{Villaescusa-Navarro} F.,  {Bull} P.,   {Viel} M.,  2015, \mndoi [\apj]
  {10.1088/0004-637X/814/2/146}, \href
  {http://adsabs.harvard.edu/abs/2015ApJ...814..146V} {814, 146}

\bibitem[\protect\citeauthoryear{{Villaescusa-Navarro}, {Alonso}  \&
  {Viel}}{{Villaescusa-Navarro} et~al.}{2017}]{FVN_2016}
{Villaescusa-Navarro} F.,  {Alonso} D.,   {Viel} M.,  2017, \mndoi [\mnras]
  {10.1093/mnras/stw3224}, \href
  {http://adsabs.harvard.edu/abs/2017MNRAS.466.2736V} {466, 2736}

\bibitem[\protect\citeauthoryear{{Villaescusa-Navarro}
  et~al.,}{{Villaescusa-Navarro} et~al.}{2018a}]{2018arXiv180409180V}
{Villaescusa-Navarro} F.,  et~al., 2018a, preprint, \href
  {http://adsabs.harvard.edu/abs/2018arXiv180409180V} {} (\mn@eprint {arXiv}
  {1804.09180})

\bibitem[\protect\citeauthoryear{{Villaescusa-Navarro}
  et~al.,}{{Villaescusa-Navarro} et~al.}{2018b}]{FVN_2018}
{Villaescusa-Navarro} F.,  et~al., 2018b, \mndoi [\apj]
  {10.3847/1538-4357/aadba0}, \href
  {https://ui.adsabs.harvard.edu/#abs/2018ApJ...866..135V} {866, 135}

\bibitem[\protect\citeauthoryear{Voivodic, Lima, Llinares  \& Mota}{Voivodic
  et~al.}{2017}]{Voivodic:2016kog}
Voivodic R.,  Lima M.,  Llinares C.,   Mota D.~F.,  2017, \mndoi [\prd]
  {10.1103/PhysRevD.95.024018}, D95, 024018

\bibitem[\protect\citeauthoryear{{Wang}, {Koribalski}, {Serra}, {van der
  Hulst}, {Roychowdhury}, {Kamphuis}  \& {Chengalur}}{{Wang}
  et~al.}{2016}]{2016MNRAS.460.2143W}
{Wang} J.,  {Koribalski} B.~S.,  {Serra} P.,  {van der Hulst} T.,
  {Roychowdhury} S.,  {Kamphuis} P.,   {Chengalur} J.~N.,  2016, \mndoi
  [\mnras] {10.1093/mnras/stw1099}, \href
  {http://adsabs.harvard.edu/abs/2016MNRAS.460.2143W} {460, 2143}

\bibitem[\protect\citeauthoryear{{Wilkinson}}{{Wilkinson}}{1991}]{1991ASPC...19..428W}
{Wilkinson} P.~N.,  1991, in {Cornwell} T.~J.,  {Perley} R.~A.,  eds,
  Astronomical Society of the Pacific Conference Series Vol. 19, IAU Colloq.
  131: Radio Interferometry. Theory, Techniques, and Applications. pp 428--432

\bibitem[\protect\citeauthoryear{{Wilman} et~al.,}{{Wilman}
  et~al.}{2008}]{Wilman2008}
{Wilman} R.~J.,  et~al., 2008, \mndoi [\mnras]
  {10.1111/j.1365-2966.2008.13486.x}, \href
  {http://adsabs.harvard.edu/abs/2008MNRAS.388.1335W} {388, 1335}

\bibitem[\protect\citeauthoryear{{Witzemann}, {Bull}, {Clarkson}, {Santos},
  {Spinelli}  \& {Weltman}}{{Witzemann} et~al.}{2018}]{2018MNRAS.477L.122W}
{Witzemann} A.,  {Bull} P.,  {Clarkson} C.,  {Santos} M.~G.,  {Spinelli} M.,
  {Weltman} A.,  2018, \mndoi [\mnras] {10.1093/mnrasl/sly062}, \href
  {http://adsabs.harvard.edu/abs/2018MNRAS.477L.122W} {477, L122}

\bibitem[\protect\citeauthoryear{{Wolz}, {Abdalla}, {Blake}, {Shaw}, {Chapman}
  \& {Rawlings}}{{Wolz} et~al.}{2014}]{2014MNRAS.441.3271W}
{Wolz} L.,  {Abdalla} F.~B.,  {Blake} C.,  {Shaw} J.~R.,  {Chapman} E.,
  {Rawlings} S.,  2014, \mndoi [\mnras] {10.1093/mnras/stu792}, \href
  {https://ui.adsabs.harvard.edu/#abs/2014MNRAS.441.3271W} {441, 3271}

\bibitem[\protect\citeauthoryear{{Wolz} et~al.,}{{Wolz}
  et~al.}{2015}]{2015aska.confE..35W}
{Wolz} L.,  et~al., 2015, in Advancing Astrophysics with the Square Kilometre
  Array (AASKA14). p.~35

\bibitem[\protect\citeauthoryear{{Wolz} et~al.,}{{Wolz}
  et~al.}{2017}]{2015arXiv151005453W}
{Wolz} L.,  et~al., 2017, \mndoi [\mnras] {10.1093/mnras/stw2556}, \href
  {https://ui.adsabs.harvard.edu/#abs/2017MNRAS.464.4938W} {464, 4938}

\bibitem[\protect\citeauthoryear{{Wolz}, {Murray}, {Blake}  \& {Wyithe}}{{Wolz}
  et~al.}{2018}]{2018arXiv180302477W}
{Wolz} L.,  {Murray} S.~G.,  {Blake} C.,   {Wyithe} J.~S.,  2018, preprint,
  \href {https://ui.adsabs.harvard.edu/#abs/2018arXiv180302477W} {} (\mn@eprint
  {arXiv} {1803.02477})

\bibitem[\protect\citeauthoryear{{Xavier}, {Abdalla}  \& {Joachimi}}{{Xavier}
  et~al.}{2016}]{2016MNRAS.459.3693X}
{Xavier} H.~S.,  {Abdalla} F.~B.,   {Joachimi} B.,  2016, \mndoi [\mnras]
  {10.1093/mnras/stw874}, \href
  {http://adsabs.harvard.edu/abs/2016MNRAS.459.3693X} {459, 3693}

\bibitem[\protect\citeauthoryear{Xia, Cuoco, Branchini  \& Viel}{Xia
  et~al.}{2015}]{Xia:2015wka}
Xia J.-Q.,  Cuoco A.,  Branchini E.,   Viel M.,  2015, \mndoi [\apjs]
  {10.1088/0067-0049/217/1/15}, 217, 15

\bibitem[\protect\citeauthoryear{{Xu}, {Wang}  \& {Chen}}{{Xu}
  et~al.}{2015}]{2015ApJ...798...40X}
{Xu} Y.,  {Wang} X.,   {Chen} X.,  2015, \mndoi [\apj]
  {10.1088/0004-637X/798/1/40}, \href
  {http://adsabs.harvard.edu/abs/2015ApJ...798...40X} {798, 40}

\bibitem[\protect\citeauthoryear{{Xu}, {Hamann}  \& {Chen}}{{Xu}
  et~al.}{2016}]{Xu2016}
{Xu} Y.,  {Hamann} J.,   {Chen} X.,  2016, \mndoi [\prd]
  {10.1103/PhysRevD.94.123518}, \href
  {http://adsabs.harvard.edu/abs/2016PhRvD..94l3518X} {94, 123518}

\bibitem[\protect\citeauthoryear{{Yahya}, {Bull}, {Santos}, {Silva},
  {Maartens}, {Okouma}  \& {Bassett}}{{Yahya}
  et~al.}{2015}]{2015MNRAS.450.2251Y}
{Yahya} S.,  {Bull} P.,  {Santos} M.~G.,  {Silva} M.,  {Maartens} R.,  {Okouma}
  P.,   {Bassett} B.,  2015, \mndoi [\mnras] {10.1093/mnras/stv695}, \href
  {http://adsabs.harvard.edu/abs/2015MNRAS.450.2251Y} {450, 2251}

\bibitem[\protect\citeauthoryear{{Zhang}, {Bunn}, {Karakci}, {Korotkov},
  {Sutter}, {Timbie}, {Tucker}  \& {Wandelt}}{{Zhang}
  et~al.}{2016}]{2016ApJS..222....3Z}
{Zhang} L.,  {Bunn} E.~F.,  {Karakci} A.,  {Korotkov} A.,  {Sutter} P.~M.,
  {Timbie} P.~T.,  {Tucker} G.~S.,   {Wandelt} B.~D.,  2016, \mndoi [\aaps]
  {10.3847/0067-0049/222/1/3}, \href
  {https://ui.adsabs.harvard.edu/#abs/2016ApJS..222....3Z} {222}

\bibitem[\protect\citeauthoryear{{Zuo}, {Chen}, {Ansari}  \& {Lu}}{{Zuo}
  et~al.}{2018}]{2018arXiv180104082Z}
{Zuo} S.,  {Chen} X.,  {Ansari} R.,   {Lu} Y.,  2018, preprint, \href
  {https://ui.adsabs.harvard.edu/#abs/2018arXiv180104082Z} {} (\mn@eprint
  {arXiv} {1801.04082})

\makeatother
\end{thebibliography}

\balance
\section*{Affiliations}
\affil{$^1$Institute of Cosmology \& Gravitation, University of Portsmouth, Dennis Sciama Building, Portsmouth PO1 3FX, UK} 
\affil{$^2$Jodrell Bank Centre for Astrophysics, School of Physics and Astronomy, The University of Manchester, 
Manchester M13 9PL, U.K.}
\affil{$^{3}$School of Physics \& Astronomy, Queen Mary University of London, 327 Mile End Road, London E1 4NS, U.K.}
\affil{$^4$Dipartimento di Fisica, Universit\'a degli Studi di Torino, Via P. Giuria 1, 10125 Torino, Italy}
\affil{$^5$INFN -- Istituto Nazionale di Fisica Nucleare, Sezione di Torino, Via P. Giuria 1, 10125 Torino, Italy}
\affil{$^6$INAF -- Osservatorio Astrofisico di Torino, Strada Osservatorio 20, 10025 Pino Torinese, Italy}
\affil{$^7$Department of Physics, University of Oxford, Denys Wilkinson Building, Keble Road, Oxford OX1 3RH, U.K.}
\affil{$^{8}$Korea Astronomy and Space Science Institute, Yuseong-gu, Daedeokdae-ro 776, Daejeon 34055, Korea}
\affil{$^{9}$Department of Physics \& Astronomy, University of the Western Cape, Cape Town 7535, South Africa}
\affil{$^{10}$SKA South Africa, The Park, Cape Town 7405, South Africa}
\affil{$^{11}$Instituto de Astrofisica e Ciencias do Espaco, Universidade de Lisboa, OAL, Tapada da Ajuda, PT1349-018 Lisboa, Portugal}
\affil{$^{12}$Astrophysics Group, School of Physics, The University of Melbourne, Parkville, VIC
 3010, Australia}
\affil{$^{13}$Department of Physics and Astronomy, University College London, Gower Street, London, WC1E 6BT, U.K.}
\affil{$^{14}$Department of Physics and Electronics, Rhodes University, PO Box 94, Grahamstown, 6140, South Africa}
\affil{$^{15}$Lorentz Institute for Theoretical Physics, Leiden University, P.O. Box 9506, 2300 RA Leiden, The Netherlands}
\affil{$^{16}$D\'{e}partement de Physique, \'{E}cole Normale Sup\'{e}rieure, PSL Research University, CNRS, 24 rue Lhomond, 75005 Paris, France}
\affil{$^{17}$Department of Mathematics \& Applied Mathematics, University of Cape Town, Cape Town 7701, South Africa}
\affil{$^{18}$INAF - Istituto di Radioastronomia, via Gobetti 101, 40129 Bologna, Italy}

\affil{$^{19}$Institut de Ci\`encies del Cosmos (ICCUB), Universitat de Barcelona (IEEC-UB), Mart\'{i} Franqu\`es 1, E08028 Barcelona, Spain}
\affil{$^{20}$Dept. de  F\' isica Qu\` antica i Astrof\' isica, Universitat de Barcelona, Mart\' i i Franqu\` es 1, E08028 Barcelona, Spain}
\affil{$^{21}$Argelander-Institut f\"ur Astronomie, Auf dem H\"ugel 71, 53121 Bonn, Germany}
\affil{$^{22}$SKA Organization, Lower Withington, Macclesfield, Cheshire SK11 9DL, U.K.}
\affil{$^{23}$D\`{e}partement de Physique Theorique and Center for Astroparticle Physics,, Universite de Gen\`{e}ve, 24 quai Ernest-Ansermet, 1211 Gen\`{e}ve 4, Switzerland}
\affil{$^{24}$Astrophysics Group, Imperial College London, Blackett Laboratory, Prince Consort Road, London, SW7 2AZ, United Kingdom}
\affil{$^{25}$National Astronomical Observatories, Chinese Academy of Sciences,
Beijing 100101, China}
\affil{$^{26}$Faculty of Engineering, Kanagawa University, 3-27-1, Rokkakubashi, Kanagawa-ku, Yokohama-shi, Kanagawa, 221-8686, Japan}
\affil{$^{27}$School of Mathematics and Physics, The University of Queensland, QLD 4072 Australia}
\affil{$^{28}$ASTRON, Netherlands Institute for Radio Astronomy, Postbus 2, 7990 AA, Dwingeloo, The Netherlands}
\affil{$^{29}$ETH Zurich, Wolfgang-Pauli-Strasse 27, CH 8093 Zurich, Switzerland}
\affil{$^{30}$Department of Physics and Astronomy, Uppsala University, SE-751 20 Uppsala, Sweden}
\affil{$^{31}$Fakult\"at f\"ur Physik, Universit\"at Bielefeld, Postfach 100131, 33501 Bielefeld, Germany}
\affil{$^{32}$SISSA, International School for Advanced Studies, Via Bonomea 265, 34136 Trieste
 TS, Italy}
\affil{$^{33}$Center for Computational Astrophysics, 162 5th Avenue, 10010 New York, NY, USA}
\affil{$^{34}$Faculty of Engineering, Kanagawa University, Kanagawa, 221-8686, Japan}
\affil{$^{35}$Institute for Astronomy, University of Edinburgh, Blackford Hill, Edinburgh EH9 3HJ, UK}
\affil{$^{36}$Dipartimento di Fisica e Astronomia ``G. Galilei'', Universit\'a degli Studi di Padova, Via Marzolo 8, 35131 Padova, Italy}

\end{document}